\documentclass[fleqn,usenatbib,onecolumn]{mnras}

\usepackage{newtxtext,newtxmath}

\usepackage[T1]{fontenc}
\usepackage{ae,aecompl}

\usepackage{graphicx}	
\usepackage{amsmath}	
\usepackage{amssymb}	

\usepackage{ulem}
\usepackage{pifont}
\usepackage{amsfonts}
\usepackage{mathrsfs}
\usepackage{gensymb}
\usepackage[dvipsnames]{xcolor}
\usepackage{enumitem}
\usepackage{mathtools}
\setenumerate{listparindent=\parindent}
\usepackage{xspace}
\DeclareGraphicsExtensions{.pdf,.png}
\graphicspath{{.}}
\usepackage{soul}
\usepackage{ulem}

\newcommand{\Req}{R_{\textrm{eq}}}

\newcommand{\from}{\colon}

\title[On parametrised equation of state inference]{On parametrised cold dense matter equation of state inference}

\author[T. E. Riley et al.]{
Thomas E. Riley,\thanks{E-mail: T.E.Riley@uva.nl}
Geert Raaijmakers,
Anna L. Watts
\\
Anton Pannekoek Institute for Astronomy, University of Amsterdam, Science Park 904, 1098 XH Amsterdam, The Netherlands\\
}

\date{Accepted 2018 April 23. Received 2018 April 17; in original form 2018 January 17}

\pubyear{2017}

\begin{document}
\label{firstpage}
\pagerange{\pageref{firstpage}--\pageref{lastpage}}
\maketitle

\begin{abstract}
Constraining the equation of state of cold dense matter in compact stars is a major science goal for observing programmes being conducted using X-ray, radio, and gravitational wave telescopes. We discuss Bayesian hierarchical inference of parametrised dense matter equations of state. In particular we generalise and examine two inference paradigms from the literature: (i) direct posterior equation of state parameter estimation, conditioned on observations of a set of rotating compact stars; and (ii) indirect parameter estimation, via transformation of an intermediary joint posterior distribution of exterior spacetime parameters (such as gravitational masses and coordinate equatorial radii). We conclude that the former paradigm is not only tractable for large-scale analyses, but is principled and flexible from a Bayesian perspective whilst the latter paradigm is not. The thematic problem of Bayesian prior definition emerges as the crux of the difference between these paradigms. The second paradigm should in general only be considered as an ill-defined approach to the problem of utilising archival posterior constraints on exterior spacetime parameters; we advocate for an alternative approach whereby such information is repurposed as an approximative likelihood function. We also discuss why conditioning on a piecewise-polytropic equation of state model -- currently standard in the field of dense matter study -- can easily violate conditions required for transformation of a probability density distribution between spaces of exterior (spacetime) and interior (source matter) parameters.
\end{abstract}




\section{Introduction}\label{sec:intro}
Compact stars are defined by extreme physical regimes which cannot be reproduced in terrestrial laboratories. Radiative compact star phenomena therefore provide a unique window onto fundamental physics \citep[for a recent reviews see][]{Miller2016,OzelReview}. The microphysical behaviour of matter under extreme conditions is poorly understood and difficult to model from first principles \citep[for recent reviews see][]{Hebeler15,Lattimer16,Chatterjee16}. Nevertheless, microphysical interactions manifest statistically in the thermodynamics of particle ensembles. Typically, compact stars are treated as uniformly rotating fluids bounded by a crust \citep[for a review see][]{Chamel2008}, whose local thermodynamic properties are described by some barotropic.\footnote{We limit the scope of the discussion in this work to compact stars whose interior temperatures and magnetic field strengths are effectively zero with respect to the stress-energy tensor. This is a reasonable approximation for most neutron stars, except in the immediate aftermath of a supernova or during the final phase of a neutron star merger, when finite temperature effects are important \citep[see][and references therein]{Lattimer16}} Equation of State (EOS) for hadronic, quark, or hybrid source matter. The EOS appears in the general relativistic stress-energy tensor: in principle one can therefore statistically infer microphysical and or thermodynamical properties of matter via its effect on global spacetime structure.

A modelling assumption we invoke throughout this work is that a single core EOS is \textit{shared} by all stars which are observed: it is plausible that the cores of all compact stars in the Universe are constituted by the same set of deconfined fundamental particles. Given our state of knowledge in the present, it is rational to invoke the assumption that a single EOS is shared globally over the density domain. Model EOS complexity can be incrementally increased over time, driven by informative data and analysis of model predictive performance \citep[][]{Gelman_book}. The crust composition, for instance, is in reality variable from star-to-star due to distinct accretion histories \citep[][]{Chamel2008}: whilst the core EOS could be defined as shared, the EOS of the crust could in principle be defined as \textit{unshared} (which may also support self-consistency when modelling radiative phenomena influenced by crusts of distinct stars). Furthermore, if multiple families of compact stars coexist in reality \citep[for recent papers on this topic see][]{Alford13,Drago2016a,Drago2016b,Alford17, Bhattacharyya2017}, then based on informative data and the predictive performance of simpler models, complexity could be increased by invoking multiple core EOSs (one per subpopulation), where it is unknown \textit{a priori} which subpopulation each observed star belongs to (thus requiring, e.g., the definition of both continuous and discrete parameters).

Constraining the dense matter EOS via astrophysical measurements of neutron stars is a major goal for current facilities: particularly for the X-ray telescope NICER \citep{Arzoumanian14} and the gravitational wave telescopes LIGO and VIRGO \citep{Abbott17}.  It is also a major science driver for the planned Square Kilometre Array radio telescope \citep{Watts15} and proposals for the next generation of large-area X-ray timing telescopes \citep{Watts16}, such as the {\it Enhanced X-ray Timing and Polarimetry mission} \citep[eXTP,][]{Zhang16} and the NASA Probe Study STROBE-X \citep{STROBEX}. The phenomena (to be) observed by these telescopes are distinct probes of the EOS. In this work we aim to remain general: whilst it is necessary to consider technique-specific spacetime modelling, we do not focus on ulterior details (such as, e.g., astrophysical and instrumental nuisance parameters) beyond offering several simple examples to improve clarity.

Consider isolated compact stars in equilibrium (stable, uniformly-rotating, axisymmetric, equatorially reflection symmetric) whose exterior spacetimes are stationary and asymptotically-flat \citep[][]{Misner1973}. General exact closed-form exterior solutions to the Einstein field equations\footnote{Throughout this work the continuity constraint is implicit in this system of equations.} which match smoothly and exactly to numerical interior solutions are not known for such objects \citep[see, e.g.,][]{Berti2004}. An observer may decompose an exterior spacetime solution into a parameter set $\mathscr{P}$ (of countably infinite cardinality) whose values uniquely relate manifold structure to the global distribution of source stress-energy \citep[see, e.g.,][]{Thorne1980}; the set $\mathscr{P}$ may be constructed from source multipole moments (e.g, the gravitational mass, the spin angular momentum, the spin-induced mass quadrupole, and so on) and parameters which describe the closed inner-boundary of the exterior domain over which the solution is valid (e.g., the coordinate equatorial radius, the surface coordinate ellipticity, and so on).

For inspiralling compact object binaries (containing at least one compact star) the exterior spacetime of the binary is non-stationary because the system is rotating and non-axisymmetric: gravitational wave radiation drives (secular) evolution of the exterior spacetime structure \citep[e.g.,][]{Misner1973}. Nevertheless, once finite-size effects become important to post-Newtonian wave generation, the relevant tidally-induced source moments can be calculated via an external quadrupolar tidal field perturbation on a stationary spacetime \citep[see][and references therein]{YagiYunes2017_tidal}. Indeed, the definition of the set $\mathscr{P}$ can be expanded to include global (wave) source properties such as binary tidal deformabilities; we direct the reader to \citet[][and references therein]{Lackey2015} for both a theoretical discussion of Bayesian EOS parameter estimation from a population of gravitational wave BNS inspiral events (including detector forecasts), and a discussion on computational practices. The reader can also refer, e.g., to the work on the recent aLIGO/Virgo BNS merger detection \citep[see][and references therein]{Abbott17}.

An observer may make measurements of photons and or material particles whose worldlines encode information about $\mathscr{P}$. Such worldlines can provide either a direct connection to the source (a past subset of the worldline is contained in the near-field region) or an indirect connection to the source (via, e.g., the photon-mediated manifestation of gravitational waves in non-stationary spacetimes, or Cherenkov radiation). Further, the observer may then statistically infer a subset of $\mathscr{P}$, conditioned on some informative set of such observations and all model assumptions (including, e.g., general relativistic gravity), as an indirect probe of microphysical processes which are manifest in the thermodynamic nature of the source matter.

Our (the authors') primary research focus is the modelling of (high-energy) electromagnetic radiation from compact stars which are not deformed by the external tidal field of another compact object and whose spacetimes are effectively stationary. For this reason we opt \textit{not} to discuss tidal deformabilities in any detail; this would require discussion on tidal heating during inspiral, and thus the validity of the assumption of a shared, cold EOS. By restricting ourselves to stars which are not deformed by the tidal fields of other compact objects, we lose generality but enable a simpler discussion.

The driving motivation for this work is the need for a theoretical generalisation of high-energy astrophysical EOS inference work in the literature, in particular that of: \citet[][]{Ozel2009, Steiner2010, SteinerLattimerBrown2013, LattimerSteiner2014, LattimerSteiner2014_LMXB, Ozel16, Nattila2016, Raithel2017, Alsing2017}. Collectively, these authors addressed the need to express knowledge of model parameters probabilistically, conditioned on observational data sets. However, there remains work to be done to rigorously understand the relations between various approaches from a probability-theoretic perspective. At its heart, this work contains an examination of the \textit{space} on which we should define a joint prior probability (density) distribution when sets of parameters defined within a theory (general relativistic gravity) are deterministically related. Crucially, the \citet[][]{Ozel2009} framework is based on post-processing an intermediary, (marginal) joint (posterior) probability density distribution of parameters which control analytic exterior spacetime solutions: continuous probabilistic knowledge of exterior spacetime parameters is transformed into continuous probabilistic knowledge of interior (EOS) parameters. In generalising the work of \citet[][]{Ozel2009} however, we encounter difficulties which imply that it is more principled to perform Bayesian EOS parameter estimation \textit{directly,} in a manner consistent with the later literature cited at the beginning of this paragraph. In this work we do \textit{not} discuss specific, direct choices of prior on the space of EOS parameters (e.g., in regards to sensitivity of inferences), for which the reader is directed to \citet{Steiner2016}.

The discussion in this work complements that of \citet[][]{Lackey2015}, to the extent that their framework can be straightforwardly and cleanly merged with the general \textit{Interior-Prior} EOS inference paradigm we outline in Sections \ref{sec:inference paradigms} and \ref{sec:model selection}. Together with \citet[][]{Lackey2015}, this work may help bridge the gap between the EOS inference work of gravitational wave community and the high-energy electromagnetic community. Common to both this work and \citet[][]{Lackey2015} is the choice to define a prior on a space of EOS parameters; by defining a prior on a space of such local parameters assumed to be shared by all observed sources, synergistic multi-messenger constraints may be derived.\footnote{Note that, as stated above, one important caveat regards the assumption of BNS components sharing the same cold EOS -- during inspiral -- as stars which do not exist in compact relativistic binaries. In principle, if there is evidence that model complexity needs to be increased, these two classes of star could be assumed to only share a subset of (continuous, phenomenological) EOS parameters, whilst another subset of EOS parameters are used to emulate temperature dependence.} For the binary components, \citet[][]{Lackey2015} then match numerical interior solutions and analytic exterior solutions to the tidally-perturbed Einstein field equations \citep[as described by, e.g.,][]{YagiYunes2013,YagiYunes2017_tidal}, enabling a wave-generation model for likelihood definition and evaluation; this is consistent with the ideas adopted in this work for direct EOS inference conditioned on electromagnetic data sets. Furthermore, for practicality given tested existing software, \citet[][]{Lackey2015} enable a two-phase parameter estimation process whereby a numerical (or approximative analytical) joint likelihood function is first computed on a space of exterior spacetime parameters (component gravitational masses and a linear combination of component tidal deformabilities) of a set of binaries; this function is subsequently used to evaluate a likelihood function on a space of EOS parameters during \textit{posterior} sampling. This two-phase process -- with a prior in part defined on a space of EOS parameters shared by all sources -- is a useful notion that appears in the high-energy literature \citep[see the dedicated review in Section~\ref{sec:Review} and the references therein, e.g.,][]{Steiner2010} and which is highlighted in more detail this work.

Bayesian statistical analyses in astrophysics often present formidable computational problems; supercomputers or large clusters are required to: apply generative models which are numerical or depend on a large number of parameters; analyse vast quantities of data; or more generically, to sample from highly-structured joint posterior distributions. Inference of an EOS model can be a remarkably resource intensive astrophysical application and this justifies a detailed discussion and review. Our treatment is deliberately general: in the main body of this work we condition on Bayesian statistics for modelling observable reality and general relativistic gravity, but in general do not discuss details of ulterior models (involving, e.g., instruments and particular astrophysical observables). Where appropriate, simple examples of models are given to improve clarity.

In Section \ref{sec:inference paradigms} we describe two paradigms for EOS inference, and in comparing these paradigms the thematic problem of Bayesian prior definition emerges as the crucial distinction. We also consider principles for large-scale collaborative analyses as a scientific community which are required to synergistically constrain the EOS. In Section \ref{sec:Review} we review a number of approaches to EOS inference presented in the literature, contextualising each with respect to the general paradigms detailed in Section \ref{sec:inference paradigms}.  We summarise in Section \ref{sec:Summary}.

\section{Bayesian Inference of a Parametrised Equation of State}\label{sec:inference paradigms}
\subsection{Executive summary}
In this section we distinguish two paradigms for formulating a marginal joint posterior probability distribution on some space $T$ orthogonally spanned by EOS parameters\footnote{We use bold font to denote both a set of abstract parameters, and a point in the associated parameter space. For instance, $\boldsymbol{\theta}\in T$ where $T\subset\mathbb{R}^{n}$, where the coordinates of point $\boldsymbol{\theta}$ represent parameter values.} $\boldsymbol{\theta}$ defined as Bayesian random variables. Throughout, let the parameters $\boldsymbol{\theta}$ be necessarily associated with some functional form for the EOS, $P=P(\varepsilon;\boldsymbol{\theta})$, where $P$ is the local comoving pressure and $\varepsilon$ the local comoving energy density.

First, in Section \ref{sec:implicit}, we define the \textit{Interior-Prior} paradigm for Bayesian inference: a posterior probability distribution is defined as an update to a prior probability distribution on a space of interior parameters (of a set of stars), and it is known on-the-fly (by definition) during posterior sampling which (exact or approximate) exterior spacetime solutions to the Einstein field equations are permitted conditioned on the EOS model -- those that \textit{match} to (exact or approximate) stable interior spacetime solutions. We define interior parameters as the joint set of EOS parameters, central densities, and angular rotation frequencies. We discuss the matching of interior and exterior solutions -- and the `exterior' parameters which control those exterior solutions -- in Section \ref{sec:solutions for generative models}.

Second, in Section~\ref{sec:explicit}, we define the \textit{Exterior-Prior} paradigm for Bayesian inference: an intermediary \textit{posterior} distribution of parameters controlling a set of analytic exterior spacetimes is transformed into a posterior distribution of parameters (as defined above) controlling the source interiors (and thus the global spacetime solution). The intermediary distribution is conditioned on some observational data set and may span, for instance, the joint space of gravitational masses, coordinate equatorial radii, and angular rotation frequencies of a model ensemble of stars (we give more detailed examples in Section \ref{sec:mapping definition}). In this case it is \textit{not} known during sampling of the posterior on exterior parameters whether a particular exterior spacetime is permitted conditioned on the EOS model -- i.e., whether an exterior spacetime \textit{matches} to an interior solution -- nor whether a particular \textit{ensemble} of exterior spacetimes is permitted conditioned on a \textit{shared} EOS.

\subsection{The Interior-Prior paradigm}\label{sec:implicit}
In this section we define and discuss the \textit{Interior-Prior} (IP) inference paradigm.

\subsubsection{Definition}\label{sec:IM definition}
For a given EOS, (stress-isotropic) equilibrium solutions to the Einstein field equations form a two-parameter family which can be naturally understood in terms of the central \textit{energy} density $\rho$ and the coordinate rotation frequency $\Omega$ \citep[][]{HT1968, Berti2004}. These local properties, together with (a functional form for) the EOS, everywhere determine the components of the source stress-energy tensor in general relativistic gravity via integration. Exterior spacetime solutions are controlled by this pair of parameters, and these solutions enter in parametrised sampling distributions (defined below) of observed radiation originating from systems hosting compact stars.

It is known that a continuous set of stable non-rotating interior solutions to the Einstein field equations parametrised, solely by central energy density $\rho$ for a fixed EOS $\boldsymbol{\theta}$, maps to a unique curve $M(R;\boldsymbol{\theta})$, where $M$ is the exterior Schwarzschild gravitational mass and $R$ is the circumferential stellar radius \citep[see, e.g.,][]{Misner1973,ST1983}. \citet{Lindblom92} describes an algorithm for numerically approximating the EOS given a discrete set of Schwarzschild solutions. In practice however, the \citet{Lindblom92} approach is not useful: there are fundamental problems to be addressed. As \citet{Lindblom92} notes, such a non-parametric inversion algorithm is not pragmatic: there does not exist an effectively infinite, observable set of stars which share an EOS. The accuracy of the resultant EOS given scalar representations of the mass and radius of each static star cannot, therefore, be perfect. Moreover, because the approach is non-parametric and not statistical, one cannot define the nature of the deviations of the derived EOS from reality.

From a Bayesian perspective \citep[refer, e.g., to][and references therein]{MacKay2003,Robert2007,Robert2009,Hogg2010,Gelman_philosophy,Gelman_book}, the exterior spacetime structure (e.g., the gravitational mass and circumferential radius) of a single (assumedly static) star is fundamentally probabilistic\footnote{Probabilistic knowledge is mathematically represented by a probability distribution (a scalar mass function or density field) operating on a parameter space.} because a data set acquired through radiation detection is a vector drawn randomly from some joint probability distribution. Stochastic processes are intrinsic to the physical system being observed (e.g., randomness in a source radiation field, such as Poisson noise for a thermal field) and intrinsic to imperfect radiation detection technologies (e.g., thermal noise). Further, one's probabilistic knowledge of the exterior spacetime structure is conditioned on a global model;\footnote{A global model is constituted by a generative model (a parametrised sampling distribution of some data set, requiring parameters controlling both deterministic and stochastic processes which describe data generation) and priors (including assumptions about parameter hierarchies).} there may exist however, a space populated by such models, each with a finite associated prior probability. Moreover, to a Bayesian, the model parameters themselves are random variables distributed according to some joint prior probability distribution which is to be updated given some data set. Finally, astrophysical stars rotate and thus a statistical inference framework is required which accounts for rotation.

Hereafter we adopt a Bayesian perspective.\footnote{Generally without favouring objectivism or subjectivism. We more closely adopt the view of a generative model as a set of assumptions defined by the union of the prior and the likelihood, and that a generative model is to be rigorously tested via posterior predictive checking, and developed if there are grounds to falsify it.} The marginal joint posterior probability distribution on the space of EOS parameters may be written via Bayes' theorem as
\begin{equation}
\mathcal{P}\left(\boldsymbol{\theta}\;|\;\mathcal{D},\mathcal{M},\mathcal{I}\right)
\propto
\mathop{\int}
\underbrace{\mathcal{P}\left(\mathcal{D}\;|\;\boldsymbol{\theta},\boldsymbol{\rho},\boldsymbol{\Omega},\boldsymbol{\eta},\mathcal{M}\right)}_{\mathcal{L}(\boldsymbol{\theta},\boldsymbol{\rho},\boldsymbol{\Omega},\boldsymbol{\eta})}
\underbrace{\mathcal{P}\left(\boldsymbol{\theta},\boldsymbol{\rho},\boldsymbol{\Omega},\boldsymbol{\alpha},\boldsymbol{\eta},\boldsymbol{\beta}\;|\;\mathcal{M},\mathcal{I}\right)}_{\pi(\boldsymbol{\theta},\boldsymbol{\rho},\boldsymbol{\Omega},\boldsymbol{\alpha},\boldsymbol{\eta},\boldsymbol{\beta})}
d\boldsymbol{\eta}d\boldsymbol{\beta}d\boldsymbol{\rho}d\boldsymbol{\Omega}d\boldsymbol{\alpha},
\label{eqn:exact posterior}
\end{equation}
where: $\mathcal{M}$ is a global hierarchical generative model; $\boldsymbol{\rho}$ and $\boldsymbol{\Omega}$ are respectively the central densities and coordinate (or asymptotic) angular rotation frequencies of stars belonging to some model ensemble $\mathcal{E}$; $\boldsymbol{\alpha}$ and $\boldsymbol{\beta}$ both denote sets of parameters which control population-level distributions of random variables (it is convention to identify such parameters as \textit{hyperparameters}); $\boldsymbol{\alpha}$ denotes a set of hyperparameters controlling the joint prior probability density distribution of $(\rho,\Omega)$ for every star;\footnote{A notable difference between this work and \citet{Lackey2015} is that we focus on defining a hierarchical population-level EOS-conditional prior on the joint space central density and rotation frequency, where stability conditions impose where the prior density falls to zero. \citet[][]{Lackey2015} however, define for subsets of sources, non-hierarchical population-level priors on the joint space of binary component masses; these priors are not conditional on the EOS and thus assign finite density to unstable solutions. In Section~\ref{sec:Review} we revisit the choice of central density priors versus mass priors in the context of high-energy inference. Stars may be considered as belonging to distinct subpopulations, however, with different physical formation channels and observational selection effects; for example, we may distinguish between subpopulations in terms of the observational phenomena by which they are detected and by which they are applied to EOS inference. To a Bayesian, subpopulations may each be associated with a distinct joint parameter distribution, whereby the global (intrinsic and observed) population is represented by a weighted mixture model. Here we merely note that when synergistically combining EOS constraints from the gravitational wave and (high-energy) electromagnetic communities there also needs to be discussion on this aspect of the prior model.} $\boldsymbol{\eta}$ denotes a set of nuisance parameters whose joint prior probability distribution is parametrised by the nuisance hyperparameters $\boldsymbol{\beta}$; $\mathcal{D}$ denotes some observational data set; and $\mathcal{I}$ denotes Bayesian prior \textit{information} (a union of data sets) conditioned on in preceding analyses to calculate the global joint (hyper)prior  probability distribution $\pi(\boldsymbol{\theta},\boldsymbol{\rho},\boldsymbol{\Omega},\boldsymbol{\alpha},\boldsymbol{\eta},\boldsymbol{\beta})\coloneq\mathcal{P}\left(\boldsymbol{\theta},\boldsymbol{\rho},\boldsymbol{\Omega},\boldsymbol{\alpha},\boldsymbol{\eta},\boldsymbol{\beta}\;|\;\mathcal{M},\mathcal{I}\right)$ on (hyper)parameter space.

We use the term \textit{model stars} (or stars in a \textit{model ensemble}) here to remind the reader that we are considering objects which exist in some abstract model and are associated with some set of model parameters. These mathematical compact stars are assumed to be the \textit{real} sources of detected radiation (which constitutes $\mathcal{D}$) on our celestial sphere.

From a Bayesian perspective, the parameters defined in a theory to model observations of a system are random variables distributed according to some joint probability distribution. Suppose we condition on observations of an \textit{ensemble} of stars: fundamental physical theories (such as general relativistic gravity) are invoked to model observational phenomena across that ensemble, and therefore each star may be associated with random variables distributed according to a population-level (or hierarchical) prior. Further, we may parametrise a prior with hyperparameters (which may or may not be nuisances); the hyperparameters also require a prior, termed a \textit{hyperprior}. We may then attempt to learn \textit{a posteriori} both the system parameters and population hyperparameters by sampling from a posterior on the (hyper)parameter space. The (hyper)prior is a necessary model component, but can be somewhat subjective in nature and a principal argument against Bayesianism \citep[e.g.,][]{Robert2007,Hogg2010}. Parametrising the prior distribution of interior conditions (e.g, central densities and rotation frequencies): (i) has physical implications in terms of choosing a prior consistently for all observed stars comprising the ensemble; and (ii) in principle increases prior robustness when hyperparameters are marginalised over \citep[e.g., Chapter 3 of][]{Robert2007}. The model $\mathcal{M}$ is hierarchical if one defines hyperparameters and their associated hyperprior distributions, not because one imposes that the EOS is shared by all stars belonging to $\mathcal{E}$.

To define the global model $\mathcal{M}$, one requires a generative model which describes the statistical properties of $\mathcal{D}$ \citep[e.g.,][]{Hogg2010}: abstractly, a generative model is any parametrised (conditional) joint probability distribution (on the space of the data) from which the set $\mathcal{D}$ (a vector) is assumed to be randomly drawn -- i.e., a conditional sampling distribution. A conditional sampling distribution evaluated at $\mathcal{D}$ is the likelihood $\mathcal{L}(\boldsymbol{\theta},\boldsymbol{\rho},\boldsymbol{\Omega},\boldsymbol{\eta})\coloneq\mathcal{P}\left(\mathcal{D}\;|\;\boldsymbol{\theta},\boldsymbol{\rho},\boldsymbol{\Omega},\boldsymbol{\eta},\mathcal{M}\right)$; the likelihood \textit{function} is a scalar field on parameter space. Equation~(\ref{eqn:exact posterior}) thus represents an updated joint (hyper)prior probability distribution on the space of all model parameters, given a parametrised generative model which defines a distribution of probability (densities or masses or both) on the joint space of the data $\mathcal{D}$.

The process of mapping the distribution of posterior probabilities $\mathcal{P}\left(\boldsymbol{\theta}\;|\;\mathcal{D},\mathcal{M},\mathcal{I}\right)$ on parameter space may be referred to as EOS \textit{parameter estimation}.\footnote{Alternatively, one might choose to express this process in terms of EOS \textit{model comparison} on a continuous model space.} More generally, one might describe the process as EOS \textit{inference}, especially if there exists a discrete set of EOS models which cannot be continuously related on a space. In Section \ref{sec:solutions for generative models} though Section \ref{sec:computational tractability} we develop the notions introduced above. In Appendix \ref{sec:appendix example} we offer an example of a model $\mathcal{M}$; we recommend that the reader uses this Appendix if the general definitions of model properties adopted in main text are deemed insufficient. Appendix \ref{sec:appendix example} may be especially useful for readers who model observations of high-energy electromagnetic radiation from compact stars.

\subsubsection{Exterior spacetime solutions for modelling observed radiation}\label{sec:solutions for generative models}
We now remark on a general requirement for construction of a model of observable (electromagnetic) radiation originating from the near-field vicinity of a compact star. In order to model the radiation incident on a detector, one is usually interested in  a suitable \textit{analytic} exterior spacetime solution which may be an approximate or exact solution to the Einstein field equations. Given such a solution and an appropriate coordinate chart, one can compute (exact or approximate) null (and timelike) geodesics in vacuum\footnote{Or in some ambient plasma \citep[see, e.g.,][]{Rogers2015,Rogers2016}.} and invoke a geometrical optics approximation for atmospheric radiation transfer and propagation to a distant observer \citep[see, e.g.,][]{Misner1973,Pechenick1983,Schneider1992,Miller98,Beloborodov2002,Poutanen2003,Bhattacharyya2005,Cadeau2005,Cadeau2007,Poutanen2006,Morsink2007,Agol2009,Bauboeck2012,Psaltis2014,Nattila2017}. Such a procedure is a critical component of any model for (high-energy) electromagnetic radiation: it is required for parametrised calculation of statistical properties of the incident radiation field (e.g., the spectral \textit{mean} photon flux, where the photon flux itself is stochastic).

Finding an analytic exterior solution and determining its suitability for a particular application, on the other hand, is a non-trivial problem. If one's aim is to construct a generative model using a parametrised exterior solution, one requires this solution to be realistic in the sense that it should meaningfully match to an interior solution to the field equations given local source properties \citep[see, e.g.,][]{Berti2004}. Otherwise, one cannot sensibly transform parameters of an EOS model into exterior spacetime parameters for the purpose of parameter estimation.

A posterior sampling process applied to Equation~(\ref{eqn:exact posterior}) does not in principle require an \textit{analytic} solution: numerical solutions to the exact field equations may be computed via 3+1 numerical relativity \citep[using some open-source code library such as \textsc{RNS},][]{Stergioulas1995}, and the relevant geodesics in turn computed from numerical data for components of the exterior metric \citep[see, e.g.,][and references therein]{Vincent2018}. In principle, a numerical exterior solution can be decomposed via an approximate, truncated series expansion whose coefficients are multipole moments \citep[][]{Thorne1980,Berti2004}: this can for example be achieved in principle via the \citet[][]{Ryan1995} extreme mass-ratio inspiral approach to evaluation of the Geroch-Hansen moments \citep{Hansen1974} and thus, equivalently, the \citet{Thorne1980} ACMC moments \citep[see][and references therein]{Cardoso2016}. A multipolar decomposition is not necessary if the model is not parametrised in terms of global spacetime properties appearing in an exterior solution, but is instead parametrised in terms of local source properties from which exact numerical solutions are computed. Any parameter estimation analysis using numerical solutions to the exact field equations is likely to be computationally prohibitive at present; fortunately, the exact field equations are certainly not necessary to make progress on EOS inference given observations of rotating stars.

There do exist exact \textit{closed-form} exterior solutions, such as the \citet[][]{Manko2000} solution which is the focus of \citet[][]{Berti2004}. As \citet[][]{Berti2004} comment however, the solution is not generally applicable because it does not continuously reduce to the Schwarzschild solution in the limit of zero rotation, but to non-rotating solutions with finite mass quadrupole moment due for instance to anisotropic source matter stresses. Whilst \citet[][]{Berti2004} show applicability of the solution to rapidly rotating numerical spacetimes \citep[those whose rotation-induced mass quadrupole moment exceeds the non-rotating quadrupole moment of the solution of][]{Manko2000}, it is not useful for slowly-rotating stars if one strictly conditions on stress-isotropic source matter, or one assigns finite \textit{a priori} support (on the space of interior parameters) to isotropy when defining the global model $\mathcal{M}$.

The physical and mathematical details of multipolar decompositions in general relativity are not fundamentally necessary to understand the arguments presented in this work; we only require the following axiom. \textit{Given the local properties of \textit{a rotating source in equilibrium}, there exists some unique, discrete representation of the corresponding exact asymptotically-flat stationary vacuum solution to the Einstein field equations, whose elements manifest in asymptotic spatial series expansions of metric components.}

\citet{Hartle1967} derived an analytic exterior solution to the Einstein field equations using a perturbative technique \citep[see also][]{HT1968}. This solution is pragmatic for the purpose of EOS inference, and we therefore discuss it in more detail. \citet[][]{Hartle1967} imposes axisymmetric spatial expansions of the metric components at second-order in (a small dimensionless) rotation frequency and then integrates the resulting ordinary first-order linear differential system of perturbation equations given a central density $\rho$ and an angular rotation frequency $\Omega$.\footnote{Strictly, the perturbative field equations are integrated in practice by specifying an arbitrary initial condition at the centre of the star for the difference between the local rate of rotation of inertial frames, $\omega$, and the angular rotation frequency, $\Omega$. Only at the surface is the value of $\Omega$ calculated via matching to the analytic form of the exterior solution. The differential equations for the perturbations are linear and thus straightforwardly scalable with respect to powers of $\Omega$ for a fixed central density and EOS. For details refer to \citet[][]{Hartle1967} and \citet[][]{HT1968}.} The exterior solution is found in terms of the non-zero rotational metric and surface deformations up to quadrupole order, and is exact at this order, whilst all higher deformations vanish; the solution is thus equivalent to a truncation of the exact exterior solution for an isolated star rotating slowly. The exterior parameters are: the rotationally perturbed gravitational mass (monopole moment) $M$; the rotationally perturbed coordinate equatorial radius $\Req$ (of the stellar surface, where the pressure of source matter is effectively zero); the rotational angular momentum\footnote{In terms of which a moment of inertia $I$ can be defined \citep[see, e.g.,][]{Hartle1967}.} (or mass-current dipole moment) $J$; the mass quadrupole moment $Q$, a second-order rotational moment; and the boundary coordinate ellipticity $e$. Imposing regularity of a numerical interior solution at the centre of the star, and matching at the surface to an analytic exterior solution which is asymptotically flat, yields values for the exterior spacetime parameters provided the interior solution is stable \citep{Hartle1967,YagiYunes2013,YagiYunes2017}.

The parameters $M$, $\Req$, $J$, $Q$, and $e$ may be sufficient to describe the exterior spacetimes of stable rotating stars: higher-order rotational metric deformations decay increasingly rapidly with the radial coordinate of the exterior field point, and although the deformation of the boundary (of the source matter distribution) scales in magnitude with the angular rotational frequency ${\Omega}$, typical astrophysical stars are observed to rotate sufficiently slowly for higher-order spacetime structure to be neglected. That is, for a star with a given EOS and central energy-density, there exists an approximate limiting (Keplerian) frequency $\Omega^2\sim GM/3R^3$ (where $M$ and $R$ are respectively the Schwarzschild gravitational mass and circumferential radius of a non-rotating star with the same central density) which delimits the domain of stability \citep[see, e.g.,][]{Haensel2009}; the slow-rotation regime is defined by non-relativistic ($R\Omega\ll c$) rotation of source matter below this limit \citep[][]{Hartle1967}. Perturbative solutions become less accurate at second-order if rotation is rapid \citep[those which do not rotate \textit{slowly},][]{Hartle1967} because third-order metric structure is non-negligible.

One can construct a generative model explicitly in terms of local source properties. Integration of the interior solution to the perturbation equations is relatively inexpensive because the equations form a coupled ordinary first-order linear system \citep{Hartle1967}. Although numerical integration of the interior perturbation equations is fast, we are ultimately interested in calculating properties of a radiation field incident on a telescope so as to evaluate a likelihood function (Section \ref{sec:IM definition}). A matched stable analytic exterior solution permits faster geodesic computation than do numerical exterior solutions because computation of the affine connection requires numerical post-processing of metric data \citep[as in, e.g.,][]{GYOTO, Vincent2012}. Moreover, an approximate slow-rotation symmetry can be exploited \citep[see][and references therein]{Nattila2017}, which is useful because (CPU-bound) null geodesic computation is expensive for non-Kerr spacetimes due to the coupled system of ordinary non-linear second-order differential geodesic equations \citep[see, e.g.,][]{Agol2009,Chan2013}.

The validity of truncation at low-order in rotation frequency \citep[][]{Hartle1967,HT1968} will in general depend on: (i) how \textit{informative} a given data set is; and (ii) whether any statistical exterior-parameter bias which arises from neglecting higher-order rotational effects is negligible relative to biases incurred by other modelling inaccuracies. It is plausible that modelling inaccuracies (other than rotational truncation) will dominate the statistical effects of conditioning on the slow-rotation approximation for many realistic systems, albeit such a statement will always require explicit proof in practice. As an example, \citet[][and references therein]{Vincent2018} discuss the importance of at least invoking the slow-rotation approximation, but also the importance of modelling X-ray observations of bursting sources with numerical, directional atmospheres (which are weakly sensitive to surface effective gravity); beyond this, however, inaccuracies are inherent to, e.g., the modelling of radiation from the near-vicinities of such sources, and in the phase-folded modelling of the dynamic surface radiation fields \citep[see, e.g., the reviews of][]{Watts16,Miller2016,OzelReview}.

\subsubsection{Practical interior-exterior parameter mappings}\label{sec:mapping definition}
Let us first define a vector\footnote{Or, interchangeably, a point in a space of parameters defined as random variables.} of random variables $\boldsymbol{y}=(\boldsymbol{\theta},\boldsymbol{\rho},\boldsymbol{\Omega})$ such that $\boldsymbol{y}\in Y$, where $Y\subset\mathbb{R}^{n+2s}$, $n$ is the number of EOS parameters $\boldsymbol{\theta}\in T$, and $T\subset\mathbb{R}^{n}$. The points $\boldsymbol{\rho}$ and $\boldsymbol{\Omega}$ exist in $s$-dimensional subspaces of $\mathbb{R}^{n+2s}$ (where $s$ is the cardinality of the ensemble $\mathcal{E}$, i.e., the number of stars).

Now let us define a parameter vector $\boldsymbol{x}\in X$, where $X\subset\mathbb{R}^{d}$, as values assumed by a set (or ensemble) of exterior spacetime parameters (the number of exterior parameters is thus $d\in\mathbb{N}$). The exterior parameters of each star form a subset of the set $\mathscr{P}$ introduced in Section \ref{sec:intro}: axisymmetric rotating exterior spacetime solutions to the Einstein field equations for compact stars do not in general exhibit closed forms, but can be decomposed into a unique (countably infinite) set of (scalar) multipole moments. The action of solving (approximations to) the field equations -- via numerical methods -- given a model EOS shared by all stars belonging a model ensemble $\mathcal{E}$, manifests as a map $\boldsymbol{y}\mapsto\boldsymbol{x}$ requiring a matching procedure between interior and exterior spacetime solutions. Let us define this map as ${f}\from Y\to X$, $\boldsymbol{y}\mapsto\boldsymbol{x}$; further, let us identify $Y$ as the \textit{domain} (of ${f}$), and $X$ the \textit{image} (of $Y$ under ${f}$) which is \textit{equal by definition} to the codomain for map ${f}$. If point $\boldsymbol{x}$  exists in the (general relativistic) image of $Y$ under ${f}$, it can be said to represent an ensemble of stable (rotating) solutions to the field equations conditioned on the shared model EOS.

In theory, unconstrained by both stochasticism in an incident radiation field and technological limitations (both in terms of compute resources and radiation detection systems), we could define $X\subset\mathbb{R}^{\infty}$ for the purpose of generative modelling.\footnote{That is, all elements of $X$ are ordered countably infinite tuples, but each coordinate is restricted to a compact subdomain of $\mathbb{R}$, and a tuple of coordinates is a point in $\mathbb{R}^{\infty}$.} We should loosely appreciate that (without imposing a spherically symmetric and thus closed-form solution) the image of a point $\boldsymbol{y}\in\mathbb{R}^{n+2}$ -- where $n$ is finite -- can exist in an infinite-dimensional codomain (representing a non-closed-form solution). This is because the map itself contains information in the form of a continuum of pairs of thermodynamic state variables which are integrated over, and the $n$ parameters of the functional form are a (finite) set of scalars which control this set of pairs. In other words, for some finite degree of rotation, ${f}(\boldsymbol{y})$ can -- depending on the method of integration -- generate a countably infinite tuple of scalars. Provided that the map $\boldsymbol{y}\mapsto\boldsymbol{x}$ is invertible, then for some fixed EOS $\boldsymbol{\theta}$ and rotation frequency ${\Omega}$, the central density $\rho$ parametrises a continuous one-dimensional set of points in $\mathbb{R}^{\infty}$; moreover, $\rho$ and ${\Omega}$ can be said to parametrise a 2-surface of unique\footnote{Although not \textit{necessary}, we here used the axiom from Section~\ref{sec:solutions for generative models} regarding the uniqueness of rotating exterior spacetime solutions given local interior source matter properties (an EOS \textit{function}) and conditions (central density and rotation frequency). We also required that the image in the EOS \textit{function}-space of any point $\boldsymbol{\theta}\in T$ is not the image of any other point in $T\subset\mathbb{R}^{n}$, such that the mapping between EOS parameter space and function space is \textit{invertible}. These conditions can be trivially relaxed in the IP-paradigm, but are critically relevant to the EP-paradigm (as discussed in Section~\ref{sec:explicit} and Appendix~\ref{app:pathologies}).} rotating single-star solutions in $\mathbb{R}^{\infty}$ for some fixed EOS \citep[see, e.g.,][]{HT1968,Berti2004}. We then have $n$ degrees of freedom in the space of EOS functions, meaning that an $(n+2)$-dimensional surface is generated in $\mathbb{R}^{\infty}$.\footnote{The field-equation mapping is here considered continuous in the sense that an open neighbourhood of a point in $\mathbb{R}^{n+2}$ is a subset of an open neighbourhood $\mathcal{U}$ of the image of that point in $\mathbb{R}^{\infty}$, where $\mathcal{U}$ is restricted to the $(n+2)$-dimensional surface (of stable and unstable solutions) embedded in $\mathbb{R}^{\infty}$.}

We now consider two practical interior-exterior mappings: (i) a model ensemble $\mathcal{E}\coloneqq\mathcal{R}$ of slowly-rotating spacetimes \citep[][]{Hartle1967} such that $\boldsymbol{y}=(\boldsymbol{\theta},\boldsymbol{\rho},\boldsymbol{{\Omega}})$ and the perturbative solutions to $\boldsymbol{x}={f}(\boldsymbol{y})$ are points $\boldsymbol{x}=(\boldsymbol{M},\boldsymbol{R}_{\textrm{eq}},\boldsymbol{J},\boldsymbol{Q},\boldsymbol{e})$; and (ii) a model ensemble $\mathcal{E}\coloneqq\mathcal{S}$ of static spacetimes such that $\boldsymbol{y}=(\boldsymbol{\theta},\boldsymbol{\rho})$ and the solutions to $\boldsymbol{x}={f}(\boldsymbol{y})$ are points $\boldsymbol{x}=(\boldsymbol{M},\boldsymbol{R})$. We note that the second case is the $\boldsymbol{\Omega}=\boldsymbol{0}$ limit of the first case (at least for stress-isotropic source matter) because the rotational perturbations to the gravitational mass and surface vanish ($\boldsymbol{e}\to\boldsymbol{0}$, $\boldsymbol{R}_{\textrm{eq}}\to\boldsymbol{R}$), and the first- and second-order multipole moments (metric perturbations) vanish ($\boldsymbol{J},\boldsymbol{Q}\to\boldsymbol{0}$).

The fundamental reason to model observations of multiple stars for EOS inference is that messenger radiation may encode insufficient mineable information on the deterministic signature of rotational metric and surface deformations (e.g., the angular momentum, the mass quadrupole, and the surface ellipticity) of an exterior solution away from spherical symmetry. Parametrised sampling distributions on the space of data (see Section \ref{sec:IM definition} for definitions) are often sensitive to nuisance parameters but relatively insensitive to exterior spacetime structure. That is, the likelihood function on a (sub)space of exterior parameters may only be weakly informative due to stochasticism (in the incident radiation field incident and in the detection process itself), and said stochasticism may primarily be controlled by nuisance parameters, not exterior spacetime parameters. Even signatures of zeroth-order parameters (such as the rotationally unperturbed gravitational mass and circumferential radius) on observables can be dominated by stochasticism if observations are not sufficiently informative, which can be the case if, e.g., a submodel with high predictive complexity is invoked in order to treat nuisance systematics.

Further, rotating exterior spacetime solutions have also been proven to exhibit \textit{approximate} universal behaviours \citep{Morsink2007,YagiYunes2013,Bauboeck2013,SteinYagi2014,YagiStein2014,AlGendy2014,YagiYunes2017_tidal,YagiYunes2017}. Universalities are emergent from approximate interior symmetries \citep[][]{YagiStein2014}; for isolated\footnote{There exist other approximate universalities, such as the universal behaviour of the tidal Love number, a deformation parameter relevant to compact object binary inspiral when extended-source effects become important \citep[][]{YagiYunes2017_tidal}, but before the source matter experiences appreciable tidal heating. For isolated compact stars however -- including those in binaries with non-compact stars whose gravitational fields are by definition weak -- tidal deformation will be entirely negligible.} rotating stars these symmetries manifest in part as approximate relations for the rotational metric (e.g., the $J$ and $Q$ multipole moments) and surface deformations (e.g., the surface boundary ellipticity, $e$) parametrised in terms of the (rotationally perturbed) gravitational mass $M$, the coordinate equatorial radius $\Req$, and the coordinate angular rotation frequency $\Omega$ \citep[e.g.,][]{AlGendy2014}. Whilst the EOS controls the continuous sequence(s) of stable $(M,\Req)$-pairs for some ${\Omega}$, the aforementioned authors have demonstrated that in the slow-rotation regime \citep[and beyond,][]{YagiYunes2013}, and within certain families of EOS, the rotational metric and surface deformations of a stationary spacetime solution are only weakly sensitive to the EOS beyond dependence on ${\Omega}$ and relation to both $M$ and $\Req$. Thus if we are to statistically distinguish -- conditioned on observations of a single star -- between multi-parameter EOSs whose exterior solutions are (for certain densities and rotation frequencies) degenerate with respect to $M$ and $\Req$ (see Appendix~\ref{app:pathologies} for a related discussion), and which exhibit \textit{approximately} universal metric and surface deformations (e.g., $J$, $Q$, and $e$), the likelihood function would need to be \textit{highly} sensitive to higher-order variations of the exterior spacetime. For single-parameter EOS models, if the single-star rotation frequency also enters in a likelihood function independently of controlling $M$ and $R_{\textrm{eq}}$, and is tightly constrained, degeneracy between interior parameters is effectively broken.

Computational limitations (e.g., likelihood evaluation speed in general determines the timescale on which samplers converge to a posterior distribution with appropriate properties for application of that sampling algorithm) and theoretical limitations (e.g., in terms of models for phenomena on the star or in its near vicinity which affect the statistical properties of the incident radiation field) support the notion that it may be impractical for us apply rotational \textit{metric} deformations of analytical exterior spacetime solutions away from spherical symmetry. Nevertheless, we opine that if likelihood functions \textit{are} somewhat sensitive to rotational truncation -- i.e., sensitive to whether or not rotational metric perturbations of some order are neglected -- but are insensitive to small variations in those deformations (as required to statistically distinguish between EOS which exhibit almost universal deformations), a (marginal) posterior distribution of interior parameters may be biased under rotational spacetime truncation.

The outermost source matter (the crust, exterior to which the ocean and atmosphere exist) exhibits the greatest susceptibility to rotational deformation, but the matter in the so-called outer-core dominates rotational metric deformation of the exterior solution away from spherical symmetry \citep[][]{YagiStein2014}. If a given likelihood function is far more sensitive to inclusion of a rotationally deformed surface than to exterior rotational metric perturbations, it is pragmatic to include surface deformation but neglect the metric perturbations \citep[][]{Morsink2007}, especially if likelihood evaluation is slow when $\ell>0$ metric perturbations are included. Therefore, in pursuit of pragmatism, let us consider a truncation to the first case stated above \citep[for details on terminology refer to][]{Hartle1967}: neglect rotational $\ell=1$ and $\ell=2$ \textit{metric} perturbations (multipole moments $\boldsymbol{J}$ and $\boldsymbol{Q}$ respectively), but feed rotational $\ell=0$ metric perturbations (to $\boldsymbol{M}$) and $\ell=2$ \textit{surface} deformations ($\boldsymbol{R}_{\textrm{eq}}$ and $\boldsymbol{e}$) to some (null) geodesic calculator (given an appropriate coordinate transformation if required). In other words, \textit{embed} a rotationally deformed 2-surface in a spherically symmetric ambient spacetime \citep[][]{Morsink2007}.\footnote{Note that the coordinate rotation frequencies hold the same asymptotic meaning for static and asymptotically flat stationary spacetime solutions.} The exterior solutions are thus reduced to dependence on $\boldsymbol{x}=(\boldsymbol{M},\boldsymbol{R}_{\textrm{eq}},\boldsymbol{e})$. To be clear, it is necessary to define \textit{truncation} for both the metric and the surface in terms of the highest rotational order $\Omega^{r}$, and the highest multipolar order, $\ell\leq r$, where for the metric $\ell\in\mathbb{N}$ but for the surface only $\{\ell=2k\from k\in\mathbb{N}\}$ do not vanish. Immediately above, the truncation of the \textit{metric} is $(r,\ell)=(2,0)$, whilst the surface truncation is the usual slow-rotation $(r,\ell)=(2,2)$.

The rotation frequencies\footnote{The local coordinate rotation frequency -- expressed with respect to a global spacetime foliation adapted to stationarity and axisymmetry -- of source matter is assumed to be everywhere uniform.} $\boldsymbol{\Omega}$ may also be defined as exterior parameters -- i.e., as a subset of the parameters $\boldsymbol{x}$. The rotation frequency of a star (e.g., a pulsar) may be statistically inferred from (electromagnetic) time-domain information: a mode of periodicity in a signal may be identified as (or related to) the rotation frequency, given some model for frequency modulation due to orbital dynamics of both star and detector. In this case the exterior solutions could be considered to depend on $\boldsymbol{x}=(\boldsymbol{\Omega},\boldsymbol{M},\boldsymbol{R}_{\textrm{eq}},\boldsymbol{e})$, where each rotation frequency is also required compute the metric and surface given an EOS and central density.

In summary, detailed modelling of the exterior spacetime of a single rotating star will not in general yield high statistical constraining power on higher-order spacetime structure (and thus on interior parameters). Instead it is practical to: (i) solve the field equations perturbatively to reduce the computational load; and (ii) define $\boldsymbol{x}$ as a union -- over the ensemble -- of parameters appearing in the perturbative exterior solutions, where for each star the solution is truncated (with respect to rotational metric and surface deformations). The degree of truncation need not be identical for all stars nor be identical for all submodels invoked to describe (assumedly) independent observations of a given star.

We now briefly consider modelling an ensemble $\mathcal{S}$ of static spacetimes as a limiting case of the model ensemble of rotating spacetimes described above. In reality, all stars should exhibit some finite degree of rotation: if all observed stars are sufficiently slowly rotating, however, we may choose to define the map ${f}$ in the limit that the stars are static -- i.e., $\boldsymbol{\Omega}\coloneqq\boldsymbol{0}$, so all rotational metric and surface deformations identically vanish and spacetime is spherically symmetric (if the local comoving source stresses are everywhere isotropic).\footnote{For static sources, the field equations -- together with the continuity constraint -- are the Tolman-Oppenheimer-Volkoff equations \citep{Tolman1939,Oppenheimer1939}.} The (exact) exterior solutions are Schwarzschild solutions, parametrised by $M$, the gravitational mass, and $R$, the coordinate radius of the spherical 2-surface; it follows that for a model ensemble of $s$ exterior solutions, $d=2s$.

We refer to the formulation of the posterior distribution given by Equation~(\ref{eqn:exact posterior}) as the \textit{Interior-Prior} paradigm because posterior estimation of the parameters of an EOS model is \textit{direct}: for an ensemble of model stars, one only assigns a finite prior probability density to sets of interior solutions which are stable and permitted by a shared EOS, and exterior solutions are then matched to numerical interior solutions on-the-fly as $\boldsymbol{y}\mapsto\boldsymbol{x}$ during posterior sampling. Crucially, there is no definition of probability on the joint space of all exterior spacetime parameters appearing in those analytic exterior solutions.  Alternatively, one can consider writing a generative model explicitly in terms of exterior parameters which are free Bayesian random variables, and are thus distributed \textit{a priori} according to some probability density distribution defined on the $\mathbb{R}^{d}$ space, and are to be sampled (Section~\ref{sec:explicit}).

\subsubsection{Posterior tractability}\label{sec:computational tractability}
Equation of state parameter estimation requires detection of (electromagnetic) radiation emitted by an ensemble of stars and the coordinated effort of an ensemble of scientists. The data $\mathcal{D}$ and the model $\mathcal{M}$ in Equation~(\ref{eqn:exact posterior}) are general and can assume complicated forms. Naturally, it is desirable for evaluation of the posterior $\mathcal{P}\left(\boldsymbol{\theta}\;|\;\mathcal{D},\mathcal{M},\mathcal{I}\right)$ to be parallelisable in terms of the application of both cognitive resources (scientists constructing models of observable reality) and compute resources. 

Let us define an enumerator $\varg\in\mathbb{N}_{>0}$ which will distinguish independent data subsets to be analysed by distinct \textit{research groups}: $\mathcal{D}\coloneqq\bigcup_{\varg}\mathcal{D}_{\varg}$. Typically, a generative \textit{submodel} will be conditioned on to describe the statistical properties of each $\mathcal{D}_{\varg}$; a submodel is a parametrised joint probability distribution from which a subset $\mathcal{D}_{\varg}$ (a vector) is assumed to be randomly drawn, together with a (hyper)prior. Submodels \textit{share} the parameters $\boldsymbol{\theta}$ and hyperparameters $\boldsymbol{\alpha}$, and in general can also share parameters in $(\boldsymbol{\rho},\boldsymbol{\Omega})$, nuisance parameters in $\boldsymbol{\eta}$, and nuisance hyperparameters in $\boldsymbol{\beta}$. For instance, the nuisance parameter subsets $\boldsymbol{\eta}_{\varg}$ and $\boldsymbol{\eta}_{\varg+1}$ may have a common subset of nuisance parameters.

The likelihood $\mathcal{L}(\boldsymbol{\theta},\boldsymbol{\rho},\boldsymbol{\Omega},\boldsymbol{\eta})$ appearing in Equation~(\ref{eqn:exact posterior}) is fundamentally separable with respect to observations of stars assumed to be independent; moreover, the likelihood is separable with respect to the data subsets $\mathcal{D}_{\varg}$. The nuisance parameters $\boldsymbol{\eta}$ are formed from a union over the discrete set of submodels defined under $\mathcal{M}$. Equation~(\ref{eqn:exact posterior}) may then be rewritten as
\begin{equation}
\mathcal{P}\left(\boldsymbol{\theta}\;|\;\mathcal{D},\mathcal{M},\mathcal{I}\right)
\propto\mathop{\int}
\underbrace{\mathcal{P}\left(\boldsymbol{\theta},\boldsymbol{\rho},\boldsymbol{\Omega},\boldsymbol{\alpha},\boldsymbol{\beta}\;|\;\mathcal{M},\mathcal{I}\right)}_{\pi(\boldsymbol{\theta},\boldsymbol{\rho},\boldsymbol{\Omega},\boldsymbol{\alpha},\boldsymbol{\beta})}
\underbrace{\mathcal{P}\left(\boldsymbol{\eta}\;|\;\boldsymbol{\theta},\boldsymbol{\rho},\boldsymbol{\Omega},\boldsymbol{\beta},\mathcal{M},\mathcal{I}\right)}_{\pi(\boldsymbol{\eta}\;|\;\boldsymbol{\theta},\boldsymbol{\rho},\boldsymbol{\Omega},\boldsymbol{\beta})}
\mathop{\prod}_{\varg}\underbrace{\mathcal{P}\left(\mathcal{D}_{\varg}\;|\;\boldsymbol{\theta},\boldsymbol{\rho}_{\varg},\boldsymbol{\Omega}_{\varg},\boldsymbol{\eta}_{\varg},\mathcal{M}\right)}_{\mathcal{L}_{\varg}(\boldsymbol{\theta},\boldsymbol{\rho}_{\varg},\boldsymbol{\Omega}_{\varg},\boldsymbol{\eta}_{\varg})}
d\boldsymbol{\eta}d\boldsymbol{\beta}d\boldsymbol{\rho}d\boldsymbol{\Omega}d\boldsymbol{\alpha},
\label{eqn:exact posterior groups no nuisance separation}
\end{equation}
with the ansatz that the nuisance prior can be straightforwardly written as a conditional prior. The parameters $(\boldsymbol{\rho}_{\varg},\boldsymbol{\Omega}_{\varg})$ are a subset of $(\boldsymbol{\rho},\boldsymbol{\Omega})$: if each data subset $\mathcal{D}_{\varg}$ derives from observations of a single star in the model ensemble $\mathcal{E}$, then $(\boldsymbol{\rho}_{\varg},\boldsymbol{\Omega}_{\varg})=(\rho_{i},\Omega_{i})$, where $i\in\mathbb{N}_{>0}, i\leq s$ enumerates stars.

The (hyper)prior is defined on the joint space of: the parameters $\boldsymbol{\theta}$, $\boldsymbol{\rho}$, and $\boldsymbol{\Omega}$ describing local conditions; the nuisance parameters, $\boldsymbol{\eta}$; and the hyperparameters, $\boldsymbol{\alpha}$ and $\boldsymbol{\beta}$. Consider a global model in which the nuisance parameters $\boldsymbol{\eta}_{\varg}$ are defined as \textit{unshared between groups}. Hyperparameters may be shared between the subsets $\boldsymbol{\beta}_{\varg}$ if, e.g., those groups handle instances of the same underlying nuisance parameter in a physical theory. We may then write
\begin{equation}
\mathcal{P}\left(\boldsymbol{\theta}\;|\;\mathcal{D},\mathcal{M},\mathcal{I}\right)
\propto\mathop{\int}
\underbrace{\mathcal{P}\left(\boldsymbol{\theta},\boldsymbol{\rho},\boldsymbol{\Omega},\boldsymbol{\alpha},\boldsymbol{\beta}\;|\;\mathcal{M},\mathcal{I}\right)}_{\pi(\boldsymbol{\theta},\boldsymbol{\rho},\boldsymbol{\Omega},\boldsymbol{\alpha},\boldsymbol{\beta})}
\mathop{\prod}_{\varg}\mathop{\int}\underbrace{\mathcal{P}\left(\mathcal{D}_{\varg}\;|\;\boldsymbol{\theta},\boldsymbol{\rho}_{\varg},\boldsymbol{\Omega}_{\varg},\boldsymbol{\eta}_{\varg},\mathcal{M}\right)}_{\mathcal{L}_{\varg}(\boldsymbol{\theta},\boldsymbol{\rho}_{\varg},\boldsymbol{\Omega}_{\varg},\boldsymbol{\eta}_{\varg})}
\underbrace{\mathcal{P}\left(\boldsymbol{\eta}_{\varg}\;|\;\boldsymbol{\theta},\boldsymbol{\rho}_{\varg},\boldsymbol{\Omega}_{\varg},\boldsymbol{\beta}_{\varg},\mathcal{M},\mathcal{I}\right)}_{\pi(\boldsymbol{\eta}_{\varg}\;|\;\boldsymbol{\theta},\boldsymbol{\rho}_{\varg},\boldsymbol{\Omega}_{\varg},\boldsymbol{\beta}_{\varg})}
d\boldsymbol{\eta}_{\varg}d\boldsymbol{\beta}
d\boldsymbol{\rho}d\boldsymbol{\Omega}d\boldsymbol{\alpha}.
\label{eqn:exact posterior groups}
\end{equation}
Each integral factor in the product over $\varg$ has an integrand consisting of the product of a likelihood $\mathcal{L}_{\varg}(\boldsymbol{\theta},\boldsymbol{\rho}_{\varg},\boldsymbol{\Omega}_{\varg},\boldsymbol{\eta}_{\varg})$ and a prior $\pi(\boldsymbol{\eta}_{\varg}\;|\;\boldsymbol{\theta},\boldsymbol{\rho}_{\varg},\boldsymbol{\Omega}_{\varg},\boldsymbol{\beta}_{\varg})$. To ensure clarity, the global likelihood function in Equation~(\ref{eqn:exact posterior}) is given by
\begin{equation}
\mathcal{L}(\boldsymbol{\theta},\boldsymbol{\rho},\boldsymbol{\Omega},\boldsymbol{\eta})
=\mathop{\prod}_{\varg}\mathcal{L}_{\varg}(\boldsymbol{\theta},\boldsymbol{\rho}_{\varg},\boldsymbol{\Omega}_{\varg},\boldsymbol{\eta}_{\varg}),
\end{equation}
and the global (hyper)prior is given by
\begin{equation}
\pi(\boldsymbol{\theta},\boldsymbol{\rho},\boldsymbol{\Omega},\boldsymbol{\alpha},\boldsymbol{\eta},\boldsymbol{\beta})
=\pi(\boldsymbol{\theta},\boldsymbol{\rho},\boldsymbol{\Omega},\boldsymbol{\alpha},\boldsymbol{\beta})\mathop{\prod}_{\varg}\pi(\boldsymbol{\eta}_{\varg}\;|\;\boldsymbol{\theta},\boldsymbol{\rho}_{\varg},\boldsymbol{\Omega}_{\varg},\boldsymbol{\beta}_{\varg}).
\end{equation}

The (hyper)prior may further be written as
\begin{equation}
\mathcal{P}\left(\boldsymbol{\theta}\;|\;\mathcal{D},\mathcal{M},\mathcal{I}\right)
\propto\mathop{\int}
\underbrace{\mathcal{P}\left(\boldsymbol{\theta},\boldsymbol{\alpha},\boldsymbol{\beta}\;|\;\mathcal{M},\mathcal{I}\right)}_{\pi(\boldsymbol{\theta},\boldsymbol{\alpha},\boldsymbol{\beta})}
\mathop{\prod}_{i=1}^{s}\underbrace{\mathcal{P}\left(\rho_{i},\Omega_{i}\;|\;\boldsymbol{\alpha},\boldsymbol{\theta},\mathcal{M},\mathcal{I}\right)}_{\pi(\rho_{i},\Omega_{i}\;|\;\boldsymbol{\alpha},\boldsymbol{\theta})}
\mathop{\prod}_{\varg}
\underbrace{\mathcal{P}\left(\mathcal{D}_{\varg}\;|\;\boldsymbol{\theta},\boldsymbol{\rho}_{\varg},\boldsymbol{\Omega}_{\varg},\boldsymbol{\beta}_{\varg},\mathcal{M}\right)}_{\mathcal{L}_{\varg}(\boldsymbol{\theta},\boldsymbol{\rho_{\varg}},\boldsymbol{\Omega}_{\varg})}
d\boldsymbol{\beta}d\boldsymbol{\rho}d\boldsymbol{\Omega}d\boldsymbol{\alpha}.
\label{eqn:joint prior distribution of densities and frequencies}
\end{equation}
The conditional argument $\boldsymbol{\theta}$ in each of the prior factors $\mathcal{P}\left(\rho_{i},\Omega_{i}\;|\;\boldsymbol{\alpha},\boldsymbol{\theta},\mathcal{M},\mathcal{I}\right)$ is required because there exist physical bounds on the joint space of $(\rho, \Omega)$ beyond which solutions to the Einstein field equations are unstable for a given EOS \citep[e.g,][sections 6.8 and 6.9]{ST1983}. Moreover, we have assumed that the prior is separable with respect to the subsets of nuisance parameters $\boldsymbol{\eta}_{\varg}$, and that those nuisance parameters have been marginalised over by each respective group -- cf. Equation~(\ref{eqn:exact posterior groups}). The subset (enumerated by $\varg$) of hyperparameters $\boldsymbol{\beta}_{\varg}$ is associated with the subset of nuisance parameters $\boldsymbol{\eta}_{\varg}$ appearing in the generative submodel applied by the $\varg^{th}$ group.

In general however, it is pragmatic for a given group -- who are, e.g., experts in modelling a particular observational phenomenon -- to analyse data from multiple stars. In this case distinct groups -- e.g., the $\varg^{th}$ and $(\varg+1)^{th}$ groups -- may \textit{both} analyse data from the $i^{th}$ star, where the subset $\mathcal{D}_{\varg,i}\subseteq\mathcal{D}_{\varg}$ and the subset $\mathcal{D}_{\varg+1,i}\subseteq\mathcal{D}_{\varg+1}$ are independent. Consider, therefore, a scenario in which some nuisance parameters are defined as shared between groups. For instance, suppose groups $\varg$ and $(\varg+1)$ both analyse data from a given star and there are thus nuisance parameters on which the sampling distributions of $\mathcal{D}_{\varg}$ and $\mathcal{D}_{\varg+1}$ depend; alternatively, nuisance parameters may be associated with a particular instrument for the handling of systematic effects in the acquisition of both $\mathcal{D}_{\varg}$ and $\mathcal{D}_{\varg+1}$. In this case the prior distribution of the nuisance parameters cannot be written as a product over $\varg$ (because otherwise priors would be raised to some integer power greater than unity). On the other hand, we may write
\begin{equation}
\mathcal{P}\left(\boldsymbol{\theta}\;|\;\mathcal{D},\mathcal{M},\mathcal{I}\right)
\propto\mathop{\int}
\underbrace{\mathcal{P}\left(\boldsymbol{\theta},\boldsymbol{\rho},\boldsymbol{\Omega},\boldsymbol{\alpha},\boldsymbol{\beta}\;|\;\mathcal{M},\mathcal{I}\right)}_{\pi(\boldsymbol{\theta},\boldsymbol{\rho},\boldsymbol{\Omega},\boldsymbol{\alpha},\boldsymbol{\beta})}
\mathop{\prod}_{i=1}^{s}\underbrace{\mathcal{P}\left(\boldsymbol{\eta}_{i}\;|\;\boldsymbol{\theta},\boldsymbol{\rho}_{i},\boldsymbol{\Omega}_{i},\boldsymbol{\beta}_{i},\mathcal{M},\mathcal{I}\right)}_{\pi(\boldsymbol{\eta}_{i}\;|\;\boldsymbol{\theta},\boldsymbol{\rho}_{i},\boldsymbol{\Omega}_{i},\boldsymbol{\beta}_{i})}
\mathop{\prod}_{\varg}\underbrace{\mathcal{P}\left(\mathcal{D}_{\varg}\;|\;\boldsymbol{\theta},\boldsymbol{\rho}_{\varg},\boldsymbol{\Omega}_{\varg},\boldsymbol{\eta}_{\varg},\mathcal{M}\right)}_{\mathcal{L}_{\varg}(\boldsymbol{\theta},\boldsymbol{\rho}_{\varg},\boldsymbol{\Omega}_{\varg},\boldsymbol{\eta}_{\varg})}
d\boldsymbol{\eta}d\boldsymbol{\beta}d\boldsymbol{\rho}d\boldsymbol{\Omega}d\boldsymbol{\alpha}
\label{eqn:exact posterior stars and groups}
\end{equation}
if subsets $\boldsymbol{\eta}_{i}$ are defined as \textit{unshared between stars of the ensemble}. Note that nuisance hyperparameters may be shared between the subsets $\boldsymbol{\beta}_{i}$.

\begin{figure}
\centering
\includegraphics[width=0.85\textwidth]{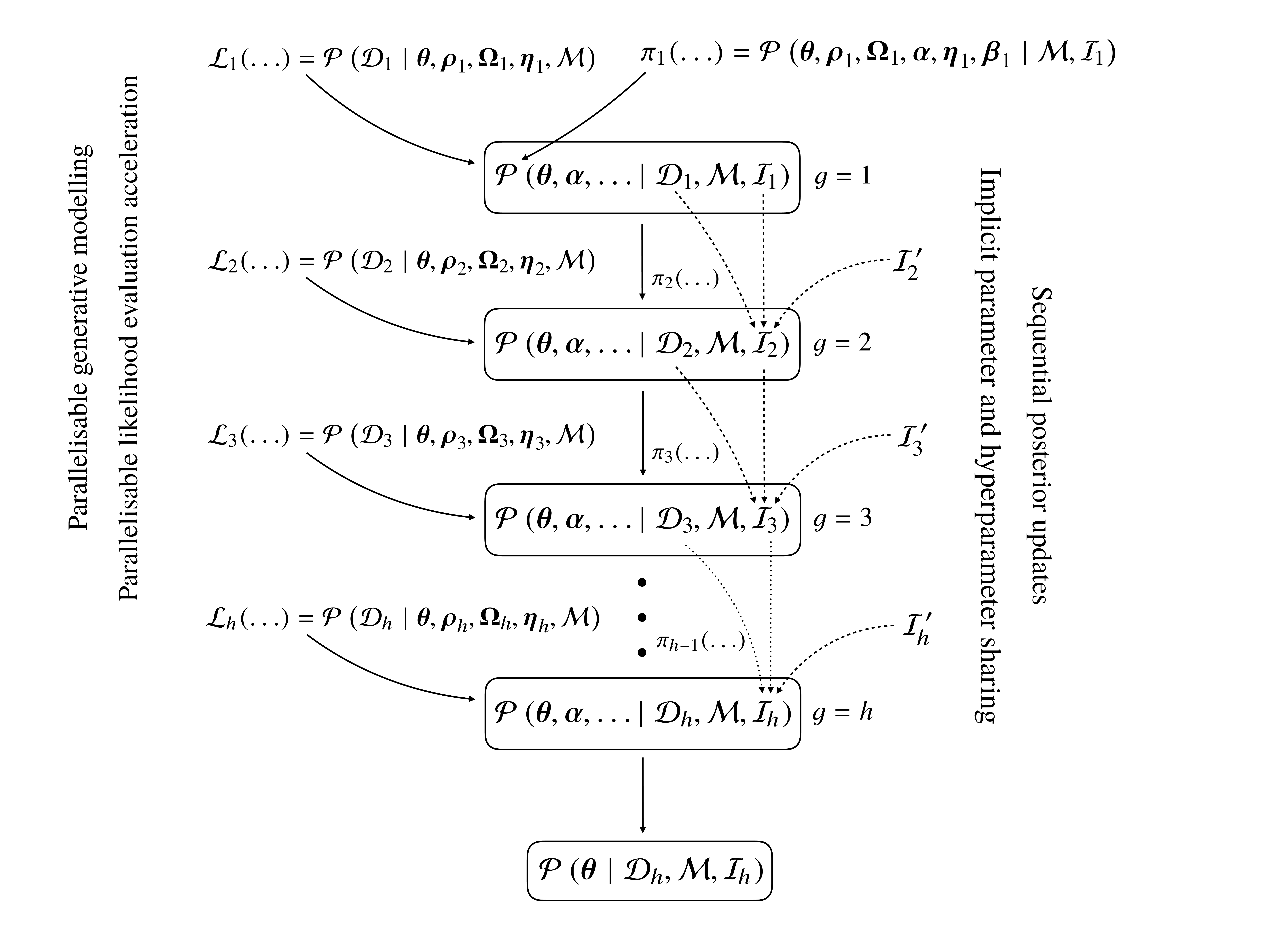}
\caption{Sequential EOS parameter estimation procedure. Cognitive workload (e.g., constructing the global model and associated likelihood \textit{evaluation} algorithms) is distributed amongst research groups in a quasi-parallel manner (whilst requiring global self-consistency of the model $\mathcal{M}$), whilst computational posterior sampling is distributed amongst their respective compute resources in an arbitrary sequential manner.}
\label{fig:IP-paradigm procedure}
\end{figure}

A fundamental remark to be made is that from a Bayesian perspective, a posterior (on some space of parameters or, equivalently, models) is defined by updating our prior knowledge conditioned on observations of reality. It is therefore natural in a Bayesian framework to distribute workload amongst independent groups via a series of posterior updates. We proceed to formulate such a procedure mathematically, and we illustrate it in Fig.~\ref{fig:IP-paradigm procedure}. We should note that the organisation of the procedure is arbitrary: in principle a single research group could perform EOS parameter estimation. In practice however, the most stringent statistical constraints will be synergistic, derived from a diverse set of observational phenomena and thus the contributions of a diverse set of research groups. Further, constraining power increases if groups globally self-organise so as to minimise marginalisation operations when (hyper)parameters are shared. Such self-organisation includes: consensus agreement on fundamental aspects of the global model (to ensure self-consistency); the distribution of data amongst groups; the ordering of groups in the posterior updating chain to optimise communication with respect to (hyper)parameter sharing (groups can perform multiple disjoint updates if appropriate); and consensus agreement on both approximate approaches to solving the field equations, and on approximations to the posteriors transferred between groups. We note that such a collaboration is clearly going to be involved, and likely requires various sources of funding. Moreover, the underlying notion is that if synergy is to be close to maximised, communication between groups may need to be more sophisticated than as is naturally the case when communication is predominantly through the literature. Thus, such a collaboration is tailored towards collective publishing (of a series of articles) once a global analysis has been completed by all participating groups, although this is not \textit{necessary}.

A second remark is that in order to obtain values for exterior spacetime parameters which appear in submodels (see Section \ref{sec:solutions for generative models}), \textit{every} group may numerically integrate (approximations to) the field equations in order to match exterior solutions to interior solutions. Groups are in principle free to apply different approximations to obtain their exterior solutions, provided each approximation is well-defined in the global model $\mathcal{M}$ and can thus be clearly and explicitly conditioned on by consensus. Moreover, each approximation needs to generate exterior solutions which are sufficiently detailed to contribute to the global constraining power conditional on $\mathcal{D}_{\varg}$ (e.g., solutions which are at least dependent on the lowest order exterior metric parameter together with the lowest order surface boundary parameter); otherwise careful justification for the use of that approximation in EOS parameter estimation is required (e.g., for constraining hyperparameters).

Coordinating the efforts of a set of research groups and collaborations is challenging; alas it is arguably necessary if we are to derive rigorous statistical constraints as a scientific community. A leading example of synergistic applied statistics is the field of precision cosmology \citep[see, e.g.,][for a recent review]{PRECISION_COSMOLOGY}: cosmologists profoundly share a single universe -- a universe that observably manifests via numerous large-scale phenomena which encode partially orthogonal information. Inference efforts are thus most powerful in collaboration as a scientific community, given some canonical set of underlying cosmologies, one of which is the well-known $\Lambda$CDM parametrisation. Parallels to the field of astrophysical dense matter study are clear: whilst the focus is on small (strong force) scales, it is plausible that the fundamental particles in our Universe are shared by all compact stars. This assumption, together with our empirical knowledge of the rich array of observable compact star phenomena, lends credence to the notion of collaboratively applying statistics to dense matter EOS study, given some canonical set of underlying EOS models. If such an approach is not realised, constraining power will not be maximal because potential synergy is not exploited.

It is important to note that at each stage in the chain illustrated in Fig.~\ref{fig:IP-paradigm procedure}, the output is an (approximation to the) joint posterior distribution of EOS parameters and other model (hyper)parameters; typically the output will assume the form of a set of (approximately) independent and identically drawn posterior samples. It follows that the output at each stage can be straightforwardly post-processed to obtain a \textit{marginal} joint posterior distribution of the EOS parameters alone. Further, each group can archive a snapshot of the posterior evaluation process which can be reverted to and updated if the consensus agreement is that the subsequent analysis should be reiterated or modified in some manner -- effectively a form of \textit{version control}.

A fundamentally important facet of such a coordinated analysis is \textit{open-sourcing}. The following should be made available to all groups during the analysis, and, arguably, made publicly available upon publication: detailed generative model documentation, including nuisance (hyper)prior and marginalisation choices; likelihood function codes (implicit in which is execution of the interior-exterior parameter mapping for a subset of stars); the applied sampling algorithm and implementation if not already publicly available; and posterior samples (e.g., entire Markov chains or nested samples and their approximate weights). Note that these samples may also be from an integrable (e.g., bounded) likelihood function depending on the collaborative analysis structure (see Section~\ref{sec:computational tractability}). Furthermore, there already exist a diverse set of open-source sampling implementations in various languages which have community support, and are more appropriate for certain problems than others. The following is a clearly non-exhaustive list of examples of open-source software packages which allow generic likelihood specification and are used in astrophysics and cosmology: \citet[][]{MultiNest, BAMBI, emcee, PolyChord_1, DNest4, PyMC3}.\footnote{Some well-known and powerful projects such as \citet[][]{Stan} require that models be specified in a more restrictive language and are thus not as generally applicable to EOS inference where the likelihood is of a complicated numerical nature.} It is even advisable for realistic problems to apply at least two distinct algorithms (via their open-source implementations) in order to robustly learn about likelihood function structure and thus make accurate numerical inferences conditioned on a given model \citep[e.g.,][]{LALInference}.

Let us consider Equation~(\ref{eqn:exact posterior groups}) from the perspective of the first group to update a prior in the analysis chain suggested above:
\begin{equation}
\mathcal{P}\left(\boldsymbol{\theta}\;|\;\mathcal{D}_{1},\mathcal{M},\mathcal{I}_{1}\right)
\propto\mathop{\int}
\underbrace{\mathcal{P}\left(\boldsymbol{\theta},\boldsymbol{\rho}_{1},\boldsymbol{\Omega}_{1},\boldsymbol{\alpha},\boldsymbol{\beta}_{1}\;|\;\mathcal{M},\mathcal{I}_{1}\right)}_{\pi(\boldsymbol{\theta},\boldsymbol{\rho}_{1},\boldsymbol{\Omega}_{1},\boldsymbol{\alpha},\boldsymbol{\beta}_{1})}
\underbrace{\mathcal{P}\left(\boldsymbol{\eta}_{1}\;|\;\boldsymbol{\theta},\boldsymbol{\rho}_{1},\boldsymbol{\Omega}_{1},\boldsymbol{\beta}_{1},\mathcal{M},\mathcal{I}_{1}\right)}_{\pi(\boldsymbol{\eta}_{1}\;|\;\boldsymbol{\theta},\boldsymbol{\rho}_{1},\boldsymbol{\Omega}_{1},\boldsymbol{\beta}_{1})}
\underbrace{\mathcal{P}\left(\mathcal{D}_{1}\;|\;\boldsymbol{\theta},\boldsymbol{\rho}_{1},\boldsymbol{\Omega}_{1},\boldsymbol{\eta}_{1},\mathcal{M}\right)}_{\mathcal{L}_{1}(\boldsymbol{\theta},\boldsymbol{\rho}_{1},\boldsymbol{\Omega}_{1},\boldsymbol{\eta}_{1})}
d\boldsymbol{\eta}_{1}d\boldsymbol{\beta}_{1}d\boldsymbol{\rho}_{1}d\boldsymbol{\Omega}_{1}d\boldsymbol{\alpha},
\label{eqn:first group posterior}
\end{equation}
where $\mathcal{I}_{1}$ may or may not be the empty set; if it is \textit{not} the empty set, it is constituted by independent data sets. In Fig.~\ref{fig:IP-paradigm procedure} we illustrate the general procedure whereby nuisance parameter and nuisance hyperparameter sharing is implicit. The simplest case to consider here is that in which none of $(\boldsymbol{\rho}_{1},\boldsymbol{\Omega}_{1},\boldsymbol{\eta}_{1},\boldsymbol{\beta}_{1})$ are shared with any other group $\varg>1$, and can thus be marginalised over by group $\varg=1$ to yield a (hyper)prior for group $\varg=2$:
\begin{equation}
\mathcal{P}\left(\boldsymbol{\theta}\;|\;\mathcal{D}_{2},\mathcal{M},\mathcal{I}_{2}\right)
\propto\mathop{\int}
\underbrace{\mathcal{P}\left(\boldsymbol{\theta},\boldsymbol{\rho}_{2},\boldsymbol{\Omega}_{2},\boldsymbol{\alpha},\boldsymbol{\beta}_{2}\;|\;\mathcal{M},\mathcal{I}_{2}\right)}_{\pi(\boldsymbol{\theta},\boldsymbol{\rho}_{2},\boldsymbol{\Omega}_{2},\boldsymbol{\alpha},\boldsymbol{\beta}_{2})}
\underbrace{\mathcal{P}\left(\boldsymbol{\eta}_{2}\;|\;\boldsymbol{\theta},\boldsymbol{\rho}_{2},\boldsymbol{\Omega}_{2},\boldsymbol{\beta}_{2},\mathcal{M},\mathcal{I}_{2}^{\prime}\right)}_{\pi(\boldsymbol{\eta}_{2}\;|\;\boldsymbol{\theta},\boldsymbol{\rho}_{2},\boldsymbol{\Omega}_{2},\boldsymbol{\beta}_{2})}
\underbrace{\mathcal{P}\left(\mathcal{D}_{2}\;|\;\boldsymbol{\theta},\boldsymbol{\rho}_{2},\boldsymbol{\Omega}_{2},\boldsymbol{\eta}_{2},\mathcal{M}\right)}_{\mathcal{L}_{2}(\boldsymbol{\theta},\boldsymbol{\rho}_{2},\boldsymbol{\Omega}_{2},\boldsymbol{\eta}_{2})}
d\boldsymbol{\eta}_{2}d\boldsymbol{\beta}_{2}d\boldsymbol{\rho}_{2}d\boldsymbol{\Omega}_{2}d\boldsymbol{\alpha},
\label{eqn:second group posterior unshared case}
\end{equation}
where the (hyper)prior
\begin{equation}
\mathcal{P}\left(\boldsymbol{\theta},\boldsymbol{\rho}_{2},\boldsymbol{\Omega}_{2},\boldsymbol{\alpha},\boldsymbol{\beta}_{2}\;|\;\mathcal{M},\mathcal{I}_{2}\right)\coloneqq
\mathcal{P}\left(\boldsymbol{\theta},\boldsymbol{\alpha}\;|\;\mathcal{D}_{1},\mathcal{M},\mathcal{I}_{1}\right)
\mathcal{P}\left(\boldsymbol{\rho}_{2},\boldsymbol{\Omega}_{2},\boldsymbol{\beta}_{2}\;|\;\boldsymbol{\theta},\boldsymbol{\alpha},\mathcal{M},\mathcal{I}_{2}^{\prime}\right),
\label{eqn:communicated posterior 1}
\end{equation}
and $\mathcal{I}_{2}\coloneqq\mathcal{D}_{1}\cup\mathcal{I}_{1}\cup\mathcal{I}_{2}^{\prime}$, where $\mathcal{I}_{2}^{\prime}$ represents any prior information (independent data sets) that satisfies $\mathcal{I}_{2}^{\prime}\cap(\mathcal{D}_{1}\cup\mathcal{I}_{1})=\emptyset$. To be clear, note that the conditional probability $\mathcal{P}\left(\boldsymbol{\theta},\boldsymbol{\alpha}\;|\;\mathcal{D}_{1},\mathcal{M},\mathcal{I}_{1}\right)$ is the posterior given by Equation~(\ref{eqn:first group posterior}), if group $\varg=1$ does \textit{not} marginalise over the hyperparameters $\boldsymbol{\alpha}$.

Provided that none of $(\boldsymbol{\rho}_{\varg},\boldsymbol{\Omega}_{\varg},\boldsymbol{\eta}_{\varg},\boldsymbol{\beta}_{\varg})$ are shared with any other group contributing to the analysis chain, it should be apparent that the above process is iterative and that for $\varg=1\ldots{h}$ groups, the final marginal posterior distribution of the EOS parameters is given by
\begin{equation}
\mathcal{P}\left(\boldsymbol{\theta}\;|\;\mathcal{D}_{{h}},\mathcal{M},\mathcal{I}_{{h}}\right)
\propto\mathop{\int}
\mathcal{P}\left(\boldsymbol{\theta},\boldsymbol{\alpha}\;|\;\mathcal{M},\mathcal{I}_{h}\backslash\mathcal{I}_{h}^{\prime}\right)
\mathcal{P}\left(\boldsymbol{\rho}_{h},\boldsymbol{\Omega}_{h},\boldsymbol{\eta}_{h},\boldsymbol{\beta}_{h}\;|\;\boldsymbol{\theta},\boldsymbol{\alpha},\mathcal{M},\mathcal{I}_{h}^{\prime}\right)
\mathcal{P}\left(\mathcal{D}_{{h}}\;|\;\boldsymbol{\theta},\boldsymbol{\rho}_{{h}},\boldsymbol{\Omega}_{{h}},\boldsymbol{\eta}_{{h}},\mathcal{M}\right)
d\boldsymbol{\eta}_{{h}}d\boldsymbol{\beta}_{{h}}d\boldsymbol{\rho}_{{h}}d\boldsymbol{\Omega}_{{h}}d\boldsymbol{\alpha},
\label{eqn:final group posterior unshared case}
\end{equation}
where $\mathcal{I}_{h}\coloneqq\left(\mathcal{D}_{1}\cup\ldots\cup\mathcal{D}_{h-1}\right)\cup\left(\mathcal{I}_{1}\cup\mathcal{I}_{2}^{\prime}\cup\ldots\cup\mathcal{I}_{h}^{\prime}\right)$ and $\mathcal{I}_{h}\backslash\mathcal{I}_{h}^{\prime}$ denotes the set difference. If we recursively express (hyper)priors $\mathcal{P}\left(\boldsymbol{\theta},\boldsymbol{\alpha}\;|\;\mathcal{M},\ldots\right)$ as marginal posteriors, we naturally recover Equation~(\ref{eqn:exact posterior groups}).

We now discuss communication between groups. Each group would communicate either: (i) approximately independent and identically drawn posterior samples (weighted or unweighted depending on the sampling algorithm); (ii) samples smoothed on a joint space using, e.g., Kernel Density Estimation; (iii) or some accurate analytical approximation to a posterior. Whilst the latter may be desirable, it nevertheless requires mapping of the posterior using some robust sampling algorithm and implementation, and the tractability of such a task depends on dimensionality and structural complexity (e.g., multi-modality and non-linear degeneracies, which require many distributional moments to accurately represent). A sophisticated nested sampling algorithm and implementation such as that offered by the open-source software \textsc{PolyChord} \citep[][]{PolyChord_1} could be applied by groups where appropriate (this software is also discussed in Section \ref{sec:model selection}, but is just one example).

If (hyper)parameters \textit{other} than $\boldsymbol{\theta}$ and $\boldsymbol{\alpha}$ are shared between groups (e.g., multiple groups analyse distinct observational phenomena from a given star) it is optimal if those shared parameters are \textit{not} marginalised over during communication between groups -- such operations would discard statistical information. On the other hand, the greater the number of (hyper)parameters that are shared and not marginalised over, the more difficult the problem of organising the global analysis. Such organisation includes: (i) the definition of the subsets $\mathcal{D}_{\varg}$; and (ii) the ordering of groups in the analysis chain in order to both minimise the total quantity of communicated numerical (posterior) data and minimise the number of parameter dimensions that need to be handled by each group.

A discussion on posterior tractability requires us to remark on parallelisation of EOS parameter estimation with respect to independent groups and their respective compute resources. Ultimately, each group considers a submodel describing their respective data subset $\mathcal{D}_{g}$; parameter estimation is thus at least partially parallelisable with respect to cognitive resources, however communication between groups is necessary to ensure global self-consistency of the model $\mathcal{M}$. However, if a subset of model (hyper)parameters are shared between groups, \textit{computational} parallelisation of posterior evaluation is not possible because the process must be serialised in a manner akin to above. Whilst the configuration of groups (and thus submodels applied to data subsets) can in principle be modified to eliminate \textit{parameter} sharing, and whilst we can omit hyperparameters from the global model $\mathcal{M}$, parallelisation of posterior evaluation over groups cannot be achieved because of the fundamental model assumption that the EOS parameters are shared by all stars and thus all groups.

The scope for computational parallelisation is limited to \textit{likelihood} acceleration: groups consider likelihoods $\mathcal{L}_{\varg}(\boldsymbol{\theta},\boldsymbol{\rho}_{\varg},\boldsymbol{\Omega}_{\varg},\boldsymbol{\eta}_{\varg})$ which can in principle be mapped simultaneously in order to construct approximate representations which are faster to evaluate during posterior sampling; examples include training unsupervised neural networks \citep[for demonstration and open-source software see \textsc{SkyNet},][]{BAMBI,SKYNET}, and, if appropriate, cruder analytical representations \cite[e.g.,][and references therein]{MacKay2003,Robert2009,LAPLACE_APPROX,Lackey2015} or (regular) discrete numerical representations on a mesh for the purpose of multi-dimensional interpolation.\footnote{To enable these cruder representations, we may need to stochastically sample the likelihood function with some robust posterior sampling software (possibly requiring some appropriately bounded uniform prior to enforce integrability). The approximately independent and identically drawn samples would need to be smoothed in multiple dimensions to generate a probability density distribution proportional to the likelihood function. If the normalisation of the likelihood function is important for Bayesian model comparison (see Section \ref{sec:model selection}), a nested sampling algorithm could be implemented. Such a sampler would approximate the prior predictive probability of the data (the evidence). The probability density distribution can then be rescaled to (approximately) preserve the Bayesian evidence (alternatively the marginal likelihood or, more explicitly, the joint prior predictive probability of $\mathcal{D}$).} Moreover, \textit{unshared} nuisance (hyper)parameters can be marginalised over numerically or analytically \citep[e.g.,][]{Taylor2010} during likelihood acceleration. It follows that the sequential posterior updates can in principle be globally accelerated if groups each accelerate their likelihood evaluation procedure whilst (posterior) communications are pending.

An alternative structure for the global analysis would be for each group to simultaneously generate an approximate representation of a likelihood function marginalised over \textit{group-specific} nuisance (hyper)parameters, either on a space of interior parameters (necessarily including the shared EOS parameters) or on a space of exterior parameters. Each of these likelihoods would be communicated to a single (``synthesis'') group who would apply the joint (hyper)prior distribution of the EOS parameters, all central densities, all rotation frequencies, the (interior) hyperparameters, and any nuisance (hyper)parameters shared between groups. The ``synthesis'' group would then sample from the posterior defined by the product of the likelihoods and the (hyper)prior.

If the $\varg^{th}$ group supplies a likelihood function on a space of exterior parameters, the ``synthesis'' group needs to be able to exhaustively evaluate that function at every image $\boldsymbol{x}\in X$ under a map ${f}\from Y\to X$, $\boldsymbol{y}\mapsto\boldsymbol{x}$ (see Section~\ref{sec:mapping definition}), where the domain $Y$ by definition consists only of points $\boldsymbol{y}$ with finite prior density (and thus points which generate only stable spacetime solutions, one per star). Instructions would need to be supplied by the $\varg^{th}$ group specifying the field equations to integrate (e.g., static or with low-order rotational perturbations) and the interior condition parameters required -- i.e., central densities and rotation frequencies of stars -- in order to match to analytic exterior solutions used to define the likelihood function. There are two notable advantages of supplying likelihood functions on spaces of exterior parameters: (i) a single group handles (the albeit difficult task of) field equation integration and solution matching for all sources; and (ii) these likelihood functions are can be applied to multiple EOS models. However, care needs to be taken that each likelihood function spans a sufficiently large region of the space of exterior parameters it is defined on; each EOS model includes a (hyper)prior distribution of the interior (hyper)parameters and the exterior likelihood function needs to be evaluable for all spacetime solutions with finite prior density (those that are stable).

A note to be made here (following on from Section~\ref{sec:mapping definition}) is that if the rotation frequencies $\boldsymbol{\Omega}$ are defined as free exterior parameters, a further assumption could in principle be made: universal behaviours allow a justified reduction in dimensionality of the space of exterior parameters associated with each star \citep[e.g.,][]{YagiYunes2013,Bauboeck2013}. Conditioned on the assumption that universal behaviour of the ellipticity \citep[][]{Morsink2007,Bauboeck2013,AlGendy2014} holds exactly for an EOS model, we can estimate the free exterior parameters $\boldsymbol{x}=(\boldsymbol{M},\boldsymbol{R}_{\textrm{eq}},\boldsymbol{\Omega})$ whilst applying the ellipticity constraint equation $e\coloneqq e(M,\Req,\Omega)$. In practice, universal relations are based on some \textit{precomputed} numerical library of compact star solutions to the field equations \citep[e.g.,][]{Morsink2007, YagiYunes2013}. When invoking a universality, one uses constraint equations empirically calculated for a particular microphysical EOS family (e.g., nucleonic or quark). The likelihood function distortion arising from truncation (here meaning the embedding of a spherical 2-surface, such that $R_{\textrm{eq}}\equiv R$), however, may well be of greater concern than any statistical bias incurred by invoking universalities associated with a particular EOS family to write rotational metric and or surface deformations as functions of exterior parameters such as $M$, $\Req$, and $\Omega$. If this such an approximation is not invoked, the likelihood function to be mapped on a space of exterior parameters for each star must also be a function of ellipticity, and for expensive likelihoods, learning the function along an additional slow dimension can be expensive. If, on the other hand, the likelihood function is mapped on a space of interior parameters, such a universality is needless because the ellipticity is an $(r,\ell)=(2,2)$ perturbation, and thus of the same order as the first perturbation to the equatorial radius; the ellipticity is then trivially calculated at effectively zero expense and is for the specific EOS model of interest.

The above alternative analysis structure is a form of parallel likelihood acceleration; the distinction here is that the handling of the (hyper)prior distribution of shared (hyper)parameters is limited to a single group. We note that the normalisation of the communicated likelihoods would be important for model comparison via a Bayes' factor metric (see Section \ref{sec:model selection} for detailed discussion). Given that each group's likelihood may require the application of sampling software to accurately map (see the above discussion and footnote), this global analysis structure may be viewed as consisting of two ``slow'' sampling phases connected by one ``slow'' communication phase due to a form of barrier synchronisation if the ``synthesis'' group is to perform only a single Bayesian posterior update. The ``synthesis'' group could in principle perform a series of Bayesian updates, using likelihood functions as they become available; in this case a single central group is dedicated to executing the chain of posterior updates in Fig.~\ref{fig:IP-paradigm procedure}. However, such an approach requires approximative computation to transform the posterior output of one node into an appropriate prior input for the subsequent node.

The analysis structure treated above (Bayesian sequential updates as illustrated in Fig. \ref{fig:IP-paradigm procedure}), on the other hand, consists of a sequence of ``slow'' sampling phases separated by ``fast'' communication phases. However, given that the time to sample an arbitrary distribution at some tolerance level is governed by properties such as dimensionality, modality, degeneracy, and likelihood evaluation time \citep[see, e.g.,][]{MacKay2003, Gelman_book, PolyChord_1}, it is conceivable that groups later in the analysis chain complete their ``accelerated'' sampling processes appreciably faster than earlier groups due to likelihood acceleration performed whilst communications were pending. Both structures (involving approximate likelihood representations) may well involve computation of a similar set of intermediary distributions and integrals: the main difference lies with the handling of the prior for shared (hyper)parameters -- i.e., the complexity of the communication and posterior sampling tasks performed by any one group. We cannot state which of analysis structures above is more efficient in general -- a consensus agreement amongst participating groups would be required at the outset of the collaboration.

The problem of structuring the analysis, especially when there exists freedom, lies in the realm of Bayesian sequential experimental design and optimisation \citep[e.g.,][for an early review on this field]{Chaloner1995}.  In the context of EOS inference, the global data set may, e.g., be comprised only of archival telescope data, or more realistically be comprised of both archival data and data which accumulates over the lifetime of an observing programme, and is to be analysed in real-time. Moreover, ordering of posterior updates is not necessarily invariant of prior choices; minimally informative (noninformative) references priors including Jeffreys -- if for instance deemed appropriate and tractable -- are dependent on a parametrised sampling distribution on the space of a data subset \citep[e.g.,][]{Jeffreys,Jeffreys1961,Bernardo1979,Robert2007,Robert2009}, and thus sensitive to group ordering \citep[e.g.,][]{Lewis_noninf_2013}. In this work we will not pursue a more sophisticated framework for design of the sequential update process in the context of EOS inference.

\subsubsection{Summary}\label{sec:implicit summary}
Bayesian inference of an EOS assumedly shared by an ensemble of stars can be formulated in a principled manner via the IP-paradigm. This paradigm is defined by direct EOS parameter estimation conditioned on: a global hierarchical generative model; the union of data acquired via observations of all stars in the ensemble; and any prior data sets. The work required for definition and implementation of the global model can naturally be distributed amongst research groups. The scope for computational parallelisation of EOS parameter estimation -- amongst research groups and their respective compute resources -- is limited because (hyper)parameters are shared between groups, meaning there must be a serial element to the global analysis; however, there is full scope for acceleration of sequential posterior updates via parallel learning of likelihood functions.

\subsection{The Exterior-Prior paradigm}\label{sec:explicit}
Let a prior be explicitly defined on some joint space of exterior spacetime parameters, instead of on a space of interior parameters (including the EOS parameters). Hereafter we shall refer to such an approach as the \textit{Exterior-Prior} (EP) paradigm for EOS parameter estimation.

\subsubsection{Definition}
The EP-paradigm by design consists of two distinct phases. Let the first phase be defined by posterior estimation of parameters of an (exact or approximate) analytic exterior spacetime solution for each star belonging to some abstract model ensemble. The free parameters of interest are multipole moments appearing in the solution and parameters describing the spacelike 2-surface outside of which the solution is considered valid (see Sections \ref{sec:intro} and \ref{sec:solutions for generative models}).

Let the second phase be defined by the post-processing of a marginal joint posterior distribution of an ensemble of exterior parameters, with the aim of transforming it into a marginal joint posterior distribution on the space of EOS parameters. Such an approach was first considered by \citet{Ozel2009} for general probability density distributions, which we have interpreted from a Bayesian perspective. We agree with examining this notion \textit{based on probability theory alone}: it may be intuitive to any given author that because continuous interior and continuous exterior parameters are deterministically related, probability density distributions can be transformed between spaces of exterior and interior parameters. However, there exist difficulties to address in the context of general relativistic gravity and population-level parameters.

\subsubsection{Conditions for transformation of probability density between spaces of exterior and interior parameters}\label{sec:mapping}

\begin{figure}
\centering
\includegraphics[width=0.8\textwidth]{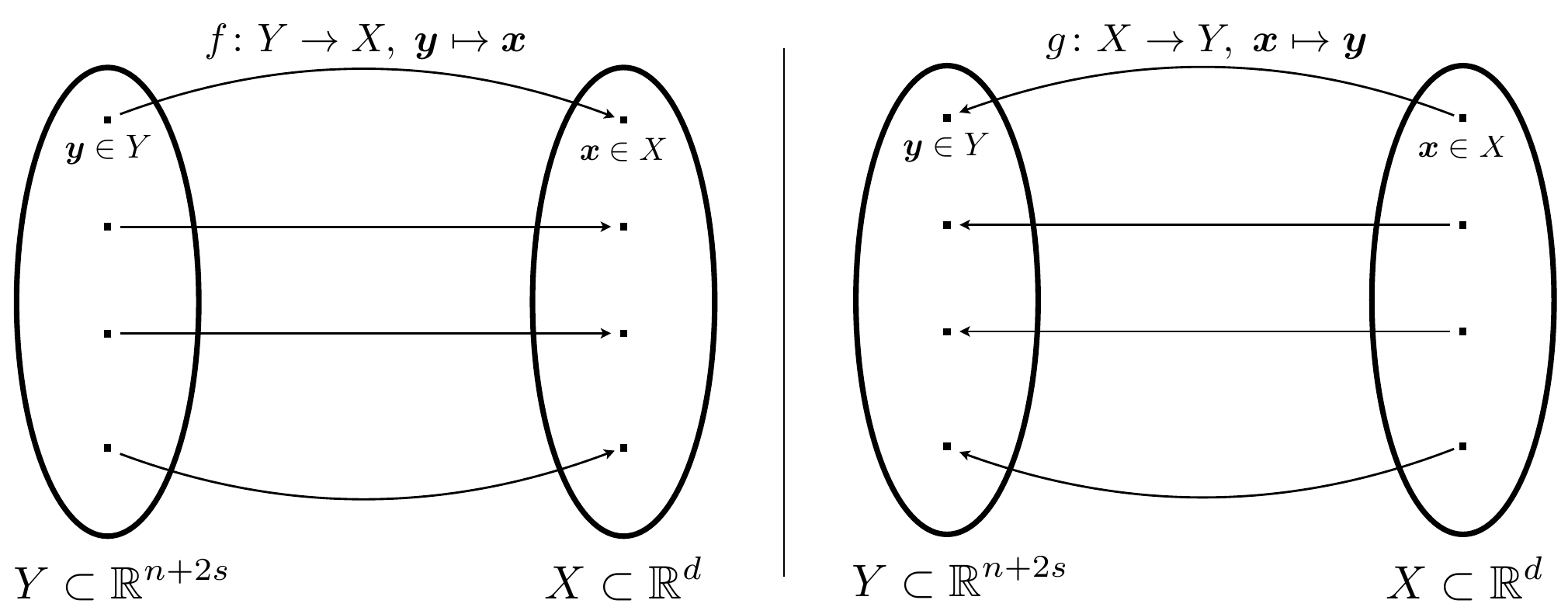}
\caption{A discrete representation of a bijection: a map which is both \textit{injective} (one-to-one) and \textit{surjective} (onto). Such a map is \textit{invertible}; the domain $Y$ of $f$ has an image $X$. Each element of $X$ is the image of exactly \textit{one} element of $Y$. In the context of EOS parameter estimation, we require $Y$ to be the (compact) subset of interior-parameter vectors with finite probabilistic support, and we require $X$ to be the (compact) subset of exterior-parameter vectors with finite probabilistic support conditioned on $Y$ and general relativistic gravity. For continuous spaces we consider such a map to be \textit{diffeomorphic}: both locally and globally invertible, and thus for continuous spaces, differentiable.}
\label{fig:invertible}
\end{figure}

In order to define a law for the transformation of a probability density distribution between spaces of exterior and interior parameters, we require the map ${f}\from Y\to X$, $\boldsymbol{y}\mapsto\boldsymbol{x}$ defined in Section~\ref{sec:mapping definition}. Let us identify the vector $\boldsymbol{x}\in X$ as a vector of \textit{random} variables, where $X\subset\mathbb{R}^{d}$ and $\mathbb{R}^{d}$ is the space of a set (or ensemble) of exterior spacetime parameters. Let us also define an \textit{inverse} function ${g}\from X\to Y$, $\boldsymbol{x}\mapsto\boldsymbol{y}$ as illustrated in Fig. \ref{fig:invertible}.\footnote{The function which is in principle evaluable (without any optimisation process or unsupervised learning process) is the general relativistic mapping $\boldsymbol{y}\mapsto\boldsymbol{x}$ from local source matter properties to a global rotating spacetime solution. Therefore we opt to denote this evaluable function by $f$, and to define the nominal inverse of ${f}$ as the function ${g}$. The function $g$ is thus associated with the `backward' transformation of the probability density distribution from exterior (global) parameters to interior (local) source matter parameters.} The relevant ansatz here is that ${f}$ (and thus, by definition, $g$) is a \textit{diffeomorphism} between the distinct continuous sets  $Y\subset\mathbb{R}^{n+2s}$ and $X\subset\mathbb{R}^{d}$ in which $\boldsymbol{y}$ and $\boldsymbol{x}$ respectively exist. To an (objectivist) Bayesian, studying the properties of such a mapping -- representing a deterministic model reparametrisation -- is crucially important for defining noninformative priors \citep[see][and references therein]{Jeffreys,Jeffreys1961,Bernardo1979,Robert2007,Robert2009}.

Let a conditional, marginal joint probability density distribution on the space of EOS parameters be denoted by $\mathcal{P}\left(\boldsymbol{\theta}\;|\ldots\right)$; posteriors are of the form $\mathcal{P}\left(\boldsymbol{\theta}\;|\;\mathcal{D},\mathcal{M},\mathcal{I}\right)$, whilst priors are of the form $\mathcal{P}\left(\boldsymbol{\theta}\;|\;\mathcal{M},\mathcal{I}\right)$. Such a distribution can be written as an integral over a derivative of a cumulative distribution function:
\begin{equation}
\begin{aligned}
\mathcal{P}\left(\boldsymbol{\theta}\;|\ldots\right)&=\mathop{\int}\mathcal{P}\left(\boldsymbol{\theta},\boldsymbol{\rho},\boldsymbol{\Omega}\;|\ldots\right)d\boldsymbol{\rho}d\boldsymbol{\Omega}\\
&=\mathop{\int}\left[\frac{\partial}{\partial y_{1}}\cdots\frac{\partial}{\partial y_{n+2s}}\mathop{\int}_{W}\mathcal{P}(\boldsymbol{x}^{\prime}\;|\ldots)d\boldsymbol{x}^{\prime}\right]d\boldsymbol{\rho}d\boldsymbol{\Omega}.
\end{aligned}
\label{eqn:fundamental posterior}
\end{equation}
where $W=\{\boldsymbol{x}^{\prime}\in X\;|\;{g}(\boldsymbol{x}^{\prime})\leq\boldsymbol{y}\}$ is the integral domain, a subset of the image $X$, defined by the coordinate-wise inequality ${g}(\boldsymbol{x}^{\prime})\leq\boldsymbol{y}$. We thus have a general law for the transformation of a probability density distribution spanning the image $X$ on the space $\mathbb{R}^{d}$ (of exterior spacetime parameters), into a probability density distribution spanning the domain $Y$ on the space $\mathbb{R}^{n}$ (of EOS parameters).

Provided that ${g}(\boldsymbol{x})\equiv{f}^{-1}(\boldsymbol{x})$ -- i.e., the map $\boldsymbol{y}\mapsto\boldsymbol{x}$ is globally (and thus locally) invertible -- we can manipulate Equation~(\ref{eqn:fundamental posterior}) using the Heaviside step $\Theta(.)$ and delta $\delta(.)$ functions as follows:
\begin{equation}
\begin{aligned}
\mathcal{P}\left(\boldsymbol{\theta}\;|\ldots\right)&
=\mathop{\int}\left[\frac{\partial}{\partial y_{1}}\cdots\frac{\partial}{\partial y_{n+2s}}\mathop{\int}\Theta\left(\boldsymbol{y}-{g}(\boldsymbol{x}^{\prime})\right)\mathcal{P}(\boldsymbol{x}^{\prime}\;|\ldots)d\boldsymbol{x}^{\prime}\right]d\boldsymbol{\rho}d\boldsymbol{\Omega}\\
&=\mathop{\int}\left[\mathop{\int}\mathcal{P}(\boldsymbol{x}^{\prime}\;|\ldots)\mathop{\prod}_{i}^{n+2s}\frac{\partial}{\partial y_{i}}\Theta\left(y_{i}-g_{i}(\boldsymbol{x}^{\prime})\right)d\boldsymbol{x}^{\prime}\right]d\boldsymbol{\rho}d\boldsymbol{\Omega}\\
&=\mathop{\int}\left[\mathop{\int}\mathcal{P}(\boldsymbol{x}^{\prime}\;|\ldots)\mathop{\prod}_{i}^{n+2s}\delta\left(y_{i}-g_{i}(\boldsymbol{x}^{\prime})\right)d\boldsymbol{x}^{\prime}\right]d\boldsymbol{\rho}d\boldsymbol{\Omega}
=\mathop{\int}\left[\mathop{\int}\delta\left(\boldsymbol{y}-{g} (\boldsymbol{x}^{\prime})\right)\mathcal{P}(\boldsymbol{x}^{\prime}\;|\ldots)d\boldsymbol{x}^{\prime}\right]d\boldsymbol{\rho}d\boldsymbol{\Omega}.
\end{aligned}
\label{eqn: fundamental posterior w/ Heaviside}
\end{equation}
Further,
\begin{equation}
\begin{aligned}
\mathcal{P}\left(\boldsymbol{\theta}\;|\ldots\right)&
=\mathop{\int}\left[\mathop{\int}\delta\left(\boldsymbol{y}-{g}(\boldsymbol{x}^{\prime})\right)\mathcal{P}(\boldsymbol{x}^{\prime}\;|\ldots)\Biggl\lvert\det\left(\frac{\partial\boldsymbol{x}^{\prime}}{\partial{g}(\boldsymbol{x}^{\prime})}\right)\Biggl\rvert d{g}\right]d\boldsymbol{\rho}d\boldsymbol{\Omega}\\
&=\mathop{\int}\mathop{\sum}_{\mathcal{X}}\mathcal{P}(\boldsymbol{x}^{\prime}\;|\ldots)\Biggl\lvert\det\left(\frac{\partial\boldsymbol{x}^{\prime}}{\partial{g}(\boldsymbol{x}^{\prime})}\right)\Biggl\rvert d\boldsymbol{\rho}d\boldsymbol{\Omega}
=\mathop{\int}\mathcal{P}({f}(\boldsymbol{y})\;|\ldots)\Biggl\lvert\det\left(\frac{\partial{f}(\boldsymbol{y})}{\partial\boldsymbol{y}}\right)\Biggl\rvert d\boldsymbol{\rho}d\boldsymbol{\Omega},
\end{aligned}
\label{eqn: fundamental posterior w inner summation}
\end{equation}
where the Jacobian linearly approximates $f$, and its determinant\footnote{Note that the ordering of $\boldsymbol{y}$ and ${f}(\boldsymbol{y})$ in terms of matrix rows and columns is unimportant: the modulus accounts for a change in sign of the Jacobian determinant upon matrix reordering.} is the generalised differential volume transformation between the $\mathbb{R}^{d}$ and $\mathbb{R}^{n+2s}$ spaces for $d\equiv n+2s$. The summation over elements of the set $\mathcal{X}=\{\boldsymbol{x}^{\prime}\from\boldsymbol{y}-{g}(\boldsymbol{x}^{\prime})=\boldsymbol{0}\}$ reduces to a single term if global injectivity is satisfied (see Figure \ref{fig:invertible}). The active equation to be solved is thus $\boldsymbol{x}={f}(\boldsymbol{y})$, via application of the general relativistic map ${f}$ from interior to exterior parameters.\footnote{The equation $\boldsymbol{x}={f}(\boldsymbol{y})$ is in principle solvable because the parameters $\boldsymbol{x}$ are global integral quantities and the parameters $\boldsymbol{y}$ are interior source matter conditions. In practice, solving $\boldsymbol{x}={f}(\boldsymbol{y})$ can be highly non-trivial, depending on the desired level of accuracy and the rate of rotation \citep[see, e.g.,][]{Stergioulas2003}.}

The post-processing phase requires us to match analytic exterior spacetime solutions to numerical interior solutions. The constraint equation $d\equiv n+2s$ relates the number of exterior spacetime parameters $d$, the number of EOS parameters $n$, and the number of stars $s$ comprising the ensemble. For an $s=1$ ensemble, the number of continuous parameters controlling the parametrised exterior solution must match the number of free continuous parameters controlling the interior solution if our aim is to transform probability densities between spaces of exterior and interior parameters. If this condition is satisfied, the exterior parameters may each be defined as continuous random variables from a Bayesian perspective. In other words, if a star (with a particular EOS, central density, and rotation frequency) is \textit{uniquely} described by some subset of parameters of an exterior solution, any remaining parameters of that solution by definition have deterministic values and thus cannot be considered random variables. As a brief example, suppose the interior parameters of a star uniquely match to: the gravitational mass $M$; the (coordinate) equatorial radius $\Req$; the surface coordinate ellipticity $e$; the angular momentum $J$; the mass quadrupole moment $Q$; and the (asymptotic) angular rotation frequency $\Omega$ (see Sections \ref{sec:solutions for generative models} and \ref{sec:mapping definition}). It is then the case that all higher-order structure (admitted by an approximate exterior solution to the field equations) -- e.g., even-parity axisymmetric multipolar surface deformation, and multipolar metric structure -- is \textit{known} as a function of $(\Omega,M,\Req,e,J,Q)$.

However, in Sections~\ref{sec:solutions for generative models} and \ref{sec:mapping definition} we acknowledged that conditioning on observations of multiple stars is necessary because likelihood functions will in general not be sufficiently informative with respect to higher-order rotational spacetime structure of a single star. Let us consider the $s>1$ ensemble of static spacetimes, $\mathcal{S}$, defined in Section~\ref{sec:mapping definition}. For static spacetimes in $\mathcal{S}$, the number of EOS parameters $n$ is likely to always be at least limiting ($n\geq1$), and therefore $s$ is required to scale with $n$. In this case we require $d\equiv n+s$, and thus $s=n$.

Let us also consider the $s>1$ ensemble of rotating spacetimes, $\mathcal{R}$, defined in Section~\ref{sec:mapping definition}. The notion was to perturbatively solve for the global spacetime solution, matched at the stellar surface. During sampling of a posterior distribution of \textit{exterior} parameters, we need to use a parametrised analytic exterior solution which can be meaningfully matched to a numerical interior solution in the second phase of EOS parameter estimation. The exterior parameters considered for practicality were $\boldsymbol{x}=(\boldsymbol{\Omega},\boldsymbol{M},\boldsymbol{R}_{\textrm{eq}},\boldsymbol{e})$, meaning that oblate surfaces with ellipticities $\boldsymbol{e}$ are embedded (everywhere exterior of a horizon) in ambient spherically symmetric spacetimes parametrised by (rotationally perturbed) gravitational masses. If $\boldsymbol{x}=(\boldsymbol{M},\boldsymbol{R}_{\textrm{eq}},\boldsymbol{\Omega})$, the cardinality $s$ of the ensemble $\mathcal{R}$ must scale with $n\geq2$ if $d\equiv n+2s$ is to be satisfied. In this case $d=3s$ and thus $s=n$.\footnote{For a fixed EOS model it is possible to have an arbitrarily sized ensemble but not an arbitrary number of exterior parameters. If we condition on rotating spacetimes, the relevant constraint equation to be satisfied is $d\equiv n+2s$. Each star increments $s$ by one, and must increment $d$ by a \textit{minimum} of two: on average, $d$ must increase faster than $2s$ if the constraint equation to be satisfied. If a subset of stars contribute only \textit{two} parameters (such as a mass and rotation frequency), these stars will in general only affect constraining power on the shared EOS parameters -- i.e., the likelihood function -- if an observed star hosts an extreme exterior spacetime relative to other observed stars, and \textit{a posteriori}, marginalised over nuisance parameters, there is strong support for an extreme exterior solution.}

Note that in practice, \textit{posterior} information about the exterior parameters will typically be in the form of (approximately) independently and identically drawn \textit{samples} (see also Section \ref{sec:EP alternatives}). However, it is not possible for us to simply reweigh exterior parameter samples with respect to a Jacobian determinant to approximate samples from a transformed distribution on a space of interior parameters. One reason is that the nature of the map $f$ (requiring numerical integration of global spacetime properties given interior conditions) means that it is a non-trivial problem to accurately evaluate an inverse map $g$ (if such a map can be defined; see Section \ref{sec:pathologies summary}). A second reason is that the exterior parameter samples will not typically be drawn on a joint space of all exterior parameters, but on disjoint subspaces on a star-by-star basis. Therefore, instead of reweighing, we need to examine use of Equation~(\ref{eqn: fundamental posterior w inner summation}), where the integrand factor $\mathcal{P}({f}(\boldsymbol{y})\;|\ldots)$ is either a numerical approximation to a posterior obtained by sample smoothing (on the subspace associated with each star), or an analytical approximation to a posterior assuming, e.g., Gaussianity of a unimodal distribution.

\subsubsection{Interior-exterior mapping pathologies}\label{sec:pathologies summary}
In Appendix~\ref{app:pathologies}, we examine interior-exterior parameter mappings in detail to highlight the difficulties in meaningfully transforming a probability density distribution between spaces of exterior and interior parameters. We term certain map properties as \textit{pathological} because they impede straightforward transformation of probability density between continuous equal-dimension spaces. These details are best reserved as supplementary material: in this work we advocate more principled statistical inference, with the IP-paradigm serving as an example. Here we will very briefly summarise Appendix~\ref{app:pathologies}.

Diffeomorphicity (Fig.~\ref{fig:invertible}) is not often not satisfied due to the map $\boldsymbol{y}\mapsto\boldsymbol{x}$ being \textit{locally non-invertible} on subdomains of $Y$. The consequence of this is that: (i) the domain $W=\{\boldsymbol{x}^{\prime}\in X\;|\;{g}(\boldsymbol{x}^{\prime})\leq\boldsymbol{y}\}$ -- of the integral over $X$ in Equation~(\ref{eqn:fundamental posterior}) -- cannot be evaluated $\forall\boldsymbol{y}\in Y$ because $\boldsymbol{x}\mapsto\boldsymbol{y}$ does not have a unique local solution; (ii) we cannot solve Equation~(\ref{eqn:fundamental posterior}); and (iii) if Equation (\ref{eqn: fundamental posterior w inner summation}) is applied regardless, there are singularities in the Jacobian -- i.e., the differential of the map \textit{loses rank}. In this case the map $\boldsymbol{y}\mapsto\boldsymbol{x}$ suffers from a severe violation of diffeomorphicity which cannot be bypassed without redefinition of the \textit{continuous} spaces -- e.g., EOS reparametrisation or redefinition of the prior to span the domain $Y\subset\mathbb{R}^{n+2s}$. An instance of this behaviour is demonstrated in Raaijmakers et al. (submitted) for the phenomenological piecewise-polytrope EOS model \citep[][]{Mueller1985, Read2009}. This type of pathology also arises in mappings for ensembles of cardinality $s>1$.

The map $\boldsymbol{y}\mapsto\boldsymbol{x}$ can also be \textit{globally} non-invertible despite being everywhere locally invertible. As above, this means that the domain $W=\{\boldsymbol{x}^{\prime}\in X\;|\;{g}(\boldsymbol{x}^{\prime})\leq\boldsymbol{y}\}$ cannot be evaluated $\forall\boldsymbol{y}\in Y$ because ${g}(\boldsymbol{x})$ does not have a unique solution. In other words, we cannot solve Equation~(\ref{eqn:fundamental posterior}) without adapting the space of exterior parameters in order to preserve information. If Equation (\ref{eqn: fundamental posterior w inner summation}) is applied regardless, probability density associated with a single image $\boldsymbol{x}\in X$ is mapped to multiple preimages $\boldsymbol{y}\in Y$. In principle we can adapt the space of exterior parameters by defining a \textit{discrete} parameter. However, the tractability of such a modification is contextual; if the nature of the discrete information loss can be understood analytically then definition of such a discrete parameter is tractable.

Finally, surjectivity (Fig. \ref{fig:invertible}) is never satisfied in practice because the image $X$ is in general only a subset of the region $Z\subset\mathbb{R}^{d}$ which a (prior and thus posterior) probability distribution of exterior parameters is defined to continuously span. Even if there exist no other pathologies, the equality of Equation~(\ref{eqn: fundamental posterior w inner summation}) is infringed because the total probability mass is not conserved under transformation between spaces.

There is a class of maps $\boldsymbol{y}\mapsto\boldsymbol{x}$ which may not violate any of the above pathologies but surjectivity. These mappings are for a single star, and an EOS model: (i) whose continuous parameters control the EOS function everywhere on the subdomain of local comoving density that the EOS function is free; and (ii) whose mapping between parameter space and function space is invertible. Such mappings, however, are highly restrictive and not pragmatic in terms of modelling.

A danger is that one could invoke Equation~(\ref{eqn: fundamental posterior w inner summation}) without fully examining the properties of both a map $\boldsymbol{y}\mapsto\boldsymbol{x}$, and of the probability density distribution (on the space of exterior parameters) to which the transformation is applied; such an approach can only be viewed as unprincipled. An ill-defined distributional transformation of a posterior between spaces of exterior and interior parameters leads to an implicit prior on the latter space; this prior may exhibit undesirable properties (such as being, e.g., informative and structured) or may simply be ill-behaved. For instance, if the Jacobian is singular at some point $\boldsymbol{y}$ but Equation~(\ref{eqn: fundamental posterior w inner summation}) is applied, a prior density of \textit{zero} is implicitly defined at that point in the space of interior parameters. We detail such an approach in Appendix~\ref{sec:prior under reparametrisation} because it is relevant to understanding the literature (see Section~\ref{sec:Review}).

\subsubsection{Archival posterior constraints on exterior spacetime parameters}\label{sec:EP alternatives}

The motivation for considering the EP-paradigm (despite the existence, in most cases, of a non-invertible interior-exterior parameter mapping) is that posterior constraints on exterior parameters already exist in the astrophysical literature \citep[the reviews on efforts to infer static masses and radii, for example see,][]{Miller2016,OzelReview}. These constraints are in some instances very expensive to derive -- due to costly likelihood function evaluations, for instance\footnote{See the discussion under Equation~(\ref{eqn:final group posterior unshared case}) for further reasons why posterior constraints can be expensive to derive \citep[see also][]{Hogg2010}.} -- and this justifies consideration of how to include this information in EOS inference.

Posterior information is in most cases in the form of (approximately) independent and identically drawn (weighted) samples which are typically used to estimate integrals \citep{MacKay2003}. In the preceding discussion we identified that if likelihood functions are identical in both paradigms (IP and EP), it is the space on which a prior is defined that is the critical distinction. Suppose we have archival posterior samples of the exterior parameters of a star. These samples would have been drawn from a distribution defined as proportional to the product of likelihood function and prior, and can trivially be marginalised over all nuisance (hyper)parameters. If the likelihood function is expensive to evaluate and our aim is to repurpose these samples for use in the IP-paradigm, we need to use the samples for likelihood evaluation when directly sampling from a posterior distribution defined on a space of interior parameters.

If the prior on the space of the exterior parameters of a given star is noninformative \textit{relative} to the data acquired by observing that star, the exterior parameter samples are approximately drawn from a nuisance-marginalised likelihood function (which is in general physically bounded). These samples can thus be smoothed on a joint space to generate a numerical approximation to the nuisance-marginalised likelihood function up to some normalisation factor for the purpose of EOS parameter estimation (but not model comparison as discussed in Section \ref{sec:model selection}); an analytic approximation to this numerical likelihood function could also be calculated if appropriate. Suppose, on the other hand, that we cannot robustly consider the posterior samples for a given star to be drawn from a likelihood function because the associated prior is not sufficiently noninformative. In this case we can divide a smoothed, numerical approximation to the (nuisance-marginalised) posterior by the known analytic prior density \citep[effectively Importance Sampling followed by smoothing for sufficiently similar priors, e.g.,][]{MacKay2003,Neiswanger2016}.

Such likelihood function approximations can then be used directly in the IP-paradigm with a prior defined on a space of interior parameters provided that the function \textit{at least} spans the image $X$ of the domain $Y$ under map $f$ (see Fig.~\ref{fig:invertible} and Section~\ref{sec:computational tractability}).

\subsubsection{Summary}\label{sec:explicit summary}
The general relativtivistic map $f\colon Y\to X$, $\boldsymbol{y}\mapsto\boldsymbol{x}$ is numerical in nature and in general expected to be non-invertible (refer to Appendix~\ref{app:pathologies}). The EP-paradigm could be considered in practice as an approach to the problem of utilising existing posterior constraints on exterior parameters, but in general is ill-defined. \textit{It is crucial to appreciate however, that under such a transformation, a prior is implicitly defined on a space of interior parameters which may exhibit undesirable properties (e.g., being informative, or worse, ill-behaved), despite being well-behaved and otherwise noninformative (e.g., invariant to bijective reparametrisation) on a space of exterior parameters}. It would be necessary to carefully consider on a case-by-case basis the sensitivity of inferences about EOS parameters to properties (including pathologies) of the mapping, and thus prove that any posterior EOS inferences derived in this way are likelihood-dominated \citep[e.g.,][]{Steiner2016}. A demonstration of implicitly prior-dominated EOS parameter estimation via the EP-paradigm may be found in Raaijmakers et al. (submitted). From a more objectivist Bayesian perspective, it is clear that we cannot construct a noninformative prior \citep[see][and references therein]{Jeffreys,Jeffreys1961,Bernardo1979,Robert2007,Robert2009} which is invariant to exterior-interior reparametrisation because in practice we are handling a non-bijective mapping. Strictly, we should formulate the EOS parameter estimation problem in the context of, e.g., the IP-paradigm (see Sections~\ref{sec:implicit} and \ref{sec:EP alternatives}).

\subsection{Bayesian equation of state model checking and comparison}\label{sec:model selection}
We now remark on Bayesian EOS model checking and comparison \citep[see, e.g., Chapters 28 and 29, and Chapters 6 and 7, respectively, of][]{MacKay2003,Gelman_book} which is crucial for development of theory. The IP-paradigm naturally supports posterior EOS model checking and model comparison, whilst the EP-paradigm offers no scope for such testing. Until stated otherwise, the body of this section regards the IP-paradigm.

The parameters of EOS models may be defined under some phenomenological functional form for thermodynamic properties of dense matter, or they may have some deeper theoretical meaning in terms of microphysical particle interactions \citep[for a recent review, see, e.g.,][]{Margueron2018}. In the latter case it may be that EOS models derived from a fundamental theory have many fixed parameters and the models clearly form some discrete set --  a discrete model space. In the former (phenomenological) case, the model space could clearly be defined as discrete, continuous, or discrete-continuous mixed, since we are dealing purely with functional forms of type $\mathbb{R}\to\mathbb{R}$. \citet{Gelman_book} recommends connecting discrete models on a continuous space where it is rational to do so, but otherwise it is not \textit{necessary} to construct continuous model spaces.

A primary use for posterior estimation of phenomenological EOS parameters is to guide development of microphysical theories of dense matter, given that the desired understanding is, ultimately, of a fundamental theoretical nature and not merely of an empirical or phenomenological nature. A \textit{synergistically} constrained phenomenological EOS model, which is by consensus preferred under some metric above other phenomenological EOS models, may prove highly useful for advancement of fundamental theory \citep[for work in this direction, see, e.g.,][]{Raithel2016,Margueron2018}. At present however, parameter estimation is arguably being performed disjointly by a number of research groups: each group produces constraints which alone are feasibly useful for advancement of fundamental theory, but can be relatively weak. Moreover, constraints published by independent groups may in principle exhibit contention (which is objectively not an issue if collaborations are conceived to improve models). Communication, open-sourcing, and rigorous testing -- such as posterior predictive checks and sensitivity analysis (see below) -- should help alleviate such contention via model development (see discussion in Section~\ref{sec:computational tractability}).

If we desire a metric for \textit{a posteriori} comparison of some (discrete) set of models which populate a space, we could opt to compute (approximate) Bayes' factors: ratios of posterior probabilities\footnote{If the model space is everywhere continuous, ratios of posterior probability densities are naturally required. If the space is discrete-continuous mixed, ratios of products of probability densities and probability masses are required. If the model space is discrete (e.g., distinct values of continuous model parameters and hyperparameters are not considered to define distinct models) as we consider in this context, ratios of posterior probability masses are required. See, e.g., \citet{FarrMandel2015} and references therein.} of models conditional on the data set $\mathcal{D}$. In this context we are particularly interested in comparing models with different EOS parametrisations but otherwise shared model structure (e.g., nuisance parameters and nuisance hyperparameters), such that the model space of interest, $\mathscr{M}$, is populated by overlapping models $\mathcal{M}_{k}$ where $k\in\mathbb{N}$. Let there exist a prior probability \textit{mass} distribution on the space $\mathscr{M}$, conditional on prior information $\mathscr{I}$ (data sets). The posterior probability of a global (hierarchical generative) model $\mathcal{M}_{k}\in\mathscr{M}$ is given by
\begin{equation}
\mathcal{P}(\mathcal{M}_{k}\;|\;\mathcal{D},\mathscr{M},\mathscr{I})=\frac{\mathcal{P}(\mathcal{D}\;|\;\mathcal{M}_{k},\mathscr{M},\mathscr{I}_{k})\mathcal{P}(\mathcal{M}_{k}\;|\;\mathscr{M},\mathscr{I})}{\mathcal{P}(\mathcal{D}\;|\;\mathscr{M},\mathscr{I})},
\label{eqn:posterior model probabilities}
\end{equation}
where: $\mathcal{P}(\mathcal{M}_{k}\;|\;\mathscr{M},\mathscr{I})$ is the prior probability of the model; $\mathcal{P}(\mathcal{D}\;|\;\mathscr{M},\mathscr{I})$ is the prior predictive probability of $\mathcal{D}$ conditional on the model space $\mathscr{M}$, given by the marginalisation
\begin{equation}
\mathcal{P}(\mathcal{D}\;|\;\mathscr{M},\mathscr{I})=\mathop{\sum}_{k}\mathcal{P}(\mathcal{D}\;|\;\mathcal{M}_{k},\mathscr{M},\mathscr{I}_{k})\mathcal{P}(\mathcal{M}_{k}\;|\;\mathscr{M},\mathscr{I});
\end{equation}
and $\mathcal{P}(\mathcal{D}\;|\;\mathcal{M}_{k},\mathscr{M},\mathscr{I}_{k})$ is the prior predictive probability of $\mathcal{D}$ conditional on the model $\mathcal{M}_{k}$ and on the prior information $\mathscr{I}_{k}$ on the continuous parameters of that model. This prior predictive probability is given by the marginalisation
\begin{equation}
\mathcal{P}(\mathcal{D}\;|\;\mathcal{M}_{k},\mathscr{M},\mathscr{I}_{k})=\mathcal{\int}\mathcal{P}(\mathcal{D}\;|\;\boldsymbol{\theta}_{k},\mathcal{M}_{k},\mathscr{M},\mathscr{I}_{k})\mathcal{P}(\boldsymbol{\theta}_{k}\;|\;\mathcal{M}_{k},\mathscr{M},\mathscr{I}_{k})d\boldsymbol{\theta}_{k},
\end{equation}
where the $\boldsymbol{\theta}_{j}$ are the EOS parameters defined under the model $\mathcal{M}_{k}$ and all other model parameters and hyperparameters (shared by $\mathcal{M}_{k}$, $\forall k$) have been implicitly marginalised over.

Explicitly, the prior predictive probability which follows from Equation~(\ref{eqn:exact posterior}) is given by
\begin{equation}
\mathcal{P}\left(\mathcal{D}\;|\;\mathcal{M}_{k},\mathscr{M},\mathscr{I}_{k}\right)
=\mathop{\int}
\underbrace{\mathcal{P}\left(\boldsymbol{\theta}_{k},\boldsymbol{\rho},\boldsymbol{\Omega},\boldsymbol{\alpha},\boldsymbol{\eta},\boldsymbol{\beta}\;|\;\mathcal{M}_{k},\mathscr{M},\mathscr{I}_{k}\right)}_{\pi(\boldsymbol{\theta}_{k},\boldsymbol{\rho},\boldsymbol{\Omega},\boldsymbol{\alpha},\boldsymbol{\eta},\boldsymbol{\beta})}
\mathop{\prod}_{\varg}\underbrace{\mathcal{P}\left(\mathcal{D}_{\varg}\;|\;\boldsymbol{\theta}_{k},\boldsymbol{\rho}_{\varg},\boldsymbol{\Omega}_{\varg},\boldsymbol{\eta}_{\varg},\mathcal{M}_{k},\mathscr{M}\right)}_{\mathcal{L}_{\varg}(\boldsymbol{\theta}_{k},\boldsymbol{\rho}_{\varg},\boldsymbol{\Omega}_{\varg},\boldsymbol{\eta}_{\varg})}
d\boldsymbol{\eta}d\boldsymbol{\beta}d\boldsymbol{\rho}d\boldsymbol{\Omega}d\boldsymbol{\alpha}d\boldsymbol{\theta}_{k}.
\label{eqn:prior predictive probability}
\end{equation}
Prior predictive probability integrals are in general difficult to compute accurately,\footnote{Thus giving rise to an entire literature on the subject, for which the reader may consider \citet{MacKay2003} and \citet{Gelman_book} as potential entry points, or, e.g., \citet[and references therein]{PolyChord_1} which details a recent nested sampling algorithm and implementation designed to handle difficult evidence integrals in moderately high-dimensional parameter spaces.} and the complexity of the integral given by Equation~(\ref{eqn:prior predictive probability}) is evident. In Section \ref{sec:computational tractability} we focussed only on evaluation of the marginal \textit{posterior} distribution of EOS parameters -- which does \textit{not} require integration -- and concluded that to increase tractability, multiple research groups with specialist knowledge and independent compute resources may need to synergistically perform EOS parameter estimation via a series of Bayesian updates.

Let us consider, therefore, the posterior distribution communicated to group $\varg=2$ by group $\varg=1$ in the example given in Section \ref{sec:computational tractability} (characterised by groups sharing only the EOS parameters $\boldsymbol{\theta}$ and the interior hyperparameters $\boldsymbol{\alpha}$). Based on Equations~(\ref{eqn:first group posterior}) and (\ref{eqn:communicated posterior 1}), group $\varg=2$ considers a (hyper)prior distribution of EOS parameters $\boldsymbol{\theta}_{k}$ and hyperparameters $\boldsymbol{\alpha}_{k}$ given by
\begin{equation}
\mathcal{P}\left(\boldsymbol{\theta}_{k},\boldsymbol{\alpha}_{k}\;|\;\mathcal{D}_{1},\mathcal{M}_{k},\mathscr{M},\mathcal{I}_{1}\right)
\propto\mathop{\int}
\underbrace{\mathcal{P}\left(\boldsymbol{\theta}_{k},\boldsymbol{\rho}_{1},\boldsymbol{\Omega}_{1},\boldsymbol{\alpha}_{k},\boldsymbol{\eta}_{1},\boldsymbol{\beta}_{1}\;|\;\mathcal{M}_{k},\mathscr{M},\mathcal{I}_{1}\right)}_{\pi(\boldsymbol{\theta}_{k},\boldsymbol{\rho}_{1},\boldsymbol{\Omega}_{1},\boldsymbol{\alpha}_{k},\boldsymbol{\eta}_{1},\boldsymbol{\beta}_{1})}
\underbrace{\mathcal{P}\left(\mathcal{D}_{1}\;|\;\boldsymbol{\theta}_{k},\boldsymbol{\rho}_{1},\boldsymbol{\Omega}_{1},\boldsymbol{\eta}_{1},\mathcal{M}_{k},\mathscr{M}\right)}_{\mathcal{L}_{1}(\boldsymbol{\theta}_{k},\boldsymbol{\rho}_{1},\boldsymbol{\Omega}_{1},\boldsymbol{\eta}_{1})}
d\boldsymbol{\eta}_{1}d\boldsymbol{\beta}_{1}d\boldsymbol{\rho}_{1}d\boldsymbol{\Omega}_{1},
\end{equation}
where the normalisation is the prior predictive probability of the data subset $\mathcal{D}_{1}$ conditional on the model $\mathcal{M}_{k}$ and prior information $\mathcal{I}_{1}$:
\begin{equation}
\mathcal{P}\left(\mathcal{D}_{1}\;|\;\mathcal{M}_{k},\mathscr{M},\mathcal{I}_{1}\right)
=\mathop{\int}
\underbrace{\mathcal{P}\left(\boldsymbol{\theta}_{k},\boldsymbol{\rho}_{1},\boldsymbol{\Omega}_{1},\boldsymbol{\alpha}_{k},\boldsymbol{\eta}_{1},\boldsymbol{\beta}_{1}\;|\;\mathcal{M}_{k},\mathscr{M},\mathcal{I}_{1}\right)}_{\pi(\boldsymbol{\theta}_{k},\boldsymbol{\rho}_{1},\boldsymbol{\Omega}_{1},\boldsymbol{\alpha}_{k},\boldsymbol{\eta}_{1},\boldsymbol{\beta}_{1})}
\underbrace{\mathcal{P}\left(\mathcal{D}_{1}\;|\;\boldsymbol{\theta}_{k},\boldsymbol{\rho}_{1},\boldsymbol{\Omega}_{1},\boldsymbol{\eta}_{1},\mathcal{M}_{k},\mathscr{M}\right)}_{\mathcal{L}_{1}(\boldsymbol{\theta}_{k},\boldsymbol{\rho}_{1},\boldsymbol{\Omega}_{1},\boldsymbol{\eta}_{1})}
d\boldsymbol{\eta}_{1}d\boldsymbol{\beta}_{1}d\boldsymbol{\rho}_{1}d\boldsymbol{\Omega}_{1}d\boldsymbol{\alpha}_{k}d\boldsymbol{\theta}_{k}.
\end{equation}
The marginal posterior distribution computed by group $\varg=2$ is thus given by
\begin{equation}
\begin{aligned}
\mathcal{P}\left(\boldsymbol{\theta}_{k},\boldsymbol{\alpha}_{k}\;|\;\mathcal{D}_{2},\mathcal{M}_{k},\mathscr{M},\mathcal{I}_{2}\right)
\propto\mathop{\int}
\mathcal{P}\left(\boldsymbol{\theta}_{k},\boldsymbol{\alpha}_{k}\;|\;\mathcal{D}_{1},\mathcal{M}_{k},\mathscr{M},\mathcal{I}_{1}\right)
&\mathcal{P}\left(\boldsymbol{\rho}_{2},\boldsymbol{\Omega}_{2},\boldsymbol{\eta}_{2},\boldsymbol{\beta}_{2}\;|\;\boldsymbol{\theta}_{k},\boldsymbol{\alpha}_{k},\mathcal{M}_{k},\mathscr{M},\mathcal{I}_{2}^{\prime}\right)\\
&\qquad\times\mathcal{P}\left(\mathcal{D}_{2}\;|\;\boldsymbol{\theta}_{k},\boldsymbol{\rho}_{2},\boldsymbol{\Omega}_{2},\boldsymbol{\eta}_{2},\mathcal{M}_{k},\mathscr{M}\right)
d\boldsymbol{\eta}_{2}d\boldsymbol{\beta}_{2}d\boldsymbol{\rho}_{2}d\boldsymbol{\Omega}_{2},
\end{aligned}
\end{equation}
with normalisation
\begin{equation}
\begin{aligned}
\mathcal{P}\left(\mathcal{D}_{2}\;|\;\mathcal{M}_{k},\mathscr{M},\mathcal{I}_{2}\right)
=\mathop{\int}
\mathcal{P}\left(\boldsymbol{\theta}_{k},\boldsymbol{\alpha}_{k}\;|\;\mathcal{D}_{1},\mathcal{M}_{k},\mathscr{M},\mathcal{I}_{1}\right)
&\mathcal{P}\left(\boldsymbol{\rho}_{2},\boldsymbol{\Omega}_{2},\boldsymbol{\eta}_{2},\boldsymbol{\beta}_{2}\;|\;\boldsymbol{\theta}_{k},\boldsymbol{\alpha}_{k},\mathcal{M}_{k},\mathscr{M},\mathcal{I}_{2}^{\prime}\right)\\
&\qquad\times\mathcal{P}\left(\mathcal{D}_{2}\;|\;\boldsymbol{\theta}_{k},\boldsymbol{\rho}_{2},\boldsymbol{\Omega}_{2},\boldsymbol{\eta}_{2},\mathcal{M}_{k},\mathscr{M}\right)
d\boldsymbol{\eta}_{2}d\boldsymbol{\beta}_{2}d\boldsymbol{\rho}_{2}d\boldsymbol{\Omega}_{2}d\boldsymbol{\alpha}_{k}d\boldsymbol{\theta}_{k},
\end{aligned}
\end{equation}
where $\mathcal{I}_{2}\triangleq\mathcal{D}_{1}\cup\mathcal{I}_{1}\cup\mathcal{I}_{2}^{\prime}$. It follows that the joint prior predictive probability of data subsets $\mathcal{D}_{1}$ and $\mathcal{D}_{2}$ is given by
\begin{equation}
\mathcal{P}\left(\mathcal{D}_{1},\mathcal{D}_{2}\;|\;\mathcal{M}_{k},\mathscr{M},\mathcal{I}_{1},\mathcal{I}_{2}^{\prime}\right)
=\mathcal{P}\left(\mathcal{D}_{1}\;|\;\mathcal{M}_{k},\mathscr{M},\mathcal{I}_{1}\right)\mathcal{P}\left(\mathcal{D}_{2}\;|\;\mathcal{M}_{k},\mathscr{M},\mathcal{I}_{2}\right).
\end{equation}

The marginal posterior distribution computed by group $\varg=h$ is given by
\begin{equation}
\begin{aligned}
\mathcal{P}\left(\boldsymbol{\theta}_{k},\boldsymbol{\alpha}_{k}\;|\;\mathcal{D}_{h},\mathcal{M}_{k},\mathscr{M},\mathcal{I}_{h}\right)
\propto\mathop{\int}
\mathcal{P}\left(\boldsymbol{\theta}_{k},\boldsymbol{\alpha}_{k}\;|\;\mathcal{D}_{h},\mathcal{M}_{k},\mathscr{M},\mathcal{I}_{h}\backslash\mathcal{I}_{h}^{\prime}\right)
&\mathcal{P}\left(\boldsymbol{\rho}_{h},\boldsymbol{\Omega}_{h},\boldsymbol{\eta}_{h},\boldsymbol{\beta}_{h}\;|\;\boldsymbol{\theta}_{k},\boldsymbol{\alpha}_{k},\mathcal{M}_{k},\mathscr{M},\mathcal{I}_{h}^{\prime}\right)\\
&\qquad\times\mathcal{P}\left(\mathcal{D}_{h}\;|\;\boldsymbol{\theta}_{k},\boldsymbol{\rho}_{h},\boldsymbol{\Omega}_{h},\boldsymbol{\eta}_{h},\mathcal{M}_{k},\mathscr{M}\right)
d\boldsymbol{\eta}_{h}d\boldsymbol{\beta}_{h}d\boldsymbol{\rho}_{h}d\boldsymbol{\Omega}_{h},
\end{aligned}
\end{equation}
with normalisation
\begin{equation}
\begin{aligned}
\mathcal{P}\left(\mathcal{D}_{h}\;|\;\mathcal{M}_{k},\mathscr{M},\mathcal{I}_{h}\right)
=
\mathop{\int}
\mathcal{P}\left(\boldsymbol{\theta}_{k},\boldsymbol{\alpha}_{k}\;|\;\mathcal{D}_{h},\mathcal{M}_{k},\mathscr{M},\mathcal{I}_{h}\backslash\mathcal{I}_{h}^{\prime}\right)
&\mathcal{P}\left(\boldsymbol{\rho}_{h},\boldsymbol{\Omega}_{h},\boldsymbol{\eta}_{h},\boldsymbol{\beta}_{h}\;|\;\boldsymbol{\theta}_{k},\boldsymbol{\alpha}_{k},\mathcal{M}_{k},\mathscr{M},\mathcal{I}_{h}^{\prime}\right)\\
&\qquad\times\mathcal{P}\left(\mathcal{D}_{h}\;|\;\boldsymbol{\theta}_{k},\boldsymbol{\rho}_{h},\boldsymbol{\Omega}_{h},\boldsymbol{\eta}_{h},\mathcal{M}_{k},\mathscr{M}\right)
d\boldsymbol{\eta}_{h}d\boldsymbol{\beta}_{h}d\boldsymbol{\rho}_{h}d\boldsymbol{\Omega}_{h}d\boldsymbol{\alpha}_{k}d\boldsymbol{\theta}_{k}.
\end{aligned}
\end{equation}
Finally, it follows that the prior predictive probability of $\mathcal{D}$ -- Equation~(\ref{eqn:prior predictive probability}) -- may be computed as
\begin{equation}
\mathcal{P}\left(\mathcal{D}\;|\;\mathcal{M}_{k},\mathscr{M},\mathscr{I}_{k}\right)=\mathcal{P}\left(\mathcal{D}\;|\;\mathcal{M}_{k},\mathscr{M},\mathcal{I}_{1},\mathcal{I}_{2}^{\prime},\ldots,\mathcal{I}_{h}^{\prime}\right)=\mathop{\prod}_{\varg=1}^{h}\mathcal{P}\left(\mathcal{D}_{\varg}\;|\;\mathcal{M}_{k},\mathscr{M},\mathcal{I}_{\varg}\right),
\end{equation}
such that $\mathscr{I}_{k}\coloneqq\mathcal{I}_{1}\cup\mathcal{I}_{2}^{\prime}\cup\ldots\cup\mathcal{I}_{h}^{\prime}$. Each group must evaluate their respective prior predictive probability integral simultaneously to sampling from their respective posterior. In order to facilitate simultaneous posterior sampling and evidence estimation, a sophisticated nested sampling algorithm such as implemented in the open-source software \textsc{PolyChord} \citep[][]{PolyChord_1} could be employed by each group.

Ratios of the posterior probabilities given by Equation~(\ref{eqn:posterior model probabilities}), such as
\begin{equation}
\mathcal{B}_{k,j}\coloneqq\frac{\mathcal{P}(\mathcal{M}_{k}\;|\;\mathcal{D},\mathscr{M},\mathscr{I})}{\mathcal{P}(\mathcal{M}_{j\neq k}\;|\;\mathcal{D},\mathscr{M},\mathscr{I})}
=\frac{\mathcal{P}(\mathcal{D}\;|\;\mathcal{M}_{k},\mathscr{M},\mathscr{I}_{k})\mathcal{P}(\mathcal{M}_{k}\;|\;\mathscr{M},\mathscr{I})}{\mathcal{P}(\mathcal{D}\;|\;\mathcal{M}_{j\neq k},\mathscr{M},\mathscr{I}_{j})\mathcal{P}(\mathcal{M}_{j\neq k}\;|\;\mathscr{M},\mathscr{I})},
\end{equation}
may thus be computed as a metric for use in model comparison (and thus decision making). We note that even distributions defined on discrete-continuous-mixed model spaces can in principle be \textit{sampled} from in order to approximate Bayes' factors of discrete models marginalised over their continuous parameters \citep[e.g.,][and references therein]{Farr2011, FarrMandel2015}. It follows that a global analysis as described above and in Section \ref{sec:implicit}, and illustrated in Fig. \ref{fig:IP-paradigm procedure}, does not in principle need to be \textit{repeated} for each discrete EOS model and associated (hyper)prior: the posterior probability masses of the discrete models can in principle be communicated between groups, together with approximate posterior density distributions of shared continuous model parameters, for sequential Bayesian updates. This is at the expense of complicating the sampling process conducted by each group contributing to the update chain.

Naturally, the process of \textit{defining} priors (which may be conditional on independent data sets) is a modelling process, and thus models can in principle be distinct due to definition of different priors \citep[see, e.g., the more detailed physical discussion in][]{Steiner2016}. \citet[][]{Gelman_book} recommend that Bayes' factors are not invoked as a metric for model comparison, one reason being potential sensitivity of prior predictive probabilities of $\mathcal{D}$ to the definition of (hyper)prior distributions of model parameters (and hyperparameters); indeed, prior choices can be somewhat arbitrary and may be untestable -- prior bounds, for instance, can be problematic if there do not exist well-defined mathematical or physical bounds within a theory. On the other hand, noninformative priors can in principle be chosen (by both objectivist and subjectivist Bayesians) which are invariant to bijective transformations \citep[][and references therein]{Jeffreys,Jeffreys1961,Bernardo1979,Robert2007,Robert2009}, provided such priors are tractable to compute and posteriors are integrable.

An advantage of invoking Bayes' factors as a metric for model comparison is that evidence integrals can be performed simultaneously to parameter estimation, as described above. Posterior predictive testing, on the other hand, must be performed given a set of samples from a posterior distribution -- that is, \textit{following} parameter estimation. Furthermore, Bayes' factors inherently invoke Occam's razor because if prior choices under a particular model lead to large predictive complexity in data space, that model will be disfavoured over a model that is compatible with the data and exhibits simpler predictive properties \citep[][]{MacKay2003,Gelman_book}. Lastly, if models which populate a model space \textit{overlap} -- e.g., due to unshared EOS parametrisations but complete sharing of nuisance (hyper)parameters -- then the prior differences are restricted to the interior (hyper)parameters. Analysis of the sensitivity of posterior distributions and prior predictive probabilities to prior definitions would be more tractable in such a case \citep[see the discussion of sensitivity analyses in][]{Gelman_book}. 

For the purpose of model checking, we may in principle calculate posterior predictive probability distributions, each conditional on a model $\mathcal{M}_{k}\in\mathscr{M}$ \citep[see, e.g.,][]{Gelman_book}. Let us define some data set $\widetilde{\mathcal{D}}$ that exists on the same space as the observed data $\mathcal{D}$; the posterior predictive sampling distribution of the random vector $\widetilde{\mathcal{D}}$ is conditional on $\mathcal{D}$ and is given by
\begin{equation}
\mathcal{P}\left(\widetilde{\mathcal{D}}\;|\;\mathcal{D},\mathcal{M}_{k},\mathscr{M},\mathscr{I}_{k}\right)
=\mathop{\int}
\mathcal{P}\left(\boldsymbol{\theta}_{k},\boldsymbol{\rho},\boldsymbol{\Omega},\boldsymbol{\alpha}_{k},\boldsymbol{\eta},\boldsymbol{\beta}\;|\;\mathcal{D},\mathcal{M}_{k},\mathscr{M},\mathscr{I}_{k}\right)
\mathop{\prod}_{\varg}\mathcal{P}\left(\widetilde{\mathcal{D}}_{\varg}\;|\;\boldsymbol{\theta}_{k},\boldsymbol{\rho}_{\varg},\boldsymbol{\Omega}_{\varg},\boldsymbol{\eta}_{\varg},\mathcal{M}_{k},\mathscr{M}\right)
d\boldsymbol{\eta}d\boldsymbol{\beta}d\boldsymbol{\rho}d\boldsymbol{\Omega}d\boldsymbol{\alpha}_{k}d\boldsymbol{\theta}_{k}.
\label{eqn:posterior predictive probability}
\end{equation}
where $\mathcal{P}\left(\boldsymbol{\theta}_{k},\boldsymbol{\rho},\boldsymbol{\Omega},\boldsymbol{\alpha}_{k},\boldsymbol{\eta},\boldsymbol{\beta}\;|\;\mathcal{D},\mathcal{M}_{k},\mathscr{M},\mathscr{I}_{k}\right)$ is the full posterior distribution of all (hyper)parameters defined under the model $\mathcal{M}_{k}$, and $\widetilde{\mathcal{D}}_{\varg}$ is the subset of $\widetilde{\mathcal{D}}$ which is relevant to group $\varg$, in analogy with $\mathcal{D}_{\varg}\subset\mathcal{D}$.  Given the existence of a set of samples from the posterior distribution of model (hyper)parameters, the approximate posterior predictive distribution of $\widetilde{\mathcal{D}}$ is given by \citep[][]{MacKay2003,Gelman_book}
\begin{equation}
\mathcal{P}\left(\widetilde{\mathcal{D}}\;|\;\mathcal{D},\mathcal{M}_{k},\mathscr{M},\mathscr{I}_{k}\right)
\approx
\frac{1}{N}\mathop{\sum}_{d=1}^{N}
\mathcal{P}\left(\widetilde{\mathcal{D}}\;|\;\left(\boldsymbol{\theta}_{k},\boldsymbol{\rho},\boldsymbol{\Omega},\boldsymbol{\eta}\right)_{d},\mathcal{M}_{k},\mathscr{M}\right)
=
\frac{1}{N}\mathop{\sum}_{d=1}^{N}
\mathop{\prod}_{\varg}
\mathcal{P}\left(\widetilde{\mathcal{D}}_{\varg}\;|\;\left(\boldsymbol{\theta}_{k},\boldsymbol{\rho}_{\varg},\boldsymbol{\Omega}_{\varg},\boldsymbol{\eta}_{\varg}\right)_{d},\mathcal{M}_{k},\mathscr{M}\right),
\end{equation}
where $d$ enumerates (approximately) independent and identical samples $\left(\boldsymbol{\theta}_{k},\boldsymbol{\rho},\boldsymbol{\Omega},\boldsymbol{\eta}\right)_{d}$ from the marginal posterior distribution of all model parameters, $\mathcal{P}\left(\boldsymbol{\theta}_{k},\boldsymbol{\rho},\boldsymbol{\Omega},\boldsymbol{\eta}\;|\;\mathcal{D},\mathcal{M}_{k},\mathscr{M},\mathscr{I}_{k}\right)$. The posterior predictive distribution of $\widetilde{\mathcal{D}}$ can be compared to $\mathcal{D}$ (graphically or otherwise) as a self-consistency check \citep[][]{Gelman_book}. However, to obtain such samples from the full posterior, each group must not marginalise over their respective nuisance parameters $\boldsymbol{\eta}_{\varg}$ as suggested in Section \ref{sec:implicit} as a way to simplify computation; these parameters would need to be \textit{sampled} by subsequent groups but, provided they do not parametrise the sampling distributions of other groups -- i.e., are unshared -- they only enter in the communicated posterior factor. Moreover, whilst such a distribution is also an indicator of unmodelled phenomena which may be included in a more sophisticated description of data generation, it is intractable to fully separate out the inaccuracies of an EOS model from inaccuracies in the submodelling of complex astrophysical phenomena and data acquisition. It may be most rational and practical for each group to instead perform such posterior predictive checking at each stage in the chain illustrated in Fig. \ref{fig:IP-paradigm procedure} in order to isolate potential \textit{submodel} inaccuracies before the next scheduled Bayesian update.

We now return briefly to discussion of the EP-paradigm. The reason that the EP-paradigm offers no scope for EOS model checking and comparison is that a prior is explicitly defined on a space of exterior parameters, not on a space of interior parameters. This means that an undesirable or poorly behaved prior is implicitly defined a space of interior parameters via an ill-defined transformation (see Appendix~\ref{sec:prior under reparametrisation}). Bayesian evidences (prior predictive probabilities) are sensitive to prior definition \citep{Gelman_book}, and in general will not be a robust metric for model comparison for priors defined via the EP-paradigm.\footnote{If the interior-exterior parameter mapping \textit{were} diffeomorphic, the evidence would be identically written as a marginalisation over either space -- e.g., Equation (25) -- but may still be sensitive to which space a prior is directly defined on unless, given the likelihood, it is proven effectively invariant to the reparametrisation given by the mapping.} However, \textit{posterior} predictive probabilities and distributions (see above discussion) will also be sensitive to EOS priors defined via the EP-paradigm if said priors are informative or ill-behaved, and if observables are sensitive to the interior source matter conditions via exterior spacetime structure; this means that posterior predictive checking (as described above) should in principle indicate that a poor choice of prior has been made. Finally, for a fixed data set and fixed definitions of exterior parameters, the EOS model must satisfy the constraint that $d\equiv n+2s$ (see Section~\ref{sec:explicit}); this restricts the \textit{space} of functional forms which may be compared, leaving only functional forms with a certain number, $n=d-2s$, of parameters.

\section{Literature review}\label{sec:Review}
In this section we compare the EOS parameter estimation paradigms detailed in Section \ref{sec:inference paradigms} to a number of approaches in the literature which have in general been implemented for smaller-scale analyses (than envisioned in this work) of electromagnetic astronomical data, and which we aimed to generalise as stated in Section \ref{sec:intro}. Authors use different notation and definitions of statistical objects, so we refrain from attempting to summarise other formalisms quantitatively, in order to avoid confusion; instead we opt to only qualitatively frame these approaches in the context of the parameter estimation paradigms offered above. We first briefly summarise critical points from Section \ref{sec:inference paradigms} which are useful for facilitating comparison.

Bayesian inference of an EOS shared by a model ensemble of stars can be formulated in a principled manner via the IP-paradigm: direct posterior EOS parameter estimation conditioned on some global hierarchical generative model and observations of an ensemble of compact stars, wherein a prior is defined on a space of interior parameters. The IP-paradigm naturally admits cognitive parallelisation of the modelling process amongst research groups. However, in practice, the scope for \textit{computational} parallelisation (over research groups) of EOS parameter estimation is limited to likelihood acceleration because (hyper)parameters are \textit{shared} amongst groups, meaning that the mapping of a global posterior distribution must be performed via a series of Bayesian distributional updates (see Section \ref{sec:computational tractability}).

The general relativistic interior-exterior parameter mapping is not diffeomorphic. If the EP-paradigm is invoked in order to leverage existing posterior constraints, it is conditional on the understanding that inferences are potentially sensitive to the implicit definition of a prior (on the space of interior parameters) with undesirable properties. In general, it is important to prove that posterior EOS inferences are likelihood-dominated, and when defining an EOS prior via the EP-paradigm this is especially true.

\subsection{\citet{Ozel2009}}
The approach to EOS parameter estimation discussed in \citet[][]{Ozel2009} for three static stars can be interpreted from a Bayesian perspective, and formed the basis for the EP-paradigm (see Section~\ref{sec:explicit}): a prior distribution is defined on a space of exterior parameters -- i.e., a space of Schwarzschild gravitational masses and circumferential radii. Notably, their posterior distribution of EOS parameters is marginalised over masses instead of central densities, which for a given EOS model can map injectively to gravitational masses of \textit{stable} spacetime solutions, but this is not always true (see the first footnote in Section~\ref{sec:Steiner2010}).

\citet{Ozel2009} did not discuss hyperparameters, but instead directly invoked the ansatz that the marginal posterior distribution of exterior parameters is separable over stars, implicit to which is the definition of a prior on a space of exterior parameters (see Section~\ref{sec:explicit}). The consequential non-surjectivity of the mapping (see Appendix~\ref{sec:surjectivity}), and the non-injectivity incurred by modelling multiple stars (see Appendix~\ref{sec:injectivity 2}), are both implicitly treated via normalisation of the transformed distribution on the EOS parameter space to define a marginal posterior.

The authors conditioned on a piecewise-polytropic EOS model, for which there exist continuous subdomains of the associated space of interior parameters in which Jacobian is singular. The particular \textit{analytic} (likelihood-function dominated) marginal posterior distribution of exterior parameters the authors considered appears to exhibit non-negligible support (probability density) only at points where the Jacobian is non-singular, thus avoiding a dominant source of distortion incurred by transforming a probability density distribution between spaces related by a non-diffeomorphic mapping (see Appendix~\ref{sec:injectivity}). Raaijmakers et al. (submitted) consider posterior distributions spanning a broader subset of the space of exterior parameters, including points at which the Jacobian is singular, in order to demonstrate sensitivity of EOS parameter inferences to pathologies in the interior-exterior parameter mapping.

We cannot, however, \textit{guarantee} that the posterior they define on their EOS parameter space can be considered robust against this pathology in the mapping. Another important point to note is that a prior defined on a space of exterior parameters may, under such an ill-defined reparametrisation (with an implicit marginalisation), manifest as informative on a space of EOS parameters -- \textit{even if singularities in the Jacobian are limited to subsets of a space (of exterior parameters) where the likelihood function is negligible}. In principle this means that only with a particular choice of prior did \citet{Ozel2009} demonstrate that one might be able to distinguish, \textit{a posteriori}, between three distinct EOS when conditioning on \textit{unbiased} constraints derived from observations with near-future high-energy telescope missions. We refer the reader to \cite{Steiner2016} for discussion on the role of the prior in Bayesian EOS inference.

The approach of \citet[][]{Ozel2009} was implemented by \citet[][]{Ozel2010}. In \citet[][]{Ozel2010}, the marginal joint posterior distribution of the exterior parameters was defined as a product over stars. The posterior associated with each star was conditional on X-ray spectral modelling. As is the case in \citet[][]{Ozel2009}, the conclusions of \citet[][]{Ozel2010} (who also used a piecewise-polytropic EOS model) may be robust to the Jacobian being singular in subdomains of parameter space. The reason for this that the joint posterior distribution of all exterior parameters appears to be non-negligible only over subdomains of the exterior parameter space where a non-singular Jacobian can be defined.

For completeness, we also remark on Raaijmakers et al. (submitted): one of the analytic posterior distributions (``Case 2'') defined on a joint space of all exterior parameters is similar to those handled by \citet[][]{Ozel2009} and \citet[][]{Ozel2010}, except that the high-density region is lower in mass (by $\sim0.3$ M$_{\odot}$) and higher in radius (by $\sim1$ km). In this case, Raaijmakers et al. (submitted) found evidence for distortion incurred due to Jacobian singularities. However the distributions considered by \citet[][]{Ozel2009} and \citet[][]{Ozel2010} are sufficiently distinct to mitigate \textit{some} concern that the conclusions are not robust to this pathology in the mapping. We make this claim on the basis of Figure 2 and the other distributions considered in Raaijmakers et al. (submitted).

We note that in both \citet[][]{Ozel2009} and \citet[][]{Ozel2010}, the joint posterior (over all stars) is defined as a product of (lower-dimensional) posterior distributions of exterior parameters, whose high-density regions \textit{overlap} (partially and fully, respectively) when projected onto a single $(M,R)$-plane (see their figures and respectively). It is important for one to be aware of the local non-invertibility of the interior-exterior parameter mapping for ensembles of more than one star, as discussed in Appendix~\ref{sec:injectivity 2}. We cannot therefore guarantee that the conclusions of \citet[][]{Ozel2009} and, in particular, \citet[][]{Ozel2010} are robust to this pathology either.

The work of \citet[][]{Ozel2015_Jacobian} is relevant to the current discussion: these authors tackled a simpler problem than the EOS inference problem, but it is nevertheless abstractly relevant. The presentation in our work is not entirely congruent with that of \citet[][]{Ozel2015_Jacobian},\footnote{A crucial point is that -- in our opinion -- it is a misnomer to refer to the comparison between these posterior formulations as \textit{Frequentist} versus \textit{Bayesian}: they are \textit{both} founded on \textit{Bayesianism}, as can be understood from the fact that probability distributions are associated with model parameters. The differences manifest due to problems with prior definition when parameter spaces within a theory are deterministically related, but not invertibly.} and thus we will summarise their solution in the context of our work. \citet[][]{Ozel2015_Jacobian} encounter the need to transform a probability density distribution between equal-dimension parameter spaces\footnote{In \citet[][]{Ozel2015_Jacobian}, the spaces of interest are: (i) the joint space of a Schwarzschild gravitational mass and circumferential radius; and (ii) two parameters derived from these exterior spacetime parameters.} between which there exists a deterministic mapping. If within a theory subsets of parameters are deterministically related but carry distinct physical meaning, one is faced with the problem of deciding which space to define a prior on. From an objectivist perspective, one might ideally find a way to define a prior which is noninformative relative to a likelihood function (which is identical for both parameter subsets) on both spaces. However, if a deterministic mapping between equal-dimension spaces is non-diffeomorphic, a probability density distribution cannot -- strictly -- be \textit{reversibly} transformed between spaces without distortion (of a non-numerical origin).

If one proceeds regardless, then in the context of Bayesianism, a prior distribution originally defined on one space is transformed into a prior on the alternate space, but this prior may well be poorly behaved and not exhibit desirable properties (e.g., continuously positive and noninformative within some boundary). In the case of \citet[][]{Ozel2015_Jacobian}, the parameter spaces in question are two-dimensional and there exists an \textit{analytic} non-diffeomorphic mapping between them; of particular concern is the existence of a singularity in the Jacobian, meaning that a finite prior density at points in one space have zero prior density at the corresponding points in another space. As discussed in Section \ref{sec:pathologies summary}, singularities in the Jacobian are also problematic for EOS parameter estimation via the EP-paradigm; in this context however, the mapping is more difficult to examine and handle because it is numerical in nature \citep[e.g., integrating the Einstein field equations in a perturbative manner, see][]{Hartle1967,HT1968}.

The solution reached by \citet[][]{Ozel2015_Jacobian} is to define the prior on the space of fundamental interest, and this is consistent with both \citet[][]{Steiner2010} -- for the problem of estimating exterior parameters -- and the principled solution we advocate in the context of EOS parameter estimation, which is to invoke the IP-paradigm (Section \ref{sec:implicit}). As discussed in Section \ref{sec:EP alternatives}, however, there may exist posterior constraints on exterior spacetime parameters which contain valuable \textit{likelihood} information (entangled with a prior defined on a space of exterior parameters) and which were computationally expensive to derive. In this case, the EP-paradigm was framed as one option for crudely utilising such information, and desirable alternatives for harnessing this information -- with minimal computational expense -- for use with the IP-paradigm were offered in Section \ref{sec:EP alternatives}.

\subsection{\citet{Steiner2010}}\label{sec:Steiner2010}
The approach of \citet[][]{Steiner2010} marked a paradigm shift relative to \citet[][]{Ozel2009}: these authors developed an EOS parameter estimation procedure for an arbitrary number of static stars, in which a prior (together with a joint likelihood function) is defined on a joint space of EOS parameters and Schwarzschild gravitational masses (instead of central densities). This procedure was applied to X-ray spectral data acquired via observations of six stars (see their section 4).

Hyperparameters of the prior distribution of gravitational masses are undefined (or alternatively may be considered as fixed), with the exception of the EOS parameters which also appear as hyperparameters required to define the domain of stable interior solutions. The authors performed direct EOS parameter estimation in a manner which we have encapsulated within the general framework of the IP-paradigm (see Section \ref{sec:implicit}); because prior distributions are not defined solely on the exterior-parameter space, it is unimportant for the purpose of EOS parameter estimation whether or not their EOS parametrisation (which is not piecewise-polytropic) has an associated diffeomorphic mapping between a space of interior parameters and a space of exterior parameters.\footnote{We note that using a (Schwarzschild) gravitational mass in lieu of a central density has notable implications. First, it is important to be aware that a conditional prior on one space may be informative under reparametrisation. Second, as noted in \citep{Steiner2010}, certain EOS parametrisations exhibit phase transitions which result, for a fixed EOS and rotation frequency, in disjoint branches of stable solutions connected by a branch of unstable solutions in central density \citep[see, e.g.,][]{ST1983, Glendenning2000, Zdunik2006, Read2009}. Therefore, whilst a prior distribution of a gravitational mass can be continuously finite between bounds (conditional on the EOS and rotation frequency), a conditional prior distribution of a central density must be zero in an intermediate branch of unstable solutions. However, if an EOS-conditional prior is defined on a joint space of gravitational masses, then for a given EOS and rotation frequency, there may be two stable solutions for a single gravitational mass if there are disjoint branches of stable solutions. Thus one must decide, e.g., which solution to use for likelihood evaluation. One way to handle this problem is to introduce a discrete parameter which distinguishes between the two solutions as detailed in Appendix~\ref{app:discrete information loss}; the associated prior is a probability mass distribution on a space of two elements. When there are not multiple solutions, both elements on the discrete space represent the same solution. When discrete nuisance parameter is marginalised over via a two-element summation, the marginalised likelihood function on the joint space of a gravitational mass and EOS parameters is as expected where there is a single stable solution, and is a prior-weighted average of the likelihoods when there are two stable solutions.} \citet[][]{Steiner2010} also comment in their discussion that the \citet{Ozel2009} formulation (the basis for the EP-paradigm) is some form of approximation to the \citet[][]{Steiner2010} Monte Carlo formulation.

Conditional on X-ray spectral data, \citet[][]{Steiner2010} formulated a set of independent normalised probability distributions: one distribution was defined per star on the joint space of its Schwarzschild gravitational mass and circumferential radius -- see their equation (30). These distributions are each the normalised product of a likelihood function of the exterior parameters (a gravitational mass $M$ and a circumferential radius $R$) and a \textit{bounded}, \textit{uniform} joint prior distribution of these exterior parameters. These distributions are thus marginal posterior distributions of the exterior parameters, and each is implicitly marginalised over any nuisance parameters and nuisance hyperparameters which enter in submodels invoked to describe the observational data subset associated with a given star. The authors then wrote their joint likelihood function of the EOS parameters and the masses as proportional to the product of these (posterior) distributions -- see their equation (31).

In other words, the authors: (i) implicitly fixed the hyperparameters of the prior distribution of exterior parameters to obtain a uniform prior distribution of each $(M,R)$ pair; (ii) constructed a marginal posterior distribution of the exterior parameters (as illustrated in Fig. \ref{fig:EP diagram}); (iii) wrote their joint likelihood function of EOS parameters and masses as \textit{proportional} to this posterior distribution of exterior parameters, where the coefficient of proportionality\footnote{Notably, this coefficient of proportionality would be useful for performing Bayesian model comparison if global prior predictive probabilities of the data are invoked as a metric for model performance (see Section \ref{sec:model selection}).} is \textit{not} dimensionless; (iv) computed a marginal posterior distribution of their EOS parameters and masses, given this likelihood function and a joint prior distribution of EOS parameters and masses.

Such a formulation is numerically valid provided the \textit{implicit} prior distributions of \textit{exterior} parameters are uniform (or sufficiently flat relative to the likelihood function; see Section \ref{sec:EP alternatives}), in which case the joint likelihood function of the EOS parameters and gravitational masses is not (appreciably) distorted. It follows that the authors did not need to transform a probability density distribution between spaces of exterior and interior parameters by everywhere multiplying by a Jacobian determinant as is required in the EP-paradigm.

The \citet[][]{Steiner2010} approach to EOS parameter estimation was implemented by \citet[][]{SteinerLattimerBrown2013}, \citet[][]{LattimerSteiner2014}, \citet[][]{LattimerSteiner2014_LMXB}, \citet[][]{Nattila2016}, and in the open-source software \texttt{bamr} \citep[][]{Steiner_bamr}.

\subsection{\citet{Ozel16}}
We now discuss the approach of \citet{Ozel16} -- later reproduced by \citet{Raithel2017} -- which is distinctly different from the approach of \citet[][]{Ozel2009}. The justification for the paradigm shift given by \citet{Ozel16} is that the approach of \citet{Ozel2009} does not permit an arbitrary number of stars (and thus exterior parameters) for some particular functional form for the EOS (with some fixed number of parameters). \citet{Steiner2010} also comment that the \citet{Ozel2009} approach to EOS parameter estimation is a simple approximation to \citet{Steiner2010}, also based on a dimensionality argument. Whilst this is a valid statement if the interior-exterior parameter mapping were diffeomorphic but subsequent Bayesian updates were not possible, it is insufficient to capture all of the problems we need to be aware of in the EP-paradigm.  \citet{Ozel16} state that their approach is similar to that of \citet[][see Section \ref{sec:Steiner2010}]{Steiner2010}: the statistical terminology used by \citet{Ozel16}, however, differs markedly from that used by \citep[][]{Steiner2010} without clarification, and their approach was not described in sufficient detail to enable us to guarantee that the two approaches are indeed the same.

In attempting to encompass the approach of \citet{Ozel16} within a general framework, we identified issues with definitions of the data. The authors described the Schwarzschild gravitational masses and circumferential radii of static stars as ``observables'': these masses and radii, however, are \textit{parameters} defined under some submodel for a \textit{fixed} (observed) data subset, and these parameters are only statistically \textit{inferable}, conditioned on all model assumptions -- i.e., parameters are statistically estimated, but not observed. The true data in this analysis formed an astronomical data set, and the joint sampling distribution of that data must be parametrised in terms of interior parameters (EOS parameters, and either central densities or Schwarzschild gravitational masses) and nuisance parameters \citep[e.g.,][]{Hogg2010}; the likelihood function of the parameters is equivalent to the conditional joint probability density of the fixed data.

The likelihood formulation is also inaccurate. The likelihood function of the EOS parameters and central densities was written as a product of the joint probability density distribution of the mass and radius of each star; each joint probability distribution is implicitly marginalised over nuisance parameters -- see equations (13) and (14) in \citet{Ozel16}. Each of these joint probability distributions (of a mass and a radius) is \textit{conditional} on the EOS parameters and the central densities: this implies that the masses and radii are \textit{not} deterministically related to the interior parameters (unless the conditional probability distributions exhibit singular support, which is not the case). There is evidence that likelihood functions might have been confused with posterior distributions -- see equation (9) of \citet{Ozel16} -- and terms such as ``posterior likelihood'' are used. This latter term is unorthodox and not explained: we infer that it denotes a likelihood function marginalised over nuisance parameters.

\citet[][]{Raithel2017} gave a more explicit definition of the likelihood: for each star the likelihood of the ``observed'' mass and radius, conditional on the EOS parameters and a central density, was written as an unbounded, normalised two-dimensional Gaussian distribution on the joint space of the mass and the radius, evaluated at the mass and radius which match to a stable interior solution to the Tolman-Oppenheimer-Volkoff equations (corresponding to values of the EOS parameters and a central density). In this formulation: (i) the ``observed'' mass-radius pair is equivalent to the maximum likelihood mass-radius pair; (ii) equation (11) in \citet[][]{Raithel2017} defines the probability density of that maximum likelihood mass-radius pair, conditional on the interior parameters $\boldsymbol{y}=(\boldsymbol{\theta},\boldsymbol{\rho})$; and (iii) the authors then maximise the corresponding probability density distribution with respect to central densities as an approximate marginalisation over central densities. This conditional probability is effectively in lieu of writing the likelihood as suggested above -- i.e., as the probability density of the data conditional on the interior parameters (and marginalised over any nuisance parameters). If our understanding is accurate, then \textit{numerically} their marginal posterior distribution of the EOS parameters is consistent with the \citet[][]{Steiner2010} formulation (and is thus encapsulated by the IP-paradigm) in that the true joint likelihood function has been normalised with respect to the exterior parameters.

\citet[][]{Raithel2017} also discuss a problem with transforming probability density between parameter spaces. In this context the transformation is (effectively) between a space of interior parameters, and a one-dimensional exterior-parameter space (a Schwarzschild circumferential radius), including a marginalisation over all but one dimension of the space of interior parameters. The authors highlight the problem of a prior distribution on one space having undesirable properties on a deterministically related space due to the Jacobian of the mapping, and thus parameter inferences on one space (in this case the space of a radius given some Schwarzschild gravitational mass) are sensitive to this implicit choice of prior. Whilst \citet[][]{Steiner2010} is critiqued, there is no discussion of this thematic problem in the context of earlier work -- e.g., \citet{Ozel2009} and \citet{Ozel2010}. Moreover, the authors highlight this issue without addressing pathologies (such as Jacobian singularities) in the mapping however, which can induce highly undesirable behaviour in an implicitly defined prior.

\subsection{\citet{Alsing2017}}
The approach of \citet{Alsing2017} is encapsulated by general framework of the IP-paradigm (see Section \ref{sec:implicit}). The authors formulated EOS parameter estimation in a novel way (relative to the literature discussed thus far): they considered the EOS parameters \textit{only} as hyperparameters which are \textit{shared} between static stars. In particular, the authors defined a joint (hyper)prior distribution of EOS parameters, Schwarzschild gravitational masses (instead of central densities; cf. Section \ref{sec:Steiner2010}), hyperparameters of the mass distribution (other than the EOS parameters), and nuisance (hyper)parameters.

In section 6.3.2 of \citet[][]{Alsing2017}, the shared hyperparameters that are marginalised over may be identified as the hyperparameters $\boldsymbol{\alpha}$ we define in the IP-paradigm (see Section \ref{sec:implicit}). The remaining hyperparameters of the mass distribution are the EOS parameters, as equivalently stated in Equation~(\ref{eqn:joint prior distribution of densities and frequencies}) where we write the prior distribution of the central densities (and rotation frequencies) as conditional on \textit{our} hyperparameters $\boldsymbol{\alpha}$ and \textit{our} EOS parameters $\boldsymbol{\theta}$. \citet[][]{Alsing2017} formulated the EOS parameters as hyperparameters by computing the \textit{maximum} gravitational mass of a stable star permitted by a given EOS; this maximum mass is the mass at which the \textit{mass distribution} (the prior distribution of the mass of a model star) is truncated. Thus the authors integrated the Tolman-Oppenheimer-Volkoff equations to map EOS parameters and initial (central) densities to interior solutions (and thus to exterior solutions) for the purpose of stability analysis. We note that, due to degeneracy, the marginal posterior distribution of the EOS parameters is sensitive only to \textit{extreme} gravitational masses which are not permitted by an arbitrary subset of the EOS parameter space but are strongly supported by the data (conditioned on the global model) and have finite support \textit{a priori} (according to the hyperparameters $\boldsymbol{\alpha}$ and the corresponding hyperprior distribution).

We now remark on likelihood function of the gravitational masses adopted from the literature that \citet[][]{Alsing2017} cite. If authors of this literature are particularly interested in estimating masses, they will: (i) numerically or analytically marginalise their likelihood functions over any nuisance (hyper)parameters;\footnote{A relevant point to note is that if we were to adopt from the literature a (normalised) likelihood function of the mass which has been marginalised over the radius of a static star (given some prior), the likelihood function of the mass is contaminated (or distorted) with exterior solutions which do not match to a stable interior solution to the field equations given an EOS parameter vector and a mass. For static stars the radius cannot be considered as a \textit{continuous} random variable if the mass and the EOS parameters are random variables because there exists a constraint given by integration of the field equations. Instead we should use a reported \textit{covariance matrix} and condition on the assumption of Gaussianity if no higher-order distributional moments are available. During EOS parameter estimation we are required to evaluate an approximation to this likelihood function, where the radius is deterministically calculated from the EOS parameters and the mass.} (ii) normalise an integrable, marginalised likelihood function on the space of a mass (implicitly invoking some bounded or unbounded flat prior distribution); and (iii) report the first few distributional moments (e.g., the expectation and variance if Gaussianity is assumed). Analytic approximations can be used to accelerate likelihood function mapping or permit fast marginalisations -- e.g., a Laplace approximation for nuisance-marginalisation and mass normalisation.

\citet[][]{Alsing2017}: (i) cited radio pulsar timing analyses, and timing analyses of X-ray and optical observations of eclipsing high-mass binary systems; and (ii) adopted the reported moments of the normalised distributions to construct approximations which are \textit{proportional} to the marginal likelihood functions of the masses. The sampling distributions of such observations are not conditional on the radii of the model stars; the gravitational masses, however, influence the binary motion, which in turn is encoded in time-domain light-curve periodicity. The EOS parameters only exert influence on the mass distribution, and thus the first footnote of Section \ref{sec:Steiner2010} is not applicable to \citet{Alsing2017}.

\subsection{General remarks}\label{sec:gen remarks}
In general, it is common for authors to perform EOS parameter estimation as a stand-alone population-level study of static stars by adopting (implicitly normalised) \textit{likelihood} functions of the masses and radii reported in the literature. These likelihood functions are typically approximated by analytic \textit{distributions} -- e.g., multivariate Gaussian distributions -- given the first few reported distributional moments. Furthermore, in the literature discussed above the authors all justifiably assume that the stars are static in order to enforce spacetime spherical symmetry; this assumption accelerates integration of the field equations and, consequentially, accelerates both likelihood evaluation and conditional-prior evaluation given that EOS parameters are also hyperparameters of a prior distribution of central densities (or gravitational masses) and rotation frequencies (if non-static) for the purpose of assigning finite probability densities only to stable solutions to the field equations.

When authors perform such stand-alone analyses, there is naturally less direct communication and collaboration both between the independent research groups who report likelihoods, and between those groups and the group estimating the EOS parameters by harvesting constraints from the literature. This means that, in general, global self-consistency and the harnessing of statistical information  -- e.g., by identifying (hyperparameters) which are shared between groups -- will be submaximal. Nevertheless, this is a valid approximation to the synergistic global analysis structures we discussed in Section \ref{sec:computational tractability} -- of which one is illustrated in Fig.~\ref{fig:IP-paradigm procedure}. In particular, it can be considered as a form of parallel acceleration of likelihood evaluation, whereby all groups communicate \textit{fast} (to evaluate) marginal likelihood function (approximations) to a single ``synthesis'' group (Section \ref{sec:computational tractability}). In the case described in this section, the likelihood functions are defined on spaces of exterior parameters, and are communicated through the literature via, e.g., distributional moments; such an analysis structure is also discussed in Section \ref{sec:EP alternatives}) as a solution for using likelihood-dominated exterior parameter constraints.

\section{Summary}\label{sec:Summary}
We have examined two paradigms for astrophysical Bayesian EOS inference. The crux of the distinction between these paradigms is the \textit{choice} of space on which a prior is defined when there exists a deterministic but non-diffeomorphic mapping in general relativistic gravity between a space of continuous interior (local source matter) parameters and a space of continuous exterior (non-local spacetime) parameters, where it is fundamentally assumed that a subset of the interior parameters are \textit{shared} between all members of some abstract model ensemble of stars.

The Interior-Prior (IP) paradigm (Section \ref{sec:implicit}) is characterised by \textit{direct} posterior EOS parameter estimation: a prior distribution is defined on a space of interior parameters, which include parameters of an \textit{assumedly universal} (core) EOS shared by all compact stars. The Exterior-Prior (EP) paradigm (Section~\ref{sec:explicit}) is characterised by direct posterior estimation of \textit{exterior} parameters, followed by an \textit{ill-defined} transformation of that (marginal) posterior distribution onto a space of EOS parameters. In the EP-paradigm a prior distribution is thus instead defined on that joint space of exterior parameters, none of which are shared between stars. Individual summaries of parameter estimation under each of these paradigms can be found in Sections \ref{sec:implicit summary} and \ref{sec:explicit summary}.

In Section \ref{sec:Review} we concluded that the IP-paradigm offers a generalisation of most studies comprising the recent literature from the (high-energy) electromagnetic astrophysics community. Moreover, although not treated in this work, we claimed that the IP-paradigm is consistent with the Bayesian EOS inference of \citet{Lackey2015}, at least under the modelling assumption that the same cold EOS is shared by all stars during the observations of both the electromagnetic and gravitational wave communities.

The IP-paradigm may be viewed as principled from the perspective of a Bayesian who is fundamentally interested in microphysical or thermodynamical EOS models and their associated parameters. Conditioned on the assumption of an EOS (and hyperparameters) which is (are) shared by all model stars, posterior estimation of parameters of an EOS model cannot be trivially parallelised with respect to distinct research groups with special expertise and independent compute resources. In an organised, collaborative analysis (Section \ref{sec:computational tractability}), groups must either: (i) sequentially update collective posterior knowledge; or (ii) communicate likelihood functions on a subspace of either the global joint interior space or global joint exterior space. Moreover, there is flexibility regarding whether all groups or a single group solve (approximations to) the Einstein field equations. Finally, there is full scope for each group to accelerate their likelihood evaluation procedure whilst communications are pending, thus accelerating the global analysis (Section \ref{sec:computational tractability}).

The EP-paradigm on the other hand -- based on older literature \citep[][]{Ozel2009} -- might be intuited as a natural approach because of the existence of a deterministic mapping (the integral form of some approximation to the Einstein field equations) between sets of local and global parameters defined in general relativistic gravity. Moreover, exterior spacetime parameters are unshared between stars and hyperparameters can be considered \textit{fixed} (Section~\ref{sec:explicit}), meaning that a global EOS inference problem can seemingly be distilled into two posterior sampling phases, wherein the first phase consists of trivially parallelised posterior sampling (by distinct research groups with independent compute resources). However, the EP-paradigm is: (i) critically dependent on properties of the mapping between interior and exterior parameter spaces satisfying certain criteria (Section~\ref{sec:explicit}); and (ii) restrictive in terms of updating posterior knowledge (Appendix~\ref{sec:updating}), and in terms of EOS model comparison and posterior predictive checking (Section \ref{sec:model selection}). We argued that priors should not be defined via the EP-paradigm if at all avoidable because this implicitly leads to a prior distribution of interior parameters which in general may be ill-behaved or have undesirable properties (such as being informative). From a more objectivist Bayesian perspective we cannot in practice construct a noninformative prior which is invariant under exterior-interior reparametrisation. The reader can refer to Appendix~\ref{app:pathologies} (in particular \ref{sec:prior under reparametrisation}) where we elucidate the relation between the paradigms. It may be for these reasons that direct EOS parameter estimation has typically been adopted in some form in the more recent literature.

In conclusion, we advocate for direct posterior inference of the parameters of interest -- microphysical or thermodynamical EOS parameters -- and thus use of a framework akin to the IP-paradigm detailed in this work, which is more principled and more flexible than the alternative EP-paradigm. We believe that authors of the literature have exhibited similar ideologies, and so we offer the IP-paradigm as a more general framework under which the electromagnetic community can collaborate on dense matter EOS inference in the future.

\section*{Acknowledgements}

We thank Daniela Huppenkothen for preliminary discussions from which a number of useful ideas for this work originated. We thank Michiel van der Klis for comments on the manuscript. We thank members of the NICER light-curve SWG for discussions and for the open sharing of perspectives and ideas; in particular we thank Fred Lamb and Cole Miller, who offered detailed comments on the topic. We are grateful to Jocelyn Read for pointing out the consistency of this work with earlier literature from the gravitational wave community on EOS inference via tidal deformabilities; this helped us better understand how the electromagnetic and gravitational wave communities can exploit synergy within an overarching statistical inference framework. In this regard we also thank Chris Van Den Broeck, Archisman Ghosh, Samaya Nissanke, Tanja Hinderer, and Tim Dietrich, amongst others, who participated in a fruitful discussion on this topic of community cross-collaboration. We are very grateful to the referee for suggesting and inspiring improvements to this work.  The authors acknowledge support from ERC Starting Grant No. 639217 CSINEUTRONSTAR (PI Watts).




\bibliographystyle{mnras}
\bibliography{references}

\begin{thebibliography}{}
\makeatletter
\relax
\def\mn@urlcharsother{\let\do\@makeother \do\$\do\&\do\#\do\^\do\_\do\%\do\~}
\def\mn@doi{\begingroup\mn@urlcharsother \@ifnextchar [ {\mn@doi@}
  {\mn@doi@[]}}
\def\mn@doi@[#1]#2{\def\@tempa{#1}\ifx\@tempa\@empty \href
  {http://dx.doi.org/#2} {doi:#2}\else \href {http://dx.doi.org/#2} {#1}\fi
  \endgroup}
\def\mn@eprint#1#2{\mn@eprint@#1:#2::\@nil}
\def\mn@eprint@arXiv#1{\href {http://arxiv.org/abs/#1} {{\tt arXiv:#1}}}
\def\mn@eprint@dblp#1{\href {http://dblp.uni-trier.de/rec/bibtex/#1.xml}
  {dblp:#1}}
\def\mn@eprint@#1:#2:#3:#4\@nil{\def\@tempa {#1}\def\@tempb {#2}\def\@tempc
  {#3}\ifx \@tempc \@empty \let \@tempc \@tempb \let \@tempb \@tempa \fi \ifx
  \@tempb \@empty \def\@tempb {arXiv}\fi \@ifundefined
  {mn@eprint@\@tempb}{\@tempb:\@tempc}{\expandafter \expandafter \csname
  mn@eprint@\@tempb\endcsname \expandafter{\@tempc}}}

\bibitem[\protect\citeauthoryear{Abbott et~al.}{Abbott et~al.}{2017}]{Abbott17}
Abbott B.~P.,  et~al., 2017, \mn@doi [Phys. Rev. Lett.]
  {10.1103/PhysRevLett.119.161101}, 119, 161101

\bibitem[\protect\citeauthoryear{{AlGendy} \& {Morsink}}{{AlGendy} \&
  {Morsink}}{2014}]{AlGendy2014}
{AlGendy} M.,  {Morsink} S.~M.,  2014, \mn@doi [ApJ]
  {10.1088/0004-637X/791/2/78}, \href
  {http://cdsads.u-strasbg.fr/abs/2014ApJ...791...78A} {791, 78}

\bibitem[\protect\citeauthoryear{{Alford} \& {Han}}{{Alford} \&
  {Han}}{2016}]{AlfordHan2016}
{Alford} M.~G.,  {Han} S.,  2016, \mn@doi [European Physical Journal A]
  {10.1140/epja/i2016-16062-9}, \href
  {http://cdsads.u-strasbg.fr/abs/2016EPJA...52...62A} {52, 62}

\bibitem[\protect\citeauthoryear{Alford \& Sedrakian}{Alford \&
  Sedrakian}{2017}]{Alford17}
Alford M.,  Sedrakian A.,  2017, \mn@doi [Phys. Rev. Lett.]
  {10.1103/PhysRevLett.119.161104}, 119, 161104

\bibitem[\protect\citeauthoryear{{Alford}, {Han}  \& {Prakash}}{{Alford}
  et~al.}{2013}]{Alford13}
{Alford} M.~G.,  {Han} S.,   {Prakash} M.,  2013, \mn@doi [\prd]
  {10.1103/PhysRevD.88.083013}, \href
  {http://adsabs.harvard.edu/abs/2013PhRvD..88h3013A} {88, 083013}

\bibitem[\protect\citeauthoryear{{Alsing}, {Silva}  \& {Berti}}{{Alsing}
  et~al.}{2017}]{Alsing2017}
{Alsing} J.,  {Silva} H.~O.,   {Berti} E.,  2017, preprint, \href
  {http://cdsads.u-strasbg.fr/abs/2017arXiv170907889A} {} (\mn@eprint {arXiv}
  {1709.07889})

\bibitem[\protect\citeauthoryear{{Arzoumanian} et~al.}{{Arzoumanian}
  et~al.}{2014}]{Arzoumanian14}
{Arzoumanian} Z.,  et~al., 2014, in Space Telescopes and Instrumentation 2014:
  Ultraviolet to Gamma Ray. p. 914420, \mn@doi{10.1117/12.2056811}

\bibitem[\protect\citeauthoryear{{Azevedo-Filho} \& {Shachter}}{{Azevedo-Filho}
  \& {Shachter}}{2013}]{LAPLACE_APPROX}
{Azevedo-Filho} A.,  {Shachter} R.~D.,  2013, preprint, \href
  {http://adsabs.harvard.edu/abs/2013arXiv1302.6782A} {} (\mn@eprint {arXiv}
  {1302.6782})

\bibitem[\protect\citeauthoryear{{Baub{\"o}ck} et~al.}{{Baub{\"o}ck}
  et~al.}{2012}]{Bauboeck2012}
{Baub{\"o}ck} M.,  et~al., 2012, ApJ, \href
  {http://adsabs.harvard.edu/abs/2012ApJ...753..175B} {753, 175}

\bibitem[\protect\citeauthoryear{{Baub{\"o}ck} et~al.}{{Baub{\"o}ck}
  et~al.}{2013}]{Bauboeck2013}
{Baub{\"o}ck} M.,  et~al., 2013, \mn@doi [ApJ] {10.1088/0004-637X/777/1/68},
  \href {http://cdsads.u-strasbg.fr/abs/2013ApJ...777...68B} {777, 68}

\bibitem[\protect\citeauthoryear{{Beloborodov}}{{Beloborodov}}{2002}]{Beloborodov2002}
{Beloborodov} A.~M.,  2002, \mn@doi [\apjl] {10.1086/339511}, \href
  {http://adsabs.harvard.edu/abs/2002ApJ...566L..85B} {566, L85}

\bibitem[\protect\citeauthoryear{Bernardo}{Bernardo}{1979}]{Bernardo1979}
Bernardo J.~M.,  1979, Journal of the Royal Statistical Society. Series B
  (Methodological), 41, 113

\bibitem[\protect\citeauthoryear{{Berti} \& {Stergioulas}}{{Berti} \&
  {Stergioulas}}{2004}]{Berti2004}
{Berti} E.,  {Stergioulas} N.,  2004, \mn@doi [\mnras]
  {10.1111/j.1365-2966.2004.07740.x}, \href
  {http://cdsads.u-strasbg.fr/abs/2004MNRAS.350.1416B} {350, 1416}

\bibitem[\protect\citeauthoryear{{Bhattacharyya} et~al.}{{Bhattacharyya}
  et~al.}{2005}]{Bhattacharyya2005}
{Bhattacharyya} S.,  et~al., 2005, \mn@doi [ApJ] {10.1086/426383}, \href
  {http://cdsads.u-strasbg.fr/abs/2005ApJ...619..483B} {619, 483}

\bibitem[\protect\citeauthoryear{{Bhattacharyya}, {Bombaci}, {Logoteta}  \&
  {Thampan}}{{Bhattacharyya} et~al.}{2017}]{Bhattacharyya2017}
{Bhattacharyya} S.,  {Bombaci} I.,  {Logoteta} D.,   {Thampan} A.~V.,  2017,
  \mn@doi [\apj] {10.3847/1538-4357/aa8b67}, \href
  {http://adsabs.harvard.edu/abs/2017ApJ...848...65B} {848, 65}

\bibitem[\protect\citeauthoryear{{Bogdanov}}{{Bogdanov}}{2013}]{Bogdanov2013}
{Bogdanov} S.,  2013, \mn@doi [\apj] {10.1088/0004-637X/762/2/96}, \href
  {http://cdsads.u-strasbg.fr/abs/2013ApJ...762...96B} {762, 96}

\bibitem[\protect\citeauthoryear{{Brewer} \& {Foreman-Mackey}}{{Brewer} \&
  {Foreman-Mackey}}{2016}]{DNest4}
{Brewer} B.~J.,  {Foreman-Mackey} D.,  2016, preprint, \href
  {http://adsabs.harvard.edu/abs/2016arXiv160603757B} {} (\mn@eprint {arXiv}
  {1606.03757})

\bibitem[\protect\citeauthoryear{{Cadeau}, {Leahy}  \& {Morsink}}{{Cadeau}
  et~al.}{2005}]{Cadeau2005}
{Cadeau} C.,  {Leahy} D.~A.,   {Morsink} S.~M.,  2005, \mn@doi [ApJ]
  {10.1086/425857}, \href {http://cdsads.u-strasbg.fr/abs/2005ApJ...618..451C}
  {618, 451}

\bibitem[\protect\citeauthoryear{{Cadeau} et~al.}{{Cadeau}
  et~al.}{2007}]{Cadeau2007}
{Cadeau} C.,  et~al., 2007, ApJ, \href
  {http://adsabs.harvard.edu/abs/2007ApJ...654..458C} {654, 458}

\bibitem[\protect\citeauthoryear{{Cardoso} \& {Gualtieri}}{{Cardoso} \&
  {Gualtieri}}{2016}]{Cardoso2016}
{Cardoso} V.,  {Gualtieri} L.,  2016, \mn@doi [Classical and Quantum Gravity]
  {10.1088/0264-9381/33/17/174001}, \href
  {http://adsabs.harvard.edu/abs/2016CQGra..33q4001C} {33, 174001}

\bibitem[\protect\citeauthoryear{Carpenter et~al.,}{Carpenter
  et~al.}{2017}]{Stan}
Carpenter B.,  et~al., 2017, \mn@doi [Journal of Statistical Software,
  Articles] {10.18637/jss.v076.i01}, 76, 1

\bibitem[\protect\citeauthoryear{Chaloner \& Verdinelli}{Chaloner \&
  Verdinelli}{1995}]{Chaloner1995}
Chaloner K.,  Verdinelli I.,  1995, Statistical Science, 10, 273

\bibitem[\protect\citeauthoryear{{Chamel} \& {Haensel}}{{Chamel} \&
  {Haensel}}{2008}]{Chamel2008}
{Chamel} N.,  {Haensel} P.,  2008, \mn@doi [Living Reviews in Relativity]
  {10.12942/lrr-2008-10}, \href
  {http://adsabs.harvard.edu/abs/2008LRR....11...10C} {11, 10}

\bibitem[\protect\citeauthoryear{{Chan}, {Psaltis}  \& {{\"O}zel}}{{Chan}
  et~al.}{2013}]{Chan2013}
{Chan} C.-k.,  {Psaltis} D.,   {{\"O}zel} F.,  2013, \mn@doi [\apj]
  {10.1088/0004-637X/777/1/13}, \href
  {http://cdsads.u-strasbg.fr/abs/2013ApJ...777...13C} {777, 13}

\bibitem[\protect\citeauthoryear{{Chatterjee} \& {Vida{\~n}a}}{{Chatterjee} \&
  {Vida{\~n}a}}{2016}]{Chatterjee16}
{Chatterjee} D.,  {Vida{\~n}a} I.,  2016, \mn@doi [European Physical Journal A]
  {10.1140/epja/i2016-16029-x}, \href
  {http://adsabs.harvard.edu/abs/2016EPJA...52...29C} {52, 29}

\bibitem[\protect\citeauthoryear{{Dexter} \& {Agol}}{{Dexter} \&
  {Agol}}{2009}]{Agol2009}
{Dexter} J.,  {Agol} E.,  2009, \mn@doi [ApJ] {10.1088/0004-637X/696/2/1616},
  \href {http://adsabs.harvard.edu/abs/2009ApJ...696.1616D} {696, 1616}

\bibitem[\protect\citeauthoryear{Drago \& Pagliara}{Drago \&
  Pagliara}{2016}]{Drago2016b}
Drago A.,  Pagliara G.,  2016, \mn@doi [The European Physical Journal A]
  {10.1140/epja/i2016-16041-2}, 52, 41

\bibitem[\protect\citeauthoryear{Drago, Lavagno, Pagliara  \& Pigato}{Drago
  et~al.}{2016}]{Drago2016a}
Drago A.,  Lavagno A.,  Pagliara G.,   Pigato D.,  2016, \mn@doi [The European
  Physical Journal A] {10.1140/epja/i2016-16040-3}, 52, 40

\bibitem[\protect\citeauthoryear{{Farr}, {Sravan}, {Cantrell}, {Kreidberg},
  {Bailyn}, {Mandel}  \& {Kalogera}}{{Farr} et~al.}{2011}]{Farr2011}
{Farr} W.~M.,  {Sravan} N.,  {Cantrell} A.,  {Kreidberg} L.,  {Bailyn} C.~D.,
  {Mandel} I.,   {Kalogera} V.,  2011, \mn@doi [\apj]
  {10.1088/0004-637X/741/2/103}, \href
  {http://cdsads.u-strasbg.fr/abs/2011ApJ...741..103F} {741, 103}

\bibitem[\protect\citeauthoryear{{Farr}, {Mandel}  \& {Stevens}}{{Farr}
  et~al.}{2015}]{FarrMandel2015}
{Farr} W.~M.,  {Mandel} I.,   {Stevens} D.,  2015, \mn@doi [Royal Society Open
  Science] {10.1098/rsos.150030}, \href
  {http://adsabs.harvard.edu/abs/2015RSOS....250030F} {2, 150030}

\bibitem[\protect\citeauthoryear{{Feroz}, {Hobson}  \& {Bridges}}{{Feroz}
  et~al.}{2009}]{MultiNest}
{Feroz} F.,  {Hobson} M.~P.,   {Bridges} M.,  2009, \mn@doi [\mnras]
  {10.1111/j.1365-2966.2009.14548.x}, \href
  {http://adsabs.harvard.edu/abs/2009MNRAS.398.1601F} {398, 1601}

\bibitem[\protect\citeauthoryear{{Foreman-Mackey}, {Hogg}, {Lang}  \&
  {Goodman}}{{Foreman-Mackey} et~al.}{2013}]{emcee}
{Foreman-Mackey} D.,  {Hogg} D.~W.,  {Lang} D.,   {Goodman} J.,  2013, \mn@doi
  [\pasp] {10.1086/670067}, \href
  {http://adsabs.harvard.edu/abs/2013PASP..125..306F} {125, 306}

\bibitem[\protect\citeauthoryear{Gelman \& Shalizi}{Gelman \&
  Shalizi}{}]{Gelman_philosophy}
Gelman A.,  Shalizi C.~R., , \mn@doi [British Journal of Mathematical and
  Statistical Psychology] {10.1111/j.2044-8317.2011.02037.x}, 66, 8

\bibitem[\protect\citeauthoryear{Gelman, Carlin, Stern, Dunson, Vehtari  \&
  Rubin}{Gelman et~al.}{2013}]{Gelman_book}
Gelman A.,  Carlin J.,  Stern H.,  Dunson D.,  Vehtari A.,   Rubin D.,  2013,
  Bayesian Data Analysis, Third Edition.
Chapman \& Hall/CRC Texts in Statistical Science, Taylor \& Francis, \url
  {https://books.google.nl/books?id=ZXL6AQAAQBAJ}

\bibitem[\protect\citeauthoryear{{Glendenning} \& {Kettner}}{{Glendenning} \&
  {Kettner}}{2000}]{Glendenning2000}
{Glendenning} N.~K.,  {Kettner} C.,  2000, \aap, \href
  {http://adsabs.harvard.edu/abs/2000A%26A...353L...9G} {353, L9}

\bibitem[\protect\citeauthoryear{{Graff}, {Feroz}, {Hobson}  \&
  {Lasenby}}{{Graff} et~al.}{2012}]{BAMBI}
{Graff} P.,  {Feroz} F.,  {Hobson} M.~P.,   {Lasenby} A.,  2012, \mn@doi
  [\mnras] {10.1111/j.1365-2966.2011.20288.x}, \href
  {http://adsabs.harvard.edu/abs/2012MNRAS.421..169G} {421, 169}

\bibitem[\protect\citeauthoryear{{Graff}, {Feroz}, {Hobson}  \&
  {Lasenby}}{{Graff} et~al.}{2014}]{SKYNET}
{Graff} P.,  {Feroz} F.,  {Hobson} M.~P.,   {Lasenby} A.,  2014, \mn@doi
  [\mnras] {10.1093/mnras/stu642}, \href
  {http://adsabs.harvard.edu/abs/2014MNRAS.441.1741G} {441, 1741}

\bibitem[\protect\citeauthoryear{{Haensel} et~al.}{{Haensel}
  et~al.}{2009}]{Haensel2009}
{Haensel} P.,  et~al., 2009, \mn@doi [A\&A] {10.1051/0004-6361/200811605},
  \href {http://cdsads.u-strasbg.fr/abs/2009A%26A...502..605H} {502, 605}

\bibitem[\protect\citeauthoryear{{Handley}, {Hobson}  \& {Lasenby}}{{Handley}
  et~al.}{2015}]{PolyChord_1}
{Handley} W.~J.,  {Hobson} M.~P.,   {Lasenby} A.~N.,  2015, \mn@doi [\mnras]
  {10.1093/mnras/stv1911}, \href
  {http://cdsads.u-strasbg.fr/abs/2015MNRAS.453.4384H} {453, 4384}

\bibitem[\protect\citeauthoryear{{Hansen}}{{Hansen}}{1974}]{Hansen1974}
{Hansen} R.~O.,  1974, \mn@doi [Journal of Mathematical Physics]
  {10.1063/1.1666501}, \href
  {http://cdsads.u-strasbg.fr/abs/1974JMP....15...46H} {15, 46}

\bibitem[\protect\citeauthoryear{{Hartle}}{{Hartle}}{1967}]{Hartle1967}
{Hartle} J.~B.,  1967, \mn@doi [\apj] {10.1086/149400}, \href
  {http://cdsads.u-strasbg.fr/abs/1967ApJ...150.1005H} {150, 1005}

\bibitem[\protect\citeauthoryear{{Hartle} \& {Thorne}}{{Hartle} \&
  {Thorne}}{1968}]{HT1968}
{Hartle} J.~B.,  {Thorne} K.~S.,  1968, \mn@doi [ApJ] {10.1086/149707}, \href
  {http://cdsads.u-strasbg.fr/abs/1968ApJ...153..807H} {153, 807}

\bibitem[\protect\citeauthoryear{{Hebeler} et~al.}{{Hebeler}
  et~al.}{2015}]{Hebeler15}
{Hebeler} K.,  et~al., 2015, \mn@doi [Annual Review of Nuclear and Particle
  Science] {10.1146/annurev-nucl-102313-025446}, \href
  {http://adsabs.harvard.edu/abs/2015ARNPS..65..457H} {65, 457}

\bibitem[\protect\citeauthoryear{{Hogg}, {Bovy}  \& {Lang}}{{Hogg}
  et~al.}{2010}]{Hogg2010}
{Hogg} D.~W.,  {Bovy} J.,   {Lang} D.,  2010, preprint, \href
  {http://adsabs.harvard.edu/abs/2010arXiv1008.4686H} {} (\mn@eprint {arXiv}
  {1008.4686})

\bibitem[\protect\citeauthoryear{{Jeffreys}}{{Jeffreys}}{1946}]{Jeffreys}
{Jeffreys} H.~F.~R.~S.,  1946, \mn@doi [Proceedings of the Royal Society of
  London A: Mathematical, Physical and Engineering Sciences]
  {10.1098/rspa.1946.0056}, 186, 453

\bibitem[\protect\citeauthoryear{Jeffreys}{Jeffreys}{1961}]{Jeffreys1961}
Jeffreys H.,  1961, Theory of Probability, third edn.
Oxford, Oxford, England

\bibitem[\protect\citeauthoryear{{Jones}}{{Jones}}{2017}]{PRECISION_COSMOLOGY}
{Jones} B.~J.~T.,  2017, {Precision Cosmology}

\bibitem[\protect\citeauthoryear{{Lackey} \& {Wade}}{{Lackey} \&
  {Wade}}{2015}]{Lackey2015}
{Lackey} B.~D.,  {Wade} L.,  2015, \mn@doi [\prd] {10.1103/PhysRevD.91.043002},
  \href {http://adsabs.harvard.edu/abs/2015PhRvD..91d3002L} {91, 043002}

\bibitem[\protect\citeauthoryear{{Lattimer} \& {Prakash}}{{Lattimer} \&
  {Prakash}}{2016}]{Lattimer16}
{Lattimer} J.~M.,  {Prakash} M.,  2016, \mn@doi [\physrep]
  {10.1016/j.physrep.2015.12.005}, \href
  {http://adsabs.harvard.edu/abs/2016PhR...621..127L} {621, 127}

\bibitem[\protect\citeauthoryear{{Lattimer} \& {Steiner}}{{Lattimer} \&
  {Steiner}}{2014a}]{LattimerSteiner2014}
{Lattimer} J.~M.,  {Steiner} A.~W.,  2014a, \mn@doi [European Physical Journal
  A] {10.1140/epja/i2014-14040-y}, \href
  {http://adsabs.harvard.edu/abs/2014EPJA...50...40L} {50, 40}

\bibitem[\protect\citeauthoryear{{Lattimer} \& {Steiner}}{{Lattimer} \&
  {Steiner}}{2014b}]{LattimerSteiner2014_LMXB}
{Lattimer} J.~M.,  {Steiner} A.~W.,  2014b, \mn@doi [\apj]
  {10.1088/0004-637X/784/2/123}, \href
  {http://cdsads.u-strasbg.fr/abs/2014ApJ...784..123L} {784, 123}

\bibitem[\protect\citeauthoryear{{Lewis}}{{Lewis}}{2013}]{Lewis_noninf_2013}
{Lewis} N.,  2013, preprint, \href
  {http://adsabs.harvard.edu/abs/2013arXiv1308.2791L} {} (\mn@eprint {arXiv}
  {1308.2791})

\bibitem[\protect\citeauthoryear{{Lindblom}}{{Lindblom}}{1992}]{Lindblom92}
{Lindblom} L.,  1992, \mn@doi [\apj] {10.1086/171882}, \href
  {http://cdsads.u-strasbg.fr/abs/1992ApJ...398..569L} {398, 569}

\bibitem[\protect\citeauthoryear{{Lo}, {Miller}, {Bhattacharyya}  \&
  {Lamb}}{{Lo} et~al.}{2013}]{Lo2013}
{Lo} K.~H.,  {Miller} M.~C.,  {Bhattacharyya} S.,   {Lamb} F.~K.,  2013,
  \mn@doi [\apj] {10.1088/0004-637X/776/1/19}, \href
  {http://cdsads.u-strasbg.fr/abs/2013ApJ...776...19L} {776, 19}

\bibitem[\protect\citeauthoryear{MacKay}{MacKay}{2003}]{MacKay2003}
MacKay D.~J.,  2003, Information Theory, Inference and Learning Algorithms.
Cambridge University Press

\bibitem[\protect\citeauthoryear{{Manko}, {Mielke}  \&
  {Sanabria-G{\'o}mez}}{{Manko} et~al.}{2000}]{Manko2000}
{Manko} V.~S.,  {Mielke} E.~W.,   {Sanabria-G{\'o}mez} J.~D.,  2000, \mn@doi
  [\prd] {10.1103/PhysRevD.61.081501}, \href
  {http://cdsads.u-strasbg.fr/abs/2000PhRvD..61h1501M} {61, 081501}

\bibitem[\protect\citeauthoryear{{Margueron}, {Hoffmann Casali}  \&
  {Gulminelli}}{{Margueron} et~al.}{2018}]{Margueron2018}
{Margueron} J.,  {Hoffmann Casali} R.,   {Gulminelli} F.,  2018, \mn@doi [\prc]
  {10.1103/PhysRevC.97.025805}, \href
  {http://adsabs.harvard.edu/abs/2018PhRvC..97b5805M} {97, 025805}

\bibitem[\protect\citeauthoryear{{Miller} \& {Lamb}}{{Miller} \&
  {Lamb}}{1998}]{Miller98}
{Miller} M.~C.,  {Lamb} F.~K.,  1998, \mn@doi [\apjl] {10.1086/311335}, \href
  {http://cdsads.u-strasbg.fr/abs/1998ApJ...499L..37M} {499, L37}

\bibitem[\protect\citeauthoryear{{Miller} \& {Lamb}}{{Miller} \&
  {Lamb}}{2015}]{Miller15}
{Miller} M.~C.,  {Lamb} F.~K.,  2015, \mn@doi [Astrophysical Journal]
  {10.1088/0004-637X/808/1/31}, \href
  {http://adsabs.harvard.edu/abs/2015ApJ...808...31M} {808, 31}

\bibitem[\protect\citeauthoryear{{Miller} \& {Lamb}}{{Miller} \&
  {Lamb}}{2016}]{Miller2016}
{Miller} M.~C.,  {Lamb} F.~K.,  2016, \mn@doi [European Physical Journal A]
  {10.1140/epja/i2016-16063-8}, \href
  {http://cdsads.u-strasbg.fr/abs/2016EPJA...52...63M} {52, 63}

\bibitem[\protect\citeauthoryear{{Misner}, {Thorne}  \& {Wheeler}}{{Misner}
  et~al.}{1973}]{Misner1973}
{Misner} C.~W.,  {Thorne} K.~S.,   {Wheeler} J.~A.,  1973, {Gravitation}

\bibitem[\protect\citeauthoryear{{Morsink}, {Leahy}, {Cadeau}  \&
  {Braga}}{{Morsink} et~al.}{2007}]{Morsink2007}
{Morsink} S.~M.,  {Leahy} D.~A.,  {Cadeau} C.,   {Braga} J.,  2007, \mn@doi
  [\apj] {10.1086/518648}, \href
  {http://cdsads.u-strasbg.fr/abs/2007ApJ...663.1244M} {663, 1244}

\bibitem[\protect\citeauthoryear{{Mueller} \& {Eriguchi}}{{Mueller} \&
  {Eriguchi}}{1985}]{Mueller1985}
{Mueller} E.,  {Eriguchi} Y.,  1985, \aap, \href
  {http://adsabs.harvard.edu/abs/1985A%26A...152..325M} {152, 325}

\bibitem[\protect\citeauthoryear{{N{\"a}ttil{\"a}} \&
  {Pihajoki}}{{N{\"a}ttil{\"a}} \& {Pihajoki}}{2017}]{Nattila2017}
{N{\"a}ttil{\"a}} J.,  {Pihajoki} P.,  2017, preprint, \href
  {http://cdsads.u-strasbg.fr/abs/2017arXiv170907292N} {} (\mn@eprint {arXiv}
  {1709.07292})

\bibitem[\protect\citeauthoryear{{N{\"a}ttil{\"a}} et~al.}{{N{\"a}ttil{\"a}}
  et~al.}{2016}]{Nattila2016}
{N{\"a}ttil{\"a}} J.,  et~al., 2016, \mn@doi [\aap]
  {10.1051/0004-6361/201527416}, \href
  {http://cdsads.u-strasbg.fr/abs/2016A%26A...591A..25N} {591, A25}

\bibitem[\protect\citeauthoryear{{Neiswanger} \& {Xing}}{{Neiswanger} \&
  {Xing}}{2016}]{Neiswanger2016}
{Neiswanger} W.,  {Xing} E.,  2016, preprint, \href
  {http://adsabs.harvard.edu/abs/2016arXiv160600787N} {} (\mn@eprint {arXiv}
  {1606.00787})

\bibitem[\protect\citeauthoryear{Oppenheimer \& Volkoff}{Oppenheimer \&
  Volkoff}{1939}]{Oppenheimer1939}
Oppenheimer J.~R.,  Volkoff G.~M.,  1939, \mn@doi [Phys. Rev.]
  {10.1103/PhysRev.55.374}, 55, 374

\bibitem[\protect\citeauthoryear{{{\"O}zel} \& {Freire}}{{{\"O}zel} \&
  {Freire}}{2016}]{OzelReview}
{{\"O}zel} F.,  {Freire} P.,  2016, \mn@doi [\araa]
  {10.1146/annurev-astro-081915-023322}, \href
  {http://adsabs.harvard.edu/abs/2016ARA%26A..54..401O} {54, 401}

\bibitem[\protect\citeauthoryear{{{\"O}zel} \& {Psaltis}}{{{\"O}zel} \&
  {Psaltis}}{2009}]{Ozel2009}
{{\"O}zel} F.,  {Psaltis} D.,  2009, \mn@doi [Physical Review D]
  {10.1103/PhysRevD.80.103003}, \href
  {http://cdsads.u-strasbg.fr/abs/2009PhRvD..80j3003O} {80, 103003}

\bibitem[\protect\citeauthoryear{{{\"O}zel} \& {Psaltis}}{{{\"O}zel} \&
  {Psaltis}}{2015}]{Ozel2015_Jacobian}
{{\"O}zel} F.,  {Psaltis} D.,  2015, \mn@doi [\apj]
  {10.1088/0004-637X/810/2/135}, \href
  {http://adsabs.harvard.edu/abs/2015ApJ...810..135O} {810, 135}

\bibitem[\protect\citeauthoryear{{{\"O}zel}, {Baym}  \& {G{\"u}ver}}{{{\"O}zel}
  et~al.}{2010}]{Ozel2010}
{{\"O}zel} F.,  {Baym} G.,   {G{\"u}ver} T.,  2010, \mn@doi [\prd]
  {10.1103/PhysRevD.82.101301}, \href
  {http://cdsads.u-strasbg.fr/abs/2010PhRvD..82j1301O} {82, 101301}

\bibitem[\protect\citeauthoryear{{{\"O}zel} et~al.}{{{\"O}zel}
  et~al.}{2016a}]{Ozel16}
{{\"O}zel} F.,  et~al., 2016a, \mn@doi [Astrophysical Journal]
  {10.3847/0004-637X/820/1/28}, \href
  {http://adsabs.harvard.edu/abs/2016ApJ...820...28O} {820, 28}

\bibitem[\protect\citeauthoryear{{{\"O}zel}, {Psaltis}, {Arzoumanian},
  {Morsink}  \& {Baub{\"o}ck}}{{{\"O}zel} et~al.}{2016b}]{Ozel2016}
{{\"O}zel} F.,  {Psaltis} D.,  {Arzoumanian} Z.,  {Morsink} S.,   {Baub{\"o}ck}
  M.,  2016b, \mn@doi [\apj] {10.3847/0004-637X/832/1/92}, \href
  {http://cdsads.u-strasbg.fr/abs/2016ApJ...832...92O} {832, 92}

\bibitem[\protect\citeauthoryear{{Pechenick}, {Ftaclas}  \&
  {Cohen}}{{Pechenick} et~al.}{1983}]{Pechenick1983}
{Pechenick} K.~R.,  {Ftaclas} C.,   {Cohen} J.~M.,  1983, ApJ, \href
  {http://adsabs.harvard.edu/abs/1983ApJ...274..846P} {274, 846}

\bibitem[\protect\citeauthoryear{{Poutanen} \& {Beloborodov}}{{Poutanen} \&
  {Beloborodov}}{2006}]{Poutanen2006}
{Poutanen} J.,  {Beloborodov} A.~M.,  2006, \mn@doi [MNRAS]
  {10.1111/j.1365-2966.2006.11088.x}, \href
  {http://cdsads.u-strasbg.fr/abs/2006MNRAS.373..836P} {373, 836}

\bibitem[\protect\citeauthoryear{{Poutanen} \& {Gierli{\'n}ski}}{{Poutanen} \&
  {Gierli{\'n}ski}}{2003}]{Poutanen2003}
{Poutanen} J.,  {Gierli{\'n}ski} M.,  2003, MNRAS, \href
  {http://adsabs.harvard.edu/abs/2003MNRAS.343.1301P} {343, 1301}

\bibitem[\protect\citeauthoryear{{Psaltis} \& {{\"O}zel}}{{Psaltis} \&
  {{\"O}zel}}{2014}]{Psaltis2014}
{Psaltis} D.,  {{\"O}zel} F.,  2014, \mn@doi [\apj]
  {10.1088/0004-637X/792/2/87}, \href
  {http://cdsads.u-strasbg.fr/abs/2014ApJ...792...87P} {792, 87}

\bibitem[\protect\citeauthoryear{{Raithel}, {{\"O}zel}  \& {Psaltis}}{{Raithel}
  et~al.}{2016}]{Raithel2016}
{Raithel} C.~A.,  {{\"O}zel} F.,   {Psaltis} D.,  2016, \mn@doi [\apj]
  {10.3847/0004-637X/831/1/44}, \href
  {http://adsabs.harvard.edu/abs/2016ApJ...831...44R} {831, 44}

\bibitem[\protect\citeauthoryear{{Raithel}, {{\"O}zel}  \& {Psaltis}}{{Raithel}
  et~al.}{2017}]{Raithel2017}
{Raithel} C.~A.,  {{\"O}zel} F.,   {Psaltis} D.,  2017, \mn@doi [\apj]
  {10.3847/1538-4357/aa7a5a}, \href
  {http://cdsads.u-strasbg.fr/abs/2017ApJ...844..156R} {844, 156}

\bibitem[\protect\citeauthoryear{{Read}, {Lackey}, {Owen}  \&
  {Friedman}}{{Read} et~al.}{2009}]{Read2009}
{Read} J.~S.,  {Lackey} B.~D.,  {Owen} B.~J.,   {Friedman} J.~L.,  2009,
  \mn@doi [\prd] {10.1103/PhysRevD.79.124032}, \href
  {http://cdsads.u-strasbg.fr/abs/2009PhRvD..79l4032R} {79, 124032}

\bibitem[\protect\citeauthoryear{Robert}{Robert}{2007}]{Robert2007}
Robert C.,  2007, The Bayesian Choice: From Decision-Theoretic Foundations to
  Computational Implementation.
Springer Texts in Statistics, Springer New York, \url
  {https://books.google.nl/books?id=6oQ4s8Pq9pYC}

\bibitem[\protect\citeauthoryear{Robert, Chopin  \& Rousseau}{Robert
  et~al.}{2009}]{Robert2009}
Robert C.~P.,  Chopin N.,   Rousseau J.,  2009, \mn@doi [Statist. Sci.]
  {10.1214/09-STS284}, 24, 141

\bibitem[\protect\citeauthoryear{{Rogers}}{{Rogers}}{2015}]{Rogers2015}
{Rogers} A.,  2015, \mn@doi [\mnras] {10.1093/mnras/stv903}, \href
  {http://adsabs.harvard.edu/abs/2015MNRAS.451...17R} {451, 17}

\bibitem[\protect\citeauthoryear{{Rogers}}{{Rogers}}{2017}]{Rogers2016}
{Rogers} A.,  2017, \mn@doi [\mnras] {10.1093/mnras/stw2829}, \href
  {http://adsabs.harvard.edu/abs/2017MNRAS.465.2151R} {465, 2151}

\bibitem[\protect\citeauthoryear{{Ryan}}{{Ryan}}{1995}]{Ryan1995}
{Ryan} F.~D.,  1995, \mn@doi [\prd] {10.1103/PhysRevD.52.5707}, \href
  {http://adsabs.harvard.edu/abs/1995PhRvD..52.5707R} {52, 5707}

\bibitem[\protect\citeauthoryear{Salvatier, Wiecki  \& Fonnesbeck}{Salvatier
  et~al.}{2016}]{PyMC3}
Salvatier J.,  Wiecki T.~V.,   Fonnesbeck C.,  2016, \mn@doi [PeerJ Computer
  Science] {10.7717/peerj-cs.55}, 2, e55

\bibitem[\protect\citeauthoryear{{Schneider}, {Ehlers}  \& {Falco}}{{Schneider}
  et~al.}{1992}]{Schneider1992}
{Schneider} P.,  {Ehlers} J.,   {Falco} E.~E.,  1992, {Gravitational Lenses},
  \mn@doi{10.1007/978-3-662-03758-4.
}

\bibitem[\protect\citeauthoryear{{Shapiro} \& {Teukolsky}}{{Shapiro} \&
  {Teukolsky}}{1983}]{ST1983}
{Shapiro} S.~L.,  {Teukolsky} S.~A.,  1983, {Black holes, white dwarfs, and
  neutron stars: The physics of compact objects}

\bibitem[\protect\citeauthoryear{{Stein}, {Yagi}  \& {Yunes}}{{Stein}
  et~al.}{2014}]{SteinYagi2014}
{Stein} L.~C.,  {Yagi} K.,   {Yunes} N.,  2014, \mn@doi [ApJ]
  {10.1088/0004-637X/788/1/15}, \href
  {http://cdsads.u-strasbg.fr/abs/2014ApJ...788...15S} {788, 15}

\bibitem[\protect\citeauthoryear{{Steiner}}{{Steiner}}{2014}]{Steiner_bamr}
{Steiner} A.~W.,  2014, {bamr: Bayesian analysis of mass and radius
  observations}, Astrophysics Source Code Library (\mn@eprint {ascl}
  {1408.020})

\bibitem[\protect\citeauthoryear{{Steiner}, {Lattimer}  \& {Brown}}{{Steiner}
  et~al.}{2010}]{Steiner2010}
{Steiner} A.~W.,  {Lattimer} J.~M.,   {Brown} E.~F.,  2010, \mn@doi [\apj]
  {10.1088/0004-637X/722/1/33}, \href
  {http://cdsads.u-strasbg.fr/abs/2010ApJ...722...33S} {722, 33}

\bibitem[\protect\citeauthoryear{{Steiner}, {Lattimer}  \& {Brown}}{{Steiner}
  et~al.}{2013}]{SteinerLattimerBrown2013}
{Steiner} A.~W.,  {Lattimer} J.~M.,   {Brown} E.~F.,  2013, \mn@doi [\apjl]
  {10.1088/2041-8205/765/1/L5}, \href
  {http://cdsads.u-strasbg.fr/abs/2013ApJ...765L...5S} {765, L5}

\bibitem[\protect\citeauthoryear{{Steiner}, {Lattimer}  \& {Brown}}{{Steiner}
  et~al.}{2016}]{Steiner2016}
{Steiner} A.~W.,  {Lattimer} J.~M.,   {Brown} E.~F.,  2016, \mn@doi [European
  Physical Journal A] {10.1140/epja/i2016-16018-1}, \href
  {http://adsabs.harvard.edu/abs/2016EPJA...52...18S} {52, 18}

\bibitem[\protect\citeauthoryear{Stergioulas}{Stergioulas}{2003}]{Stergioulas2003}
Stergioulas N.,  2003, \mn@doi [Living Reviews in Relativity]
  {10.1007/lrr-2003-3}, 6

\bibitem[\protect\citeauthoryear{{Stergioulas} \& {Friedman}}{{Stergioulas} \&
  {Friedman}}{1995}]{Stergioulas1995}
{Stergioulas} N.,  {Friedman} J.~L.,  1995, \mn@doi [\apj] {10.1086/175605},
  \href {http://adsabs.harvard.edu/abs/1995ApJ...444..306S} {444, 306}

\bibitem[\protect\citeauthoryear{{Taylor} \& {Kitching}}{{Taylor} \&
  {Kitching}}{2010}]{Taylor2010}
{Taylor} A.~N.,  {Kitching} T.~D.,  2010, \mn@doi [\mnras]
  {10.1111/j.1365-2966.2010.17201.x}, \href
  {http://cdsads.u-strasbg.fr/abs/2010MNRAS.408..865T} {408, 865}

\bibitem[\protect\citeauthoryear{{Thorne}}{{Thorne}}{1980}]{Thorne1980}
{Thorne} K.~S.,  1980, \mn@doi [Reviews of Modern Physics]
  {10.1103/RevModPhys.52.299}, \href
  {http://cdsads.u-strasbg.fr/abs/1980RvMP...52..299T} {52, 299}

\bibitem[\protect\citeauthoryear{Tolman}{Tolman}{1939}]{Tolman1939}
Tolman R.~C.,  1939, \mn@doi [Phys. Rev.] {10.1103/PhysRev.55.364}, 55, 364

\bibitem[\protect\citeauthoryear{{Veitch} et~al.,}{{Veitch}
  et~al.}{2015}]{LALInference}
{Veitch} J.,  et~al., 2015, \mn@doi [\prd] {10.1103/PhysRevD.91.042003}, \href
  {http://adsabs.harvard.edu/abs/2015PhRvD..91d2003V} {91, 042003}

\bibitem[\protect\citeauthoryear{{Vincent}, {Paumard}, {Gourgoulhon}  \&
  {Perrin}}{{Vincent} et~al.}{2011}]{GYOTO}
{Vincent} F.~H.,  {Paumard} T.,  {Gourgoulhon} E.,   {Perrin} G.,  2011,
  \mn@doi [Classical and Quantum Gravity] {10.1088/0264-9381/28/22/225011},
  \href {http://adsabs.harvard.edu/abs/2011CQGra..28v5011V} {28, 225011}

\bibitem[\protect\citeauthoryear{{Vincent}, {Gourgoulhon}  \&
  {Novak}}{{Vincent} et~al.}{2012}]{Vincent2012}
{Vincent} F.~H.,  {Gourgoulhon} E.,   {Novak} J.,  2012, \mn@doi [Classical and
  Quantum Gravity] {10.1088/0264-9381/29/24/245005}, \href
  {http://adsabs.harvard.edu/abs/2012CQGra..29x5005V} {29, 245005}

\bibitem[\protect\citeauthoryear{{Vincent} et~al.,}{{Vincent}
  et~al.}{2018}]{Vincent2018}
{Vincent} F.~H.,  et~al., 2018, \mn@doi [\apj] {10.3847/1538-4357/aab0a3},
  \href {http://adsabs.harvard.edu/abs/2018ApJ...855..116V} {855, 116}

\bibitem[\protect\citeauthoryear{{Watts}}{{Watts}}{2012}]{Watts2012}
{Watts} A.~L.,  2012, \mn@doi [ARAA] {10.1146/annurev-astro-040312-132617},
  \href {http://adsabs.harvard.edu/abs/2012ARA%26A..50..609W} {50, 609}

\bibitem[\protect\citeauthoryear{{Watts} et~al.}{{Watts}
  et~al.}{2015}]{Watts15}
{Watts} A.,  et~al., 2015, Advancing Astrophysics with the Square Kilometre
  Array (AASKA14), \href {http://adsabs.harvard.edu/abs/2015aska.confE..43W}
  {p.~43}

\bibitem[\protect\citeauthoryear{{Watts} et~al.}{{Watts}
  et~al.}{2016}]{Watts16}
{Watts} A.~L.,  et~al., 2016, \mn@doi [Reviews of Modern Physics]
  {10.1103/RevModPhys.88.021001}, \href
  {http://adsabs.harvard.edu/abs/2016RvMP...88b1001W} {88, 021001}

\bibitem[\protect\citeauthoryear{{Wilson-Hodge} et~al.,}{{Wilson-Hodge}
  et~al.}{2017}]{STROBEX}
{Wilson-Hodge} C.~A.,  et~al., 2017, \mn@doi [Results in Physics]
  {10.1016/j.rinp.2017.09.013}, \href
  {http://adsabs.harvard.edu/abs/2017ResPh...7.3704W} {7, 3704}

\bibitem[\protect\citeauthoryear{{Yagi} \& {Yunes}}{{Yagi} \&
  {Yunes}}{2013}]{YagiYunes2013}
{Yagi} K.,  {Yunes} N.,  2013, \mn@doi [\prd] {10.1103/PhysRevD.88.023009},
  \href {http://cdsads.u-strasbg.fr/abs/2013PhRvD..88b3009Y} {88, 023009}

\bibitem[\protect\citeauthoryear{{Yagi} \& {Yunes}}{{Yagi} \&
  {Yunes}}{2017a}]{YagiYunes2017_tidal}
{Yagi} K.,  {Yunes} N.,  2017a, \mn@doi [Classical and Quantum Gravity]
  {10.1088/1361-6382/34/1/015006}, \href
  {http://cdsads.u-strasbg.fr/abs/2017CQGra..34a5006Y} {34, 015006}

\bibitem[\protect\citeauthoryear{{Yagi} \& {Yunes}}{{Yagi} \&
  {Yunes}}{2017b}]{YagiYunes2017}
{Yagi} K.,  {Yunes} N.,  2017b, \mn@doi [\physrep]
  {10.1016/j.physrep.2017.03.002}, \href
  {http://adsabs.harvard.edu/abs/2017PhR...681....1Y} {681, 1}

\bibitem[\protect\citeauthoryear{{Yagi} et~al.}{{Yagi}
  et~al.}{2014}]{YagiStein2014}
{Yagi} K.,  et~al., 2014, \mn@doi [Phys. Rev. D] {10.1103/PhysRevD.90.063010},
  \href {http://cdsads.u-strasbg.fr/abs/2014PhRvD..90f3010Y} {90, 063010}

\bibitem[\protect\citeauthoryear{{Zdunik}, {Bejger}, {Haensel}  \&
  {Gourgoulhon}}{{Zdunik} et~al.}{2006}]{Zdunik2006}
{Zdunik} J.~L.,  {Bejger} M.,  {Haensel} P.,   {Gourgoulhon} E.,  2006, \mn@doi
  [\aap] {10.1051/0004-6361:20054260}, \href
  {http://adsabs.harvard.edu/abs/2006A%26A...450..747Z} {450, 747}

\bibitem[\protect\citeauthoryear{{Zhang} et~al.}{{Zhang}
  et~al.}{2016}]{Zhang16}
{Zhang} S.~N.,  et~al., 2016, preprint, \href
  {http://adsabs.harvard.edu/abs/2016arXiv160708823Z} {} (\mn@eprint {arXiv}
  {1607.08823})

\makeatother
\end{thebibliography}


\appendix

\section{Pathologies in the interior-exterior parameter mapping}\label{app:pathologies}
In this Appendix we present detailed arguments regarding the existence of pathologies in the general relativistic interior-exterior parameter mapping for the purpose of probability density transformation (see Section \ref{sec:mapping}).

\subsection{Single-star interior-exterior solution matching}\label{sec:matching}
First we discuss properties of a map ${f}\from Y\to X$, $\boldsymbol{y}\mapsto\boldsymbol{x}$ which matches analytic exterior and numerical interior solutions for a single star ($s=1$). See Sections~\ref{sec:mapping definition} and \ref{sec:explicit} for a reminder of the mapping definition.

The parameters $\boldsymbol{y}$ entirely specify the ensemble of spacetime solutions, so only the map $\boldsymbol{y}\mapsto\boldsymbol{x}$ can ever be information-losing for arbitrary $s$ (see Fig.~\ref{fig:non-injective} and Fig.~\ref{fig:non-injective discrete sets} for example diagrams of information-losing maps). It follows that the number of \textit{free} exterior spacetime parameters in an analytic \textit{single-star} exterior solution cannot be $d>n+2$. That is, we cannot: (i) define a probability density distribution on a space of $d>n+2$ free exterior parameters; (ii) transform that probability density distribution onto a joint space of $(n+2)$ interior parameters and $(d-n-2)$ exterior parameters; and (iii) then marginalise over those $(d-n-2)$ exterior parameters. Further, we cannot marginalise over $(d-n-2)$ exterior parameters \textit{before} the transformation: such an operation would project all (exterior) probability density onto a surface of the same dimensionality as the solution-surface and this operation has no analogue in the IP-paradigm discussed in Section \ref{sec:implicit}. In other words, exterior solutions that are locally associated with precisely zero probability density (conditioned on the EOS model and general relativistic gravity) in the IP-paradigm would be assigned a finite probability density and marginalised over. For inference conditioned on data $\mathcal{D}$ acquired by observing a single star, it is required that $d\equiv n+2$.

In Section \ref{sec:mapping definition} we provided practical examples of the mapping criterion $d\equiv n+2s$ for ensembles with $s>1$.

\subsection{Problematic equation of state parametrisations}\label{sec:injectivity}

In this section we discuss, in general terms, EOS parametrisations whose map ${f}\from Y\to X$, $\boldsymbol{y}\mapsto\boldsymbol{x}$ is not diffeomorphic (Fig. \ref{fig:invertible}) for a single star, and thus for an ensemble of such stars.

A generalised local volume transformation (a Jacobian determinant) is required for operating on local scalar densities. The density transformation cannot be everywhere defined if the Jacobian anywhere loses rank: (i) eigenvectors of the Jacobian of the map $\boldsymbol{y}\mapsto\boldsymbol{x}$ are lost; or (ii) partial derivatives are undefined due to loss of differentiability. Such points are \textit{singularities} at which the map $\boldsymbol{y}\mapsto\boldsymbol{x}$ is not \textit{locally} invertible -- i.e., the map is locally non-injective (or degenerate), and thus information-losing.

The origin of such behaviour can be the EOS parametrisation, and this is problematic in the EP-paradigm for estimation of interior parameters. We note that subsets of interior and exterior parameters describe a single star, and thus partial derivatives of exterior parameters of one star with respect to interior condition parameters (central density and rotation frequency) of another star naturally vanish. The clearest violations of local injectivity may manifest upon searching for vanishing partial derivatives with respect to the shared EOS parameters. If \textit{partial} derivatives of \textit{all} exterior parameters with respect to an interior parameter anywhere vanish, it is clear that the map $\boldsymbol{y}\mapsto\boldsymbol{x}$ is locally non-invertible. It follows that it is the \textit{single-star} exterior spacetime solution surface whose dependence on EOS parameters needs to be inspected; this understanding can then be applied to an ensemble of stars which are each associated with such a solution surface.

We now discuss maps for which partial derivatives of \textit{all} exterior parameters vanish with respect to one or more EOS parameters in regions of parameter space. The local physics is parametrised by $\boldsymbol{y}\in Y$; in particular, the model for the EOS is simply a function $P\from\mathbb{R}\to\mathbb{R}$, $P=P(\varepsilon;\boldsymbol{\theta})$ which is integrated over (when solving the field equations) to execute $\boldsymbol{y}\mapsto\boldsymbol{x}$. A stable solution contains source matter no denser than that at the centre of the star, and exterior parameters are  sensitive only to less energetic matter elsewhere in the stratified interior. Moreover, for a particular EOS, there will exist for some rotation frequency a bounding stable solution in central density.

Consider a functional form for the EOS with the property that the thermodynamic dependence on at least one parameter of $\boldsymbol{\theta}$ is restricted to a subdomain in (local comoving) density. Now consider an ensemble of stars: if all central densities lie below the lower-bound of such a density subdomain, their exterior spacetime solutions do not depend on the subset of parameters which control the EOS beyond that lower-bound -- i.e., there is no source matter whose thermodynamic state is controlled by a subset of EOS parameters. It follows that all partial derivatives of exterior parameters with respect to a subset of EOS parameters locally vanish (at all orders).

\begin{figure}
\centering
\includegraphics[width=0.4\textwidth]{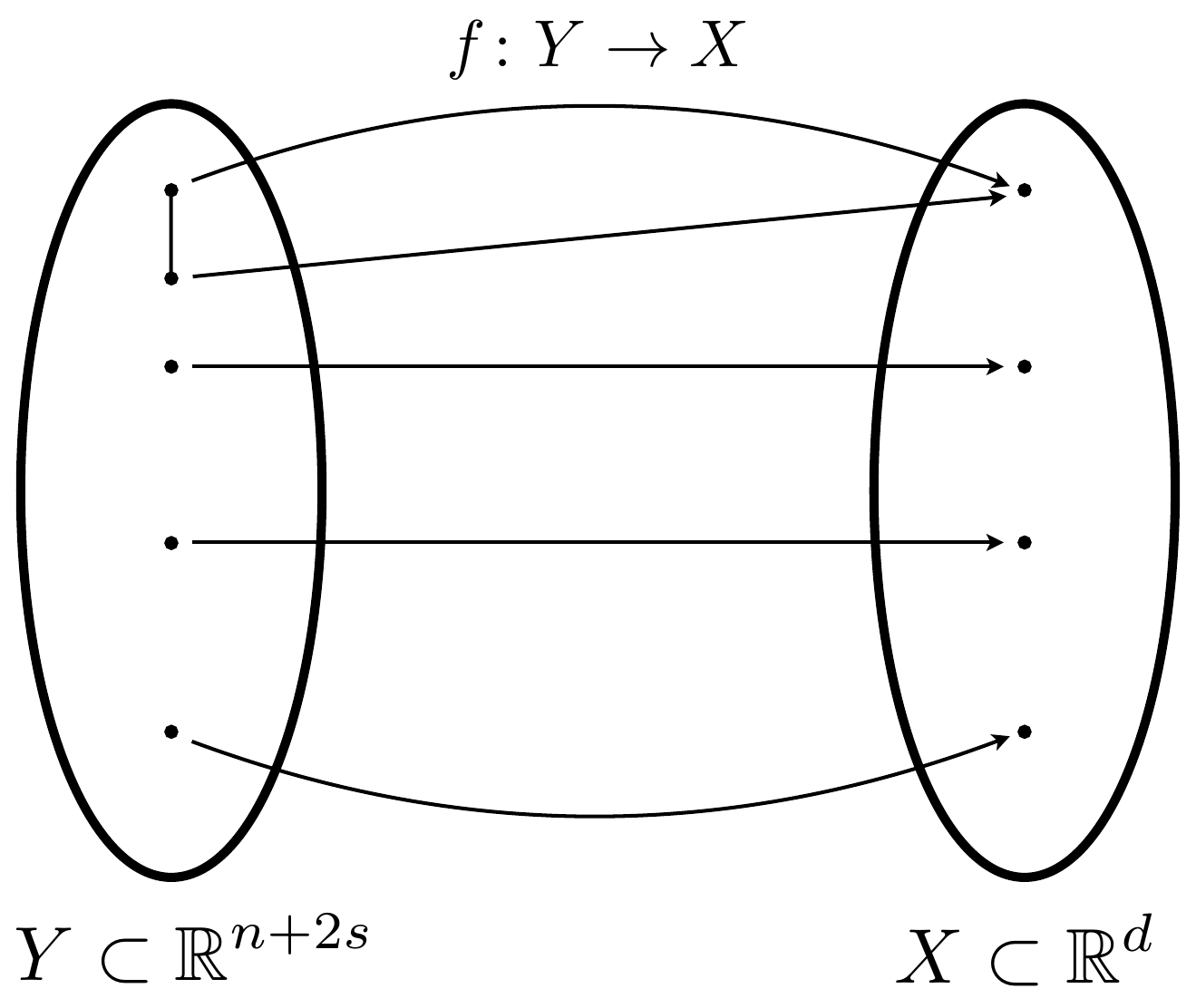}
\caption{A representation of a non-injective map $f$ under which a continuous subset of points in the domain $Y$ share an image $\boldsymbol{x}\in X$, where $X$ is the image of $Y$ under map $f$. The local dimensionality of this subset is equal to the number of eigenvectors of the differential of the mapping which locally vanish. In the diagram the degeneracy is depicted as an integral curve (which may exist in a surface of such integral curves) -- i.e., extending over a non-local sequence of points -- because the local injectivity violations we discuss below tend not to be isolated.}
\label{fig:non-injective}
\end{figure}

As we illustrate in Fig.~\ref{fig:non-injective}, this problem can also be expressed in terms of non-injectivity of the map $\boldsymbol{y}\mapsto\boldsymbol{x}$: each point in some subset of the codomain of ${f}$ is \textit{not} the image of \textit{at most} one point in the domain of ${f}$. This is true both locally and non-locally because the Jacobian is singular on surfaces. In particular, for fixed central densities and rotation frequencies, continuous subsets of points (along one or more orthogonal EOS directions) in the domain $Y$ \textit{share} single images (sets of exterior spacetime solutions) in $X$. Moreover, for fixed EOS parameters (and fixed rotation frequencies), the Jacobian is singular for a continuous subset of the subspace of central densities. Thus, given a probability density distribution over the image $X$, there does not exist a well-defined mapping of that distribution onto its preimage $Y$.

An important instance of the non-invertibility described above is exhibited by the piecewise-polytropic class of EOS models \citep[][]{Mueller1985, Read2009}; we demonstrate this behaviour and discuss the implications for EOS parameter estimation using piecewise-polytropic models in Raaijmakers et al. (submitted). Another instance of this behaviour may arise if the EOS model is calculated in the thermodynamic limit of some microphysical theory, and a subset of parameters entering that underlying theory describe source matter interactions only beyond some (local comoving) density. 

Injectivity is not violated in this manner if the thermodynamic dependence of the source matter on each parameter of $\boldsymbol{\theta}$ manifests \textit{globally} over the density domain. However, the mapping from EOS parameter space to EOS function space\footnote{That is, the space of functions $P\from\mathbb{R}\to\mathbb{R}$, $P=P(\varepsilon)$.} may in principle be locally non-invertible even if the parameters exert global control over the function, and such a pathology carries over as a property of the full map $\boldsymbol{y}\mapsto\boldsymbol{x}$. With a sensible parametrisation, however, these singularities, if they exist, may only be isolated and thus potentially less problematic.

To a Bayesian applying Equation~(\ref{eqn: fundamental posterior w inner summation}) in the presence of Jacobian singularities is commensurate with implicitly defining \textit{zero} prior density at these points. In this case of the piecewise-polytropic EOS model, this leads to the \textit{a priori} omission of entire regions of the central-density subspace; consequently, the \textit{marginal} posterior distribution of the parameters of interest, $\boldsymbol{\theta}$, may be severely distorted relative to a posterior whose prior was directly defined on a space of interior parameters and is thus well-behaved, and is, e.g., continuously positive and noninformative within some boundary. Note also that the existence of singularities in the Jacobian is one of several reasons why the normalisation of a probability density distribution of exterior parameters may not be conserved under the transformation law given by Equation~(\ref{eqn: fundamental posterior w inner summation}) -- we discuss other reasons in Appendices~\ref{app:discrete information loss} and \ref{sec:surjectivity}.

The importance of local non-invertibility for implicit prior definition will however depend on the properties of the (marginal) posterior distribution on the joint space of all exterior parameters being handled (see Raaijmakers et al., submitted). For instance, if an noninformative prior is defined on a space of exterior parameters, and a likelihood function (equivalently of interior or exterior parameters) is negligible everywhere the map is locally non-invertible, an implicitly defined prior density (on a space of interior parameters) will not be zero where the likelihood function is non-negligible -- i.e., provided the prior density is non-zero on the space of exterior parameters. Of course, a prior implicitly defined on a space of interior parameters may exhibit other undesirable properties even if it is not zero or negligible where the likelihood function is non-negligible. Moreover, likelihoods conditioned on future data may support regions where past likelihood functions are small (although this may also be indicative of modelling inaccuracies and statistical bias, and thus the need for model development).

Now we remark on mappings which permit thermodynamic phase transitions. There exist EOSs which generate disjoint sequences (or branches) of stable solutions to the field equations. Examples include the hybrid hadron-quark models with phase transitions which manifest as zeroth-order thermodynamic discontinuities \citep[][]{AlfordHan2016}, and piecewise-polytropic models with phase transitions which manifest as first-order thermodynamic discontinuities \citep[][]{Read2009}. In this case we must handle the map carefully because the domain $Y$ is defined as the set of points associated with stable spacetime solutions, and therefore derivatives need to be defined at the terminal points of stable branches. If the surface of stable and unstable solutions is everywhere first-order continuous with respect to the interior parameters $\boldsymbol{y}$, partial derivatives can be defined by differentiability of the surface (or equivalently by semi-differentiability at the boundary of the \textit{stable} solution-surface). If the surface (of stable and unstable solutions) exhibits first-order discontinuities with respect to central density due to the onset of instability, and those discontinuities are corners \citep[see, e.g., the ``cusps'' described and illustrated in][]{AlfordHan2016}, one-sided partial derivatives can be computed using the local stable solutions. If a (one-sided) partial derivative at a first-order discontinuity cannot be defined (e.g., the discontinuity is a cusp or vertical tangent), we would have to remove that point from the mapping.

\subsection{Global injectivity}\label{app:discrete information loss}

\begin{figure}
\centering
\includegraphics[width=0.4\textwidth]{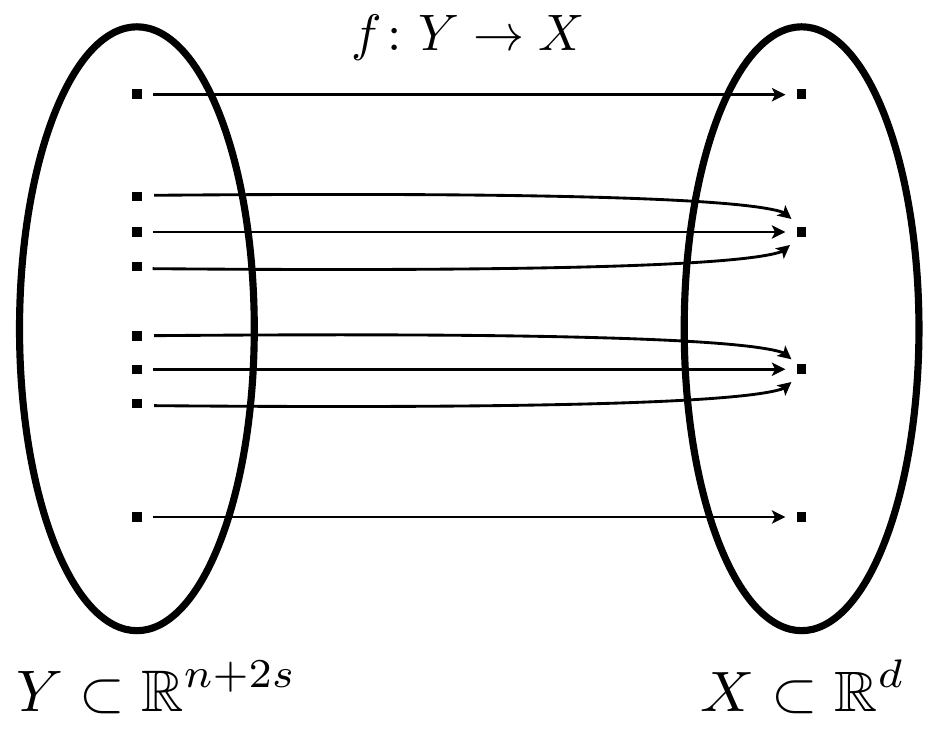}
\caption{A representation of a non-injective map $f$ which is everywhere locally invertible but information-losing -- i.e., subsets of points in the domain $Y$ share an image $\boldsymbol{x}\in X$, where $X$ is the image of $Y$ under map $f$, such that the map is not \textit{globally} invertible.}
\label{fig:non-injective discrete sets}
\end{figure}

The map $\boldsymbol{y}\mapsto\boldsymbol{x}$ is such that a \textit{single} image $\boldsymbol{x}\in X$ is generated via field equation integration given point $\boldsymbol{y}\in Y$. In this section we assume that there is no \textit{local} violation of injectivity, meaning the Jacobian is everywhere non-singular. Image $\boldsymbol{x}$ may still be generated by multiple points in the domain $Y$, however, meaning that the mapping is information-losing due to a \textit{global} violation of injectivity as illustrated in Fig.~\ref{fig:non-injective discrete sets}.\footnote{If the converse were true -- i.e., a point $\boldsymbol{x}\in X$ being the image of \textit{at most} one point in $Y$, but the preimage of some subset of $X$ (containing \textit{at least} two distinct points) being a single point in $Y$ -- the map $\boldsymbol{y}\mapsto\boldsymbol{x}$ would \textit{not} be information-losing and thus there would exist an inverse map $g\from X\to Y$ for transformation of the probability density distribution.} This might occur, for instance, if the mapping from EOS parameter space to EOS function space is not globally injective. In reality it may be the case that global violations of injectivity are intrinsically related to violations of local invertibility.

In order to highlight the problem with inappropriately applying Equation~(\ref{eqn: fundamental posterior w inner summation}), we now formulate a probability density transformation for an information-losing map $\boldsymbol{y}\mapsto\boldsymbol{x}$ which is everywhere locally invertible. Given some probability density distribution $\mathcal{P}(\boldsymbol{y}\;|\ldots)$ spanning the domain $Y$, the probability density distribution spanning the image $X$, where $f:Y\to X$, is given -- cf. Equation~(\ref{eqn:fundamental posterior}) -- by
\begin{equation}
\mathcal{P}\left(\boldsymbol{x}\;|\ldots\right)=\frac{\partial}{\partial x_{1}}\cdots\frac{\partial}{\partial x_{d}}\mathop{\int}_{V}\mathcal{P}(\boldsymbol{y}^{\prime}\;|\ldots)d\boldsymbol{y}^{\prime},
\end{equation}
where $V=\{\boldsymbol{y}^{\prime}\in Y\;|\;{f}(\boldsymbol{y}^{\prime})\leq\boldsymbol{x}\}$ is the integral domain, a subset of the domain $Y$, defined by the coordinate-wise inequality ${f}(\boldsymbol{y}^{\prime})\leq\boldsymbol{x}$. It follows -- cf. Equation~(\ref{eqn: fundamental posterior w/ Heaviside}) -- that
\begin{equation}
\mathcal{P}\left(\boldsymbol{x}\;|\ldots\right)
=\frac{\partial}{\partial x_{1}}\cdots\frac{\partial}{\partial x_{d}}\mathop{\int}\Theta\left(\boldsymbol{x}-{f}(\boldsymbol{y}^{\prime})\right)\mathcal{P}(\boldsymbol{y}^{\prime}\;|\ldots)d\boldsymbol{y}^{\prime}
=\mathop{\int}\delta\left(\boldsymbol{x}-{f}(\boldsymbol{y}^{\prime})\right)\mathcal{P}(\boldsymbol{y}^{\prime}\;|\ldots)d\boldsymbol{y}^{\prime}.
\end{equation}
If the map $\boldsymbol{y}\mapsto\boldsymbol{x}$ is everywhere \textit{locally} invertible and surjective, but not globally injective, we can write
\begin{equation}
\mathcal{P}\left(\boldsymbol{x}\;|\ldots\right)
=\mathop{\int}\delta\left(\boldsymbol{x}-{f}(\boldsymbol{y}^{\prime})\right)\mathcal{P}(\boldsymbol{y}^{\prime}\;|\ldots)\Biggl\lvert\det\left(\frac{\partial\boldsymbol{y}^{\prime}}{\partial{f}(\boldsymbol{y}^{\prime})}\right)\Biggl\rvert df
=\mathop{\sum}_{j}\mathcal{P}(\boldsymbol{y}_{j}\;|\ldots)\Biggl\lvert\det\left(\frac{\partial\boldsymbol{y}}{\partial f(\boldsymbol{y})}\right)\Biggl\rvert_{\boldsymbol{y}_{j}},
\label{eqn:interior-exterior transformation}
\end{equation}
where $j\in\mathbb{N}_{>0}$ enumerates points $\boldsymbol{y}_{j}\in Y$ which satisfy $f(\boldsymbol{y}_{j})=\boldsymbol{x}$. Whereas the summation in Equation~(\ref{eqn: fundamental posterior w inner summation}) reduced to a single term because a single image $\boldsymbol{x}$ is generated from a point $\boldsymbol{y}$, the map $\boldsymbol{y}\mapsto\boldsymbol{x}$ is information-losing if there exist single-star exterior solutions which match to multiple stable interior solutions and thus the summation in Equation~(\ref{eqn:interior-exterior transformation}) does not everywhere (in the image $X$) reduce to a single term.

Consider now the inversion of this distributional transformation to recover the distribution $\mathcal{P}(\boldsymbol{y}\;|\ldots)$. Consider a point $\boldsymbol{x}_{1}=f(\boldsymbol{y}_{1})$ which does \textit{not} have a unique preimage in $Y$. Applying the transformation implicit in Equation~(\ref{eqn: fundamental posterior w inner summation}) and substituting the expression for $\mathcal{P}(\boldsymbol{x}\;|\ldots)$ given by Equation~(\ref{eqn:interior-exterior transformation}) yields:
\begin{equation}
\begin{aligned}
\mathcal{P}(\boldsymbol{y}_{1}\;|\ldots)
&\neq
\mathcal{P}(\boldsymbol{x}_{1}\;|\ldots)\Biggl\lvert\det\left(\frac{\partial f(\boldsymbol{y})}{\partial\boldsymbol{y}}\right)\Biggl\rvert_{\boldsymbol{y}_{1}}\\
&=\Biggl\lvert\det\left(\frac{\partial f(\boldsymbol{y})}{\partial\boldsymbol{y}}\right)\Biggl\rvert_{\boldsymbol{y}_{1}}\times\mathop{\sum}_{j}\mathcal{P}(\boldsymbol{y}_{j}\;|\ldots)\Biggl\lvert\det\left(\frac{\partial\boldsymbol{y}}{\partial f(\boldsymbol{y})}\right)\Biggl\rvert_{\boldsymbol{y}_{j}}
=\mathcal{P}(\boldsymbol{y}_{1}\;|\ldots)+\Biggl\lvert\det\left(\frac{\partial f(\boldsymbol{y})}{\partial\boldsymbol{y}}\right)\Biggl\rvert_{\boldsymbol{y}_{1}}\times\mathop{\sum}_{j\neq1}\mathcal{P}(\boldsymbol{y}_{j}\;|\ldots)\Biggl\lvert\det\left(\frac{\partial\boldsymbol{y}}{\partial f(\boldsymbol{y})}\right)\Biggl\rvert_{\boldsymbol{y}_{j}}.
\end{aligned}
\label{eqn:distortion measure}
\end{equation}
Evidently, the normalisation of the distribution on $Y$ will not be conserved if the inverse transformation given by Equation~(\ref{eqn: fundamental posterior w inner summation}) is applied. Further, the distribution will be distorted at points $\boldsymbol{y}\in Y$ which do not uniquely match to an ensemble of exterior solutions; since these points are discretely distributed throughout the domain $Y$, the distortions result in local non-differentiability of the transformed distribution. Only if for all preimages $\boldsymbol{y}_{j}$ the difference between the local probability density before and after the transformation (with a renormalisation) is negligible relative to the global maximum density on $Y$, the transformation law of Section \ref{sec:mapping} will incur only small distortions of the original distribution spanning the domain $Y$. In principle, one option to quantify distortion is to transform \textit{from a space of interior parameters} followed by an inverse transformation -- e.g., Equation~(\ref{eqn:distortion measure}) -- because a diffeomorphic map would return the original distribution (to within some numerical tolerance). The original distribution would thus by definition be an \textit{exact} distribution for the purpose of comparison.

In principle, we can redefine a map $\boldsymbol{y}\mapsto\boldsymbol{x}$ which is locally by not globally invertible by defining a \textit{discrete} exterior parameter $\boldsymbol{\mathscr{D}}$ which will enforce that $f$ is information-conserving: a \textit{conditional} probability \textit{mass} distribution $\mathcal{P}(\boldsymbol{\mathscr{D}}\;|\;\boldsymbol{x},\ldots)$ controls the fractional probability \textit{density} at point $\boldsymbol{x}\in X$ (on the space $\mathbb{R}^{d}$) transformed to each preimage $\boldsymbol{y}\in Y$ which share the image $\boldsymbol{x}$, $\forall\boldsymbol{x}\in X$. The cardinality of the discrete set of preimages $\boldsymbol{y}\in Y$ which share each image $\boldsymbol{x}\in X$ is equal to the number of elements on the space of the discrete parameter with finite conditional probability mass; moreover, for a given image $\boldsymbol{x}$, there must be an unambiguous, injective mapping from elements on the space of $\mathscr{D}$ to preimages $\boldsymbol{y}$ sharing that image, such that an inverse map $g(\boldsymbol{x},\mathscr{D})\equiv f^{-1}(\boldsymbol{x},\mathscr{D})$ can be defined, where ${f}\from Y\to (X,\mathbb{N})$, $\boldsymbol{y}\mapsto(\boldsymbol{x},\mathscr{D})$ is diffeomorphic with respect to the continuous spaces $\mathbb{R}^{n+2s}$ and $\mathbb{R}^{d}$. It is also permissible for map ${g}\from (X,\mathbb{N})\to Y$, $(\boldsymbol{x},\mathscr{D})\mapsto\boldsymbol{y}$ to be information-losing. If these criteria are not satisfied, the total probability mass will not be conserved under transformation due to an ill-defined mapping. Explicitly, we require the following modification to Equation~(\ref{eqn: fundamental posterior w inner summation}):
\begin{equation}
\mathcal{P}\left(\boldsymbol{\theta}\;|\ldots\right)
=\mathop{\int}\mathop{\sum}_{\mathcal{X}}\mathcal{P}(\boldsymbol{x},\mathscr{D}\;|\ldots)\Biggl\lvert\det\left(\frac{\partial\boldsymbol{x}}{\partial{g}(\boldsymbol{x},\mathscr{D})}\right)\Biggl\rvert d\boldsymbol{\rho}d\boldsymbol{\Omega}
=\mathop{\int}\mathop{\sum}_{\mathcal{X}}\mathcal{P}(f(\boldsymbol{y})\;|\ldots)\Biggl\lvert\det\left(\frac{\partial f_{\boldsymbol{x}}(\boldsymbol{y})}{\partial\boldsymbol{y}}\right)\Biggl\rvert d\boldsymbol{\rho}d\boldsymbol{\Omega},
\label{eqn: discrete parameter}
\end{equation}
where $\boldsymbol{x}\equiv f_{\boldsymbol{x}}(\boldsymbol{y})$, and $\mathcal{X}$ is a set of elements on the space $\mathbb{N}$ of the discrete parameter $\mathscr{D}$, such that $\mathcal{X}=\{\mathscr{D}\from\boldsymbol{y}-{g}(\boldsymbol{x},\mathscr{D})=\boldsymbol{0}\}$. Again, if the map ${f}\from Y\to (X,\mathbb{N})$, $\boldsymbol{y}\mapsto(\boldsymbol{x},\mathscr{D})$ is injective, the integrand summation reduces to a single term $\forall\boldsymbol{y}$.

Difficulties arise when one must define such a discrete parameter \textit{given} a probability density distribution spanning the image $X$ and one does not possess full understanding of the interior-exterior map $\boldsymbol{y}\mapsto\boldsymbol{x}$ -- e.g, due to absence of analyticity -- which allows one to satisfy the above criteria. If the information-loss exhibited by the map $\boldsymbol{y}\mapsto\boldsymbol{x}$ was of a discrete nature, it would be intractable to define some generally-applicable discrete parameter for information conservation because the mapping can assume many forms and is numerical. In general however, the map $\boldsymbol{y}\mapsto\boldsymbol{x}$ exhibits a more severe form of information-loss, due to local non-invertibility. Moreover, it may be the case that global violations of injectivity arise due to local non-invertibility (as a trivial example consider a map $\mathbb{R}\to\mathbb{R}$ with local extrema), in which case defining an additional discrete parameter would not deal with the singularities.

\subsection{Local invertibility versus ensemble cardinality}\label{sec:injectivity 2}

We now focus on the properties of a map $\boldsymbol{y}\mapsto\boldsymbol{x}$ assuming that: (i) all parameters which constitute $\boldsymbol{\theta}$ globally control the EOS over the (local comoving) density domain; and (ii) the image in the EOS \textit{function}-space of any point $\boldsymbol{\theta}\in T$ is not the image of any other point in $T\subset\mathbb{R}^{n}$. By the second condition we mean that the map between EOS parameter and function spaces is of the form $\mathbb{R}^{n}\to\mathbb{R}^{n}$ (where the numbers in the latter space directly and uniquely enter in a function of the local comoving density $\varepsilon$) and is both locally and globally invertible; there are then no degeneracies due to information loss. Suppose that these conditions are satisfied: diffeomorphicity can be still be violated if the ensemble cardinality satisfies $s>1$ -- i.e., if we condition on observations of more than one star.


\begin{figure}
\centering
\includegraphics[width=0.7\textwidth]{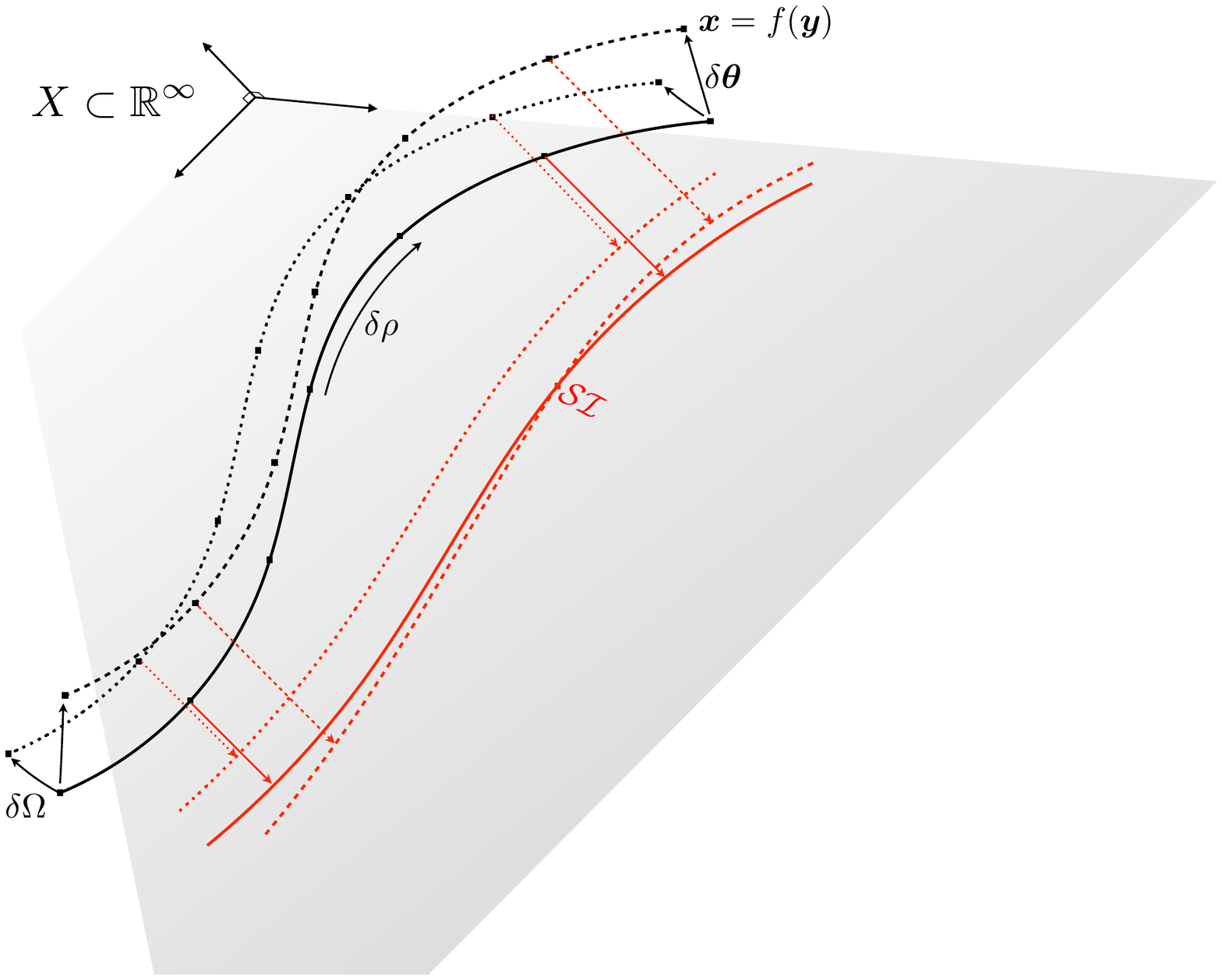}
\caption{A discrete low-dimensional cartoon representation of a projection of the solution-surface onto a subspace of $\mathbb{R}^{\infty}$, inducing a self-intersection (a degenerate truncated exterior solution) labelled $\mathcal{SI}$. The labelled self-intersection will in general be but one of a continuum of such self-intersections. Truncation is discussed in Sections~\ref{sec:solutions for generative models} and \ref{sec:mapping definition}, and in this context means truncation in multipolar order of an analytic exterior solution, whilst rotational corrections at lower-multipole-order may or may not be preserved. If the truncation is strictly in rotational order, truncation in multipolar order is implicit, and there are also small variations of the solution-surface in the lower-dimensional space because the interior-exterior map for a single star lowers in rotational approximation order.}
\label{fig:self-intersections}
\end{figure}

A digression on the single-star solution-surface (introduced in Appendix~\ref{sec:matching}) is now required if we are to address map invertibility. Projections of the solution-surface into finite-dimensional subspaces are generated by ignoring a subset of coordinates (exterior parameters). If we define $d<n+2$, continua of surface self-intersections are incurred -- a form of degeneracy due to information loss under the map $\boldsymbol{y}\mapsto\boldsymbol{x}$; we illustrate these properties in Fig.~\ref{fig:self-intersections}. In other words, for a single star, $\boldsymbol{y}\mapsto\boldsymbol{x}$ is non-injective because exterior solutions match to more than one interior solution. If $\boldsymbol{y}\mapsto\boldsymbol{x}$ is defined such that number of exterior parameters constituting $\boldsymbol{x}$ be sufficiently large so as to guarantee uniqueness of the exterior solution given values of the interior parameters, injectivity is not violated. In general, because the (single-star) domain and its image are continuous sets, the space $\mathbb{R}^{d}$ (a superset of the map codomain) must have a dimensionality greater than or equal to that of $\mathbb{R}^{n+2}$ (a superset of the map domain) for injectivity to be satisfied. If the dimensionality of the codomain is defined to be $d\geq n+2$, the codimension of single-star solution-surface is greater than or equal to zero and therefore the surface in general does not self-intersect.\footnote{If $\Omega$ is defined as a parameter in both the $\mathbb{R}^{n+2}$ and $\mathbb{R}^{d}$ spaces, self-intersections on $\Omega$-hyperslices are problematic in terms of local injectivity violation because the $\Omega$-direction is a basis vector in both spaces.}

Consider an analysis conditional on both observations of a single star and on an arbitrary set of submodels which share exterior parameters (up to their respective truncation orders). In this case the solution-surface is not projected into a subspace of dimension $d<n+2$, and thus injectivity is not violated (provided the above axioms are satisfied). However, as discussed in Sections~\ref{sec:solutions for generative models} and \ref{sec:mapping definition}, if we are to statistically distinguish between EOSs whose exterior solutions can (for certain densities and rotation frequencies) be degenerate with respect to $M$ and $\Req$, and which exhibit approximately universal metric and surface deformations (e.g., $J$, $Q$, and $e$) due to emergent interior symmetries, the likelihood function must be highly sensitive to higher-order variations of the exterior spacetime.

The natural solution to this problem in the IP-paradigm (see Section~\ref{sec:mapping definition}) was to condition on observations of multiple slowly-rotating stars, and to truncate the \textit{metric} in multipolar-order to further accelerate likelihood evaluation, provided joint likelihood functions of remaining parameters (e.g., rotationally perturbed $\boldsymbol{M}$, $\boldsymbol{R}_{\textrm{eq}}$, and $\boldsymbol{e}$, and possibly $\boldsymbol{\Omega}$) are insensitive to such approximations. In the EP-paradigm, we also adopted in Section~\ref{sec:mapping} the notion of using a universal relation for the surface ellipticity when a prior is defined on a space of exterior parameters; we will return to this example in below. In principle this can be extended to use of approximate universal relations for $J$ and $Q$, which are \textit{not} free parameters with an associated prior, but defined through a constraint equation in terms of $M$, $\Req$, and $\Omega$. The intuitive reason for this is that invoking universal relations in such a manner \textit{may} reduce likelihood function inaccuracy, whilst enabling one to condition -- in a single joint analysis -- on observations of more stars for a fixed number of EOS parameters.

A consequence of defining vector $\boldsymbol{x}$ as an ($s>1$) \textit{ensemble-solution} (a point in a space orthogonally spanned by parameters of a set of exterior spacetimes) is that if at least two rotating stars each contribute at least three exterior parameters to $\boldsymbol{x}$, the $(n+2)$-dimensional solution-surface must be projected (for each star) into a subspace in order to satisfy $d\equiv n+2s$ over the ensemble. It follows that self-intersections are induced: the exterior solution for each star is truncated at some order and matches to multiple truncated interior solutions for that star (Fig.~\ref{fig:self-intersections}).\footnote{If an analytic exterior solution is truncated at some order in rotation frequency, this means an expansion of the Einstein field equations is truncated in rotation frequency, and thus so are numerical interior solutions to which the truncated exterior solutions must match. For slow-rotation, the metric may also be further truncated in multipolar order to impose spherical symmetry and thus accelerate likelihood evaluation.} The most useful and natural case to consider here is that each rotating star represents a solution to the same field equation approximation: homogeneous projections of the $(n+2)$-dimensional solution-surface thus form the exterior-parameter spaces of stars in the ensemble.\footnote{If the field equation approximations happen to be heterogeneous across the ensemble, the situation is more complex (as hinted in the caption of Fig.~\ref{fig:self-intersections}), but less pragmatic, and the pathologies we proceed to discuss will not be appropriately bypassed.} If the approximation is that of slow-rotation with multipolar metric truncation (to impose exterior metric spherical symmetry), where the rotation frequencies are defined as both interior and exterior parameters, and surface deformation is approximated with a universality relation, the free exterior parameters are $\boldsymbol{x}=(\boldsymbol{M},\boldsymbol{R}_{\textrm{eq}},\boldsymbol{\Omega})$. There is thus similarity to a case where the ensemble consists of only \textit{static} stars (Section~\ref{sec:mapping definition}) in that the mass-radius plane is the primary focus and source of constraining power on EOS parameters and central densities.

If all exterior solutions match to stable interior solutions admitted by two or more EOS functions $\boldsymbol{\theta}$, the map $\boldsymbol{y}\mapsto\boldsymbol{x}$ is non-injective and thus non-invertible at those points $\boldsymbol{x}$ which are not unique images. For instance, if all stable exterior solutions in an ensemble (whose exterior parameters are collectively denoted by $\boldsymbol{x}$) are identical, point $\boldsymbol{x}$ clearly matches to more than one set $\boldsymbol{y}$ of stable interior solutions (e.g., Fig.~\ref{fig:self-intersections}).\footnote{We note that typically, informative observations of multiple stars with similar exterior spacetimes will not yield statistical constraining power in excess of that yielded by informative observations of stars with appreciably different exterior spacetimes. Nevertheless, the truncated exterior solutions associated with a fixed $\boldsymbol{\theta}=\boldsymbol{\theta}_{1}$ (a fixed EOS, but variable $\rho$ and $\Omega$) can be degenerate with respect to the truncated solutions of some fixed $\boldsymbol{\theta}=\boldsymbol{\theta}_{2}$ (a different fixed EOS, but variable $\rho$ and $\Omega$). That is, points $\boldsymbol{x}$ exist which are permitted by two or more EOS, and thus are not generated by \textit{unique} preimages. A clear example is any point at which the exterior solutions are all identical, and that particular single-star solution is permitted by two or more EOS (each with some values of $\rho$ and $\Omega$). This example is non-exhaustive: a second example is any point at which all the exterior solutions are equal to one of two single-star solutions, and two EOS each permit both of those single-star solutions.} It is interesting to note the nature of the injectivity violation: the map is in fact \textit{locally} non-invertible at these points. The Jacobian loses rank at any point where the central densities and rotation frequencies of \textit{two or more} stars are identical because locally there exist \textit{one or more} orthogonal interior basis vectors along which none of the exterior solutions change. Since such points form (intersecting) surfaces in $\mathbb{R}^{n+2s}$ of co-dimension $(n+2\mathbb{I}-2)$, where $\mathbb{I}\geq 2$ is the number of stars with identical interior conditions and thus spacetimes, the Jacobian singularities are evidently not (all) isolated. These surfaces do not cover the entire space of interior parameters, so only for a subset of the domain $Y$ is the map not locally invertible: for the remaining subset the $d\equiv n+2s$ free exterior parameters $\boldsymbol{x}$ are entirely sufficient to uniquely specify a preimage $\boldsymbol{y}$.

If there exists a finite probability density in the local neighbourhood of an ensemble-solution $\boldsymbol{x}\in X$ which does not uniquely specify a preimage $\boldsymbol{y}\in Y$ (e.g., a point $\boldsymbol{x}$ at which all exterior solutions are identical), it is not well-defined how to map this probability density onto the domain $Y$. If  the Jacobian (the transformation law of Section \ref{sec:mapping}) is applied, to a Bayesian the prior implicitly defined on a space of interior parameters will be ill-behaved as discussed in Appendix~\ref{sec:injectivity}: there will be continuous subsets of points in the space of interior parameters where the prior density is zero, and in the near-vicinities of these points the prior density will be small. Final points to note are that: (i) if the rotation frequencies are also considered as exterior parameters, then by definition vectors which do not change any exterior solutions are restricted to exist in $\boldsymbol{\Omega}$-hyperslices of $\mathbb{R}^{n+2s}$; and (ii) if a likelihood function is highly sensitive to rotation frequencies defined as exterior parameters, and the supported frequencies of all stars are sufficiently different, these Jacobian singularities may not prove as important to consider.

\subsection{Surjectivity}\label{sec:surjectivity}

\begin{figure}
\centering
\includegraphics[width=0.4\textwidth]{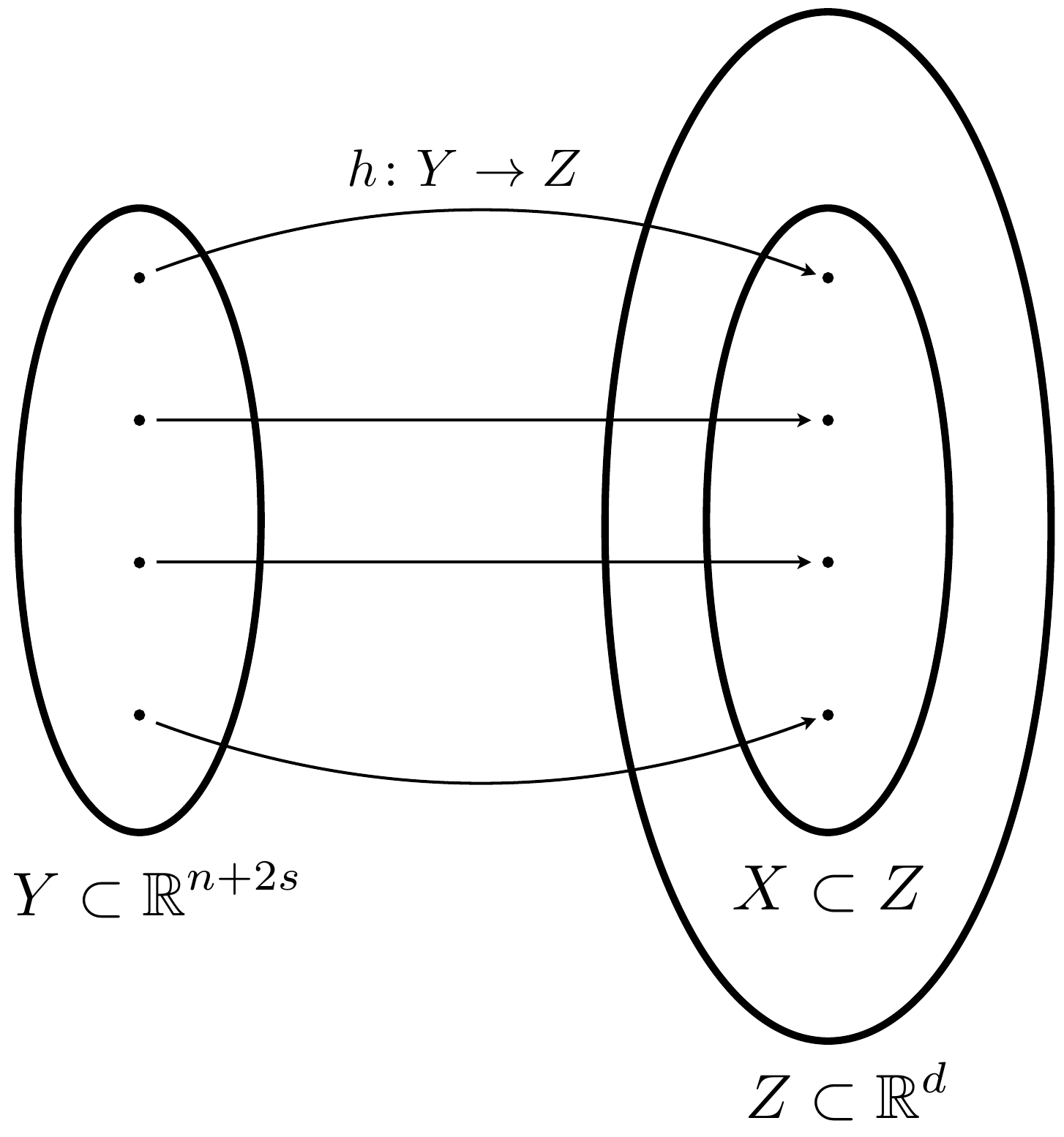}
\caption{A discrete representation of a non-surjective map ${h}\from Y\to Z$ where the codomain $Z$ is a superset of the image $X$ of the domain $Y$. The image $X$ is equivalent to the codomain $X$ of ${f}\from Y\to X$, where the maps ${f}$ and ${h}$ share a domain. An inclusion map $\iota\from X\xhookrightarrow{}Z$ is required to map $X$ onto the codomain of ${h}$, where $Z$ is defined by the subset of a $d$-dimensional space over which a probability density distribution is finite.}
\label{fig:non-surjective}
\end{figure}

Let us consider the computation of a posterior distribution of the free parameters of an ensemble of exterior spacetime solutions, where the \textit{form} of each analytic exterior solution is shared between stars (with the possible exception of the truncation order). In order to define where (in the space of exterior parameters) a marginal prior distribution of the exterior parameters is permitted to be finite conditioned on general relativistic gravity and the EOS model, we would need to precompute the image, $X\subset\mathbb{R}^{d}$, of $Y\subset\mathbb{R}^{n+2s}$ under the map ${f}$; this image would correspond to the set of stable ensemble-solutions which can have finite associated probability densities because they match to an ensemble of interior solutions (see Fig.~\ref{fig:invertible}). Both computation and implementation of the image $X$ is, however, impractical: it would in general need to be used to restrict the ensemble solutions for which a sampling algorithm is permitted to evaluate likelihoods (see Appendix \ref{sec:prior under reparametrisation} for further discussion on prior definitions for EOS parameter estimation). We proceed to discuss -- in terms of surjectivity of the map from interior to exterior parameters -- the consequences of \textit{not} precomputing the image $X$.

In Section \ref{sec:mapping} we introduced the notion of the $(n+2)$-dimensional surface of \textit{single-star} exterior solutions -- a subset of some ambient space $\mathbb{R}^{d}$ we define as the image $X$ of $Y$ under the map $f$. We then argued that only $d\equiv n+2$ parameters of an analytic exterior solution may be considered free for parameter estimation conditioned on observations of a single star. Let us now consider a point $\boldsymbol{z}\in\mathbb{R}^{d}$ which represents a \textit{possible} single-star solution: we may assign a finite local probability density to point $\boldsymbol{z}$ during exterior-parameter estimation (the first phase of the EP-paradigm). In the post-processing phase of the EP-paradigm, we must invoke an EOS model and \textit{explicitly} match stable interior solutions to exterior solutions. It follows that if we do not precompute the image $X$ in order to restrict exterior-parameter estimation, point $\boldsymbol{z}$ may not lie on the solution-surface -- i.e., point $\boldsymbol{z}\notin X$ -- either because it does not match to a \textit{stable} interior solution, or because it simply does not match to any interior solution. For the latter case, we can reason that a single functional form for the EOS is conditioned on, and the associated parameter space is of some finite dimension $n$: the space of (relativistically permitted) EOS functions is thus incomplete, whilst the exact (assumedly universal) EOS may require an infinite number of basis functions to reconstruct exactly. It follows that for $Y\subset\mathbb{R}^{n+2}$, the map ${f}\from Y\to X$ is not in general surjective with respect to the complete set of physical single-star exterior solutions which match to stable interior solutions given all permissible functional forms for the EOS; further, there may exist points $\boldsymbol{z}$ which are assigned a finite posterior probability density but which are not exterior solutions for \textit{any} relativistically permissible EOS and are thus unphysical. As an example, suppose the exterior parameters are $M$, $\Req$, and $\Omega$: a point $\boldsymbol{z}=(M,\Req,\Omega)$ associated with a finite local probability density may not match to any stable interior solution to the field equations when a certain EOS model is conditioned on.

In Appendix~\ref{sec:injectivity 2} we justified modelling an ensemble $\boldsymbol{x}\in X$ of exterior solutions, where the map ${f}\from Y\to X$ is defined for $Y\subset\mathbb{R}^{n+2s}$, with $s$ defined as the number of stars. Now let some point $\boldsymbol{z}\in\mathbb{R}^{d}$, where $d\equiv n+2s$, be a \textit{possible} ensemble-solution: as before, we may assign a finite local probability density to point $\boldsymbol{z}$ during exterior-parameter estimation. Similarly, in the post-processing phase we must invoke an EOS model and \textit{explicitly} match \textit{sets} of stable interior solutions (admitted by a \textit{single} EOS $\boldsymbol{\theta}$) to \textit{sets} of exterior solutions. If we do not precompute the image $X$, each single-star exterior solution comprising $\boldsymbol{z}$ will not in general match to a stable interior solution; on the other hand, if they \textit{all} match to stable interior solutions, those interior solutions may not be admitted by a single EOS $\boldsymbol{\theta}$. In other words, the map $f$ is not in general surjective with respect to the complete set of ensemble-solutions $\boldsymbol{x}$ generated by all functional forms of the EOS given general relativistic gravity, and further, there may exist points $\boldsymbol{z}$ which are assigned a finite (posterior) probability density but are not exterior-solution ensembles which match to \textit{any} relativistically permissible EOS -- such points are thus unphysical.

To summarise, consider Fig.~\ref{fig:non-surjective}: let us define a map ${h}\from Y\to Z$ where the codomain $Z$ is defined as the subset of a $d$-dimensional space over which a (posterior) probability density distribution is finite. The stable ensemble-solutions $X\subset\mathbb{R}^{d}$ generated under $f:Y\to X$ are but a proper subset of $Z$ because ensembles of stable interior solutions (admitted by an EOS $\boldsymbol{\theta}$) are only matched to ensembles of exterior solutions during the post-processing phase of the EP-paradigm.

The consequence of map ${h}$ being non-surjective is that the integral of the probability density over the image $X$ is not unity. Therefore, the integral of a corresponding density distribution over the domain $Y\subset\mathbb{R}^{n+2s}$ is not unity if the transformation law suggested in Section~\ref{sec:mapping} is applied, even if the map $f\from Y\to X$ is invertible -- i.e., even if the map $h$ is injective. Thus the distribution on $Y$ can only be termed a \textit{probability} density distribution if we renormalise over $Y$. We opine that it is most rational to renormalise over the subset of $\mathbb{R}^{n}$ spanned by the EOS model parameters $\boldsymbol{\theta}$, following marginalisation over $\boldsymbol{\rho}$ and $\boldsymbol{\Omega}$. Finally, we note that if map ${h}$ is injective, a distributional renormalisation is equivalent to redefinition of the codomain of ${h}$ as the image $X$, thus forcing equivalence to map ${f}$.

\subsection{The prior under an exterior-interior reparametrisation}\label{sec:prior under reparametrisation}

In this section we demonstrate the implicit definition of a prior on a space of interior parameters when the EP-paradigm is invoked. Such an approach has been implemented in the literature (see Section~\ref{sec:Review}).

The fundamental difference between the paradigms is the parameter space on which a prior is explicitly defined. In the IP-paradigm (see Section \ref{sec:IM definition} through \ref{sec:computational tractability}), a prior is defined on the space $\mathbb{R}^{n+2s}$ of interior parameters $\boldsymbol{y}=(\boldsymbol{\theta},\boldsymbol{\rho},\boldsymbol{\Omega})$. In the EP-paradigm, a prior is defined on the space $\mathbb{R}^{d}$ of exterior parameters $\boldsymbol{z}$, where we adopt the map $h\colon Y\to Z$, $\boldsymbol{y}\mapsto\boldsymbol{z}$ as introduced in Appendix~\ref{sec:surjectivity}. Indeed, if we invoke the ansatz that the map $\boldsymbol{y}\mapsto\boldsymbol{z}$ is everywhere diffeomorphic (see Section \ref{sec:mapping} through Appendix~\ref{sec:surjectivity}), the map $h$ is equivalent to map $f\colon Y\to X$, $\boldsymbol{y}\to\boldsymbol{x}$; in this special case the points $\boldsymbol{z}$ are elements of the same set as $\boldsymbol{x}$. It is important for there to exist two distinct labels for elements (points) of the sets $X$ and $Z$, which are not identical without the above ansatz because $Z\supset X$. Under the ansatz that $Z\equiv X$, the marginal posterior distribution of the EOS parameters $\boldsymbol{\theta}$ can be written -- cf. Equation~(\ref{eqn: fundamental posterior w inner summation}) -- as 
\begin{equation}
\mathcal{P}\left(\boldsymbol{\theta}\;|\;\mathcal{D},\mathcal{M},\mathcal{I}\right)
=\mathop{\int}\mathcal{P}(\boldsymbol{z}\;|\;\mathcal{D},\mathcal{M},\mathcal{I})\Biggl\lvert\det\left(\frac{\partial\boldsymbol{z}}{\partial\boldsymbol{y}}\right)\Biggl\rvert d\boldsymbol{\rho}d\boldsymbol{\Omega}
=\mathop{\int}\mathcal{P}(\boldsymbol{x}\;|\;\mathcal{D},\mathcal{M},\mathcal{I})\Biggl\lvert\det\left(\frac{\partial\boldsymbol{x}}{\partial\boldsymbol{y}}\right)\Biggl\rvert d\boldsymbol{\rho}d\boldsymbol{\Omega}.
\label{eqn:posterior ansatz}
\end{equation}
The notion of post-processing a marginal posterior distribution of the exterior parameters $\boldsymbol{x}$ to calculate a marginal posterior distribution of the EOS parameters $\boldsymbol{\theta}$ is natural and has been invoked in the astrophysical literature \citep[see, e.g.,][]{Ozel2009}.

Let us manipulate Equation~(\ref{eqn:exact posterior}) under the ansatz of map diffeomorphicity to obtain Equation~(\ref{eqn:posterior ansatz}), in order to demonstrate that the posteriors would be identical if this ansatz held true. The (hyper)prior $\pi(\boldsymbol{\theta},\boldsymbol{\rho},\boldsymbol{\Omega},\boldsymbol{\alpha},\boldsymbol{\eta},\boldsymbol{\beta})$ may be written as
\begin{equation}
\mathcal{P}(\boldsymbol{\theta},\boldsymbol{\rho},\boldsymbol{\Omega},\boldsymbol{\alpha},\boldsymbol{\eta},\boldsymbol{\beta}\;|\;\mathcal{M},\mathcal{I})
=\mathcal{P}(\boldsymbol{x},\boldsymbol{\alpha},\boldsymbol{\eta},\boldsymbol{\beta}\;|\;\mathcal{M},\mathcal{I})\Biggl\lvert\det\left(\frac{\partial\boldsymbol{x}}{\partial\boldsymbol{y}}\right)\Biggl\rvert .
\end{equation}
Further, let us write the (hyper)prior distribution of the interior parameters, nuisance parameters, and hyperparameters as follows:
\begin{equation}
\begin{aligned}
\mathcal{P}(\boldsymbol{\theta},\boldsymbol{\rho},\boldsymbol{\Omega},\boldsymbol{\alpha},\boldsymbol{\eta},\boldsymbol{\beta}\;|\;\mathcal{M},\mathcal{I})
&=
\mathcal{P}(\boldsymbol{\theta}\;|\;\mathcal{M},\mathcal{I})
\mathcal{P}(\boldsymbol{\rho},\boldsymbol{\Omega}\;|\;\boldsymbol{\theta},\boldsymbol{\alpha},\mathcal{M},\mathcal{I})
\mathcal{P}(\boldsymbol{\alpha}\;|\;\mathcal{M},\mathcal{I})
\mathcal{P}(\boldsymbol{\eta}\;|\;\boldsymbol{\theta},\boldsymbol{\rho},\boldsymbol{\Omega},\boldsymbol{\beta},\mathcal{M},\mathcal{I})
\mathcal{P}(\boldsymbol{\beta}\;|\;\mathcal{M},\mathcal{I})\\
&=\mathcal{P}(\boldsymbol{x}\;|\;\boldsymbol{\alpha},\mathcal{M},\mathcal{I})
\mathcal{P}(\boldsymbol{\alpha}\;|\;\mathcal{M},\mathcal{I})
\mathcal{P}(\boldsymbol{\eta}\;|\;\boldsymbol{\theta},\boldsymbol{\rho},\boldsymbol{\Omega},\boldsymbol{\beta},\mathcal{M},\mathcal{I})
\mathcal{P}(\boldsymbol{\beta}\;|\;\mathcal{M},\mathcal{I})
\Biggl\lvert\det\left(\frac{\partial\boldsymbol{x}}{\partial\boldsymbol{y}}\right)\Biggl\rvert ,
\end{aligned}
\end{equation}
where the prior distribution of the \textit{exterior} parameters, nuisance parameters, and hyperparameters is
\begin{equation}
\mathcal{P}(\boldsymbol{x},\boldsymbol{\alpha},\boldsymbol{\eta},\boldsymbol{\beta}\;|\;\mathcal{M},\mathcal{I})
=
\mathcal{P}(\boldsymbol{x}\;|\;\boldsymbol{\alpha},\mathcal{M},\mathcal{I})
\mathcal{P}(\boldsymbol{\alpha}\;|\;\mathcal{M},\mathcal{I})
\mathcal{P}(\boldsymbol{\beta}\;|\;\mathcal{M},\mathcal{I})
\mathcal{P}(\boldsymbol{\eta}\;|\;\boldsymbol{x},\boldsymbol{\beta},\mathcal{M},\mathcal{I}).
\end{equation}
Whilst the prior distribution of the central densities and rotation frequencies is separable over stars as (see Section \ref{sec:computational tractability})
\begin{equation}
\mathcal{P}(\boldsymbol{\rho},\boldsymbol{\Omega}\;|\;\boldsymbol{\theta},\boldsymbol{\alpha},\mathcal{M},\mathcal{I})
=
\mathop{\prod}_{i=1}^{s}
\mathcal{P}(\rho_{i},\Omega_{i}\;|\;\boldsymbol{\theta},\boldsymbol{\alpha},\mathcal{M},\mathcal{I}),
\end{equation}
the prior distribution of the exterior parameters, $\mathcal{P}(\boldsymbol{x}\;|\;\boldsymbol{\alpha},\mathcal{M},\mathcal{I})$, is non-separable over stars because the EOS parameters are shared. Note that whilst the transformed prior distribution of exterior parameters $\boldsymbol{x}$, $\mathcal{P}(\boldsymbol{x}\;|\;\boldsymbol{\alpha},\mathcal{M},\mathcal{I})$, is conditional on the hyperparameters $\boldsymbol{\alpha}$ on which control the prior $\mathcal{P}(\boldsymbol{\rho},\boldsymbol{\Omega}\;|\;\boldsymbol{\theta},\boldsymbol{\alpha},\mathcal{M},\mathcal{I})$, it is \textit{not} conditional on the EOS parameters $\boldsymbol{\theta}$ because point $\boldsymbol{x}$ represents a set of exterior solutions which \textit{deterministically} matches to a set stable interior solutions to the field equations. The EOS parameters $\boldsymbol{\theta}$ appear in the prior $\mathcal{P}(\boldsymbol{\rho},\boldsymbol{\Omega}\;|\;\boldsymbol{\theta},\boldsymbol{\alpha},\mathcal{M},\mathcal{I})$, on the other hand, to enforce that only interior conditions ($\rho$,$\Omega$) which generate stable interior solutions are assigned finite support \textit{a priori}.

Equation~(\ref{eqn:exact posterior}) thus becomes
\begin{equation}
\mathcal{P}\left(\boldsymbol{\theta}\;|\;\mathcal{D},\mathcal{M},\mathcal{I}\right)
\propto\mathop{\int}
\mathcal{P}\left(\mathcal{D}\;|\;\boldsymbol{x},\boldsymbol{\eta},\mathcal{M}\right)
\mathcal{P}\left(\boldsymbol{x},\boldsymbol{\alpha},\boldsymbol{\eta},\boldsymbol{\beta}\;|\;\mathcal{M},\mathcal{I}\right)
\Biggl\lvert\det\left(\frac{\partial\boldsymbol{x}}{\partial\boldsymbol{y}}\right)\Biggl\rvert d\boldsymbol{\eta}d\boldsymbol{\beta}d\boldsymbol{\rho}d\boldsymbol{\Omega}d\boldsymbol{\alpha},
\end{equation}
where the prior predictive probability of $\mathcal{D}$ is given by
\begin{equation}
\begin{aligned}
\mathcal{P}\left(\mathcal{D}\;|\;\mathcal{M},\mathcal{I}\right)
&=\mathop{\int}
\mathcal{P}\left(\mathcal{D}\;|\;\boldsymbol{x},\boldsymbol{\eta},\mathcal{M}\right)
\mathcal{P}\left(\boldsymbol{x},\boldsymbol{\alpha},\boldsymbol{\eta},\boldsymbol{\beta}\;|\;\mathcal{M},\mathcal{I}\right)
\Biggl\lvert\det\left(\frac{\partial\boldsymbol{x}}{\partial\boldsymbol{y}}\right)\Biggl\rvert d\boldsymbol{\eta}d\boldsymbol{\beta}d\boldsymbol{\rho}d\boldsymbol{\Omega}d\boldsymbol{\alpha}d\boldsymbol{\theta}\\
&=\mathop{\int}
\mathcal{P}\left(\mathcal{D}\;|\;\boldsymbol{x},\boldsymbol{\eta},\mathcal{M}\right)
\mathcal{P}\left(\boldsymbol{x},\boldsymbol{\alpha},\boldsymbol{\eta},\boldsymbol{\beta}\;|\;\mathcal{M},\mathcal{I}\right)
d\boldsymbol{\eta}d\boldsymbol{\beta}d\boldsymbol{x}d\boldsymbol{\alpha},
\label{eqn:transformed posterior normalisation}
\end{aligned}
\end{equation}
and thus
\begin{equation}
\begin{aligned}
\mathcal{P}\left(\boldsymbol{\theta}\;|\;\mathcal{D},\mathcal{M},\mathcal{I}\right)
&=\frac{1}{\mathcal{P}\left(\mathcal{D}\;|\;\mathcal{M},\mathcal{I}\right)}
\mathop{\int}
\mathcal{P}\left(\mathcal{D}\;|\;\boldsymbol{x},\boldsymbol{\eta},\mathcal{M}\right)
\mathcal{P}\left(\boldsymbol{x},\boldsymbol{\alpha},\boldsymbol{\eta},\boldsymbol{\beta}\;|\;\mathcal{M},\mathcal{I}\right)
\Biggl\lvert\det\left(\frac{\partial\boldsymbol{x}}{\partial\boldsymbol{y}}\right)\Biggl\rvert d\boldsymbol{\eta}d\boldsymbol{\beta}d\boldsymbol{\rho}d\boldsymbol{\Omega}d\boldsymbol{\alpha}\\
&=\mathop{\int}\left[\mathop{\int}\mathcal{P}(\boldsymbol{x},\boldsymbol{\alpha},\boldsymbol{\eta},\boldsymbol{\beta}\;|\;\mathcal{D},\mathcal{M},\mathcal{I})d\boldsymbol{\eta}d\boldsymbol{\beta}d\boldsymbol{\alpha}\right]\Biggl\lvert\det\left(\frac{\partial\boldsymbol{x}}{\partial\boldsymbol{y}}\right)\Biggl\rvert d\boldsymbol{\rho}d\boldsymbol{\Omega}
=\mathop{\int}\mathcal{P}\left(\boldsymbol{x}\;|\;\mathcal{D},\mathcal{M},\mathcal{I}\right)\Biggl\lvert\det\left(\frac{\partial\boldsymbol{x}}{\partial\boldsymbol{y}}\right)\Biggl\rvert d\boldsymbol{\rho}d\boldsymbol{\Omega},
\end{aligned}
\label{eqn:transformed posterior}
\end{equation}
as required. Note that $\mathcal{P}\left(\mathcal{D}\;|\;\mathcal{M},\mathcal{I}\right)$ is taken under the integral to as the normalisation of the posterior distribution of the exterior parameters, nuisance parameters, and hyperparameters:
\begin{equation}
\mathcal{P}(\boldsymbol{x},\boldsymbol{\alpha},\boldsymbol{\eta},\boldsymbol{\beta}\;|\;\mathcal{D},\mathcal{M},\mathcal{I})\propto\mathcal{P}(\mathcal{D}\;|\;\boldsymbol{x},\boldsymbol{\eta},\mathcal{M})\mathcal{P}(\boldsymbol{x},\boldsymbol{\alpha},\boldsymbol{\eta},\boldsymbol{\beta}\;|\;\mathcal{M},\mathcal{I}),
\end{equation}
where $\mathcal{P}(\mathcal{D}\;|\;\boldsymbol{x},\boldsymbol{\eta},\mathcal{M})$ is the likelihood function of the exterior parameters $\boldsymbol{x}$ and the nuisance parameters $\boldsymbol{\eta}$. This likelihood is \textit{identical} to the likelihood of the interior parameters and nuisance parameters, $\mathcal{P}(\mathcal{D}\;|\;\boldsymbol{\theta},\boldsymbol{\rho},\boldsymbol{\Omega},\boldsymbol{\eta},\mathcal{M})$, provided that in both paradigms precisely the same analytic exterior solutions are used under the map $\boldsymbol{y}\mapsto\boldsymbol{x}$. The marginal posterior distribution of the exterior parameters, $\mathcal{P}(\boldsymbol{x}\;|\;\mathcal{D},\mathcal{M},\mathcal{I})$, spans the image $X$ of the domain $Y$ under map $f$.

The EP-paradigm is an approximation to the IP-paradigm characterised by ignoring violations of the ansatz that the map $\boldsymbol{y}\mapsto\boldsymbol{z}$ is diffeomorphic. The prior distribution of the exterior parameters needs to be recast because in the EP-paradigm the prior is not itself the result of a distributional transformation from the space of interior parameters. In other words, information about the EOS model is not encoded by the prior distribution of the exterior parameters. It is for this reason that even if the map $h$ is injective, the equality of Equation~(\ref{eqn:posterior ansatz}) will be infringed: the prior distribution of the exterior parameters will span some set $Z\supset X$ (see Fig.~\ref{fig:non-surjective}), and thus renormalisation will be required, as discussed in Appendix~\ref{sec:surjectivity}. Indeed, in the EP-paradigm we: (i) define the hyperparameters $\boldsymbol{\alpha}$ as \textit{directly} parametrising the prior distribution of the exterior parameters $\boldsymbol{z}=h(\boldsymbol{y})$; and (ii) define the (hyper)prior $\pi(\boldsymbol{z},\boldsymbol{\alpha},\boldsymbol{\eta},\boldsymbol{\beta})\neq\pi(\boldsymbol{y},\boldsymbol{\alpha},\boldsymbol{\eta},\boldsymbol{\beta})$  as
\begin{equation}
\begin{aligned}
\mathcal{P}(\boldsymbol{z},\boldsymbol{\alpha},\boldsymbol{\eta},\boldsymbol{\beta}\;|\;\mathcal{M},\mathcal{I})
=
\mathcal{P}(\boldsymbol{\alpha}\;|\;\mathcal{M},\mathcal{I})
\mathcal{P}(\boldsymbol{\beta}\;|\;\mathcal{M},\mathcal{I})
\mathcal{P}(\boldsymbol{\eta}\;|\;\boldsymbol{z},\boldsymbol{\beta},\mathcal{M},\mathcal{I})
\mathop{\prod}_{i=1}^{s}
\mathcal{P}(\boldsymbol{z}_{i}\;|\;\boldsymbol{\alpha},\mathcal{M}).
\end{aligned}
\label{eqn:prior expansion shared nuisance}
\end{equation}

An important caveat here is that by assuming the EOS parameters $\boldsymbol{\theta}$ are shared by all stars, each spacetime is fundamentally controlled by two remaining interior parameters: the central density $\rho$ and the coordinate rotation frequency ${\Omega}$. Thus, if we desire a \textit{meaningful} hierarchical model for the exterior parameters, we can state that for each star only two exterior parameters (e.g., a gravitational mass and coordinate rotation frequency) are drawn from a joint prior distribution parametrised by hyperparameters $\boldsymbol{\alpha}$, whilst all other exterior parameters are \textit{a priori} drawn from some (noninformative) joint distribution which represents our ignorance of the \textit{ensemble} exterior spacetime solution-surface admitted by the EOS model (that is, the image $X$ of $Y$ under $f$). We note that in general, the prior distribution of the exterior parameters, nuisance parameters, and hyperparameters could in principle be defined (at least for a subset of parameters and hyperparameters) as a marginal posterior distribution conditioned on an independent data set -- we discuss this possibility in Appendix~\ref{sec:updating}.

We then \textit{define} the marginal posterior distribution of the EOS parameters via an ill-defined transformation:
\begin{equation}
\mathcal{P}_{\textrm{EP}}\left(\boldsymbol{\theta}\;|\;\mathcal{D},\mathcal{M},\mathcal{I}\right)
\propto
\mathop{\int}\mathcal{P}\left(\boldsymbol{z}\;|\;\mathcal{D},\mathcal{M},\mathcal{I}\right)\Biggl\lvert\det\left(\frac{\partial\boldsymbol{z}}{\partial\boldsymbol{y}}\right)\Biggl\rvert d\boldsymbol{\rho}d\boldsymbol{\Omega},
\label{eqn:violate equality by expanding hyperprior in terms of spacetime parameters}
\end{equation}
where the posterior distribution of the exterior parameters is denoted by $\mathcal{P}(\boldsymbol{z}\;|\;\mathcal{D},\mathcal{M},\mathcal{I})$, spans the codomain $Z$ of map $h$, and is marginal:
\begin{equation}
\mathcal{P}(\boldsymbol{z}\;|\;\mathcal{D},\mathcal{M},\mathcal{I})
=
\mathop{\int}
\mathcal{P}(\boldsymbol{z},\boldsymbol{\alpha},\boldsymbol{\eta},\boldsymbol{\beta}\;|\;\mathcal{D},\mathcal{M},\mathcal{I})
d\boldsymbol{\eta}d\boldsymbol{\beta}d\boldsymbol{\alpha}
\propto
\mathop{\int}
\mathcal{P}(\mathcal{D}\;|\;\boldsymbol{z},\boldsymbol{\alpha},\boldsymbol{\eta},\boldsymbol{\beta},\mathcal{M})
\mathcal{P}(\boldsymbol{z},\boldsymbol{\alpha},\boldsymbol{\eta},\boldsymbol{\beta}\;|\;\mathcal{M},\mathcal{I})
d\boldsymbol{\eta}d\boldsymbol{\beta}d\boldsymbol{\alpha}.
\label{eqn:marginal EP posterior}
\end{equation}

A pertinent ansatz to address is that of separability of the posterior distribution of (exterior and nuisance) parameters over stars or, more generally, over research \textit{groups} with the cognitive and computational resources to contribute to the EOS parameter estimation effort (see Section \ref{sec:computational tractability} for more detail). The likelihood function is separable with respect to observational data from distinct isolated stars. The (hyper)prior, however, is not separable over stars or groups if: (i) \textit{nuisance} parameters ($\boldsymbol{\eta}$; see Section \ref{sec:IM definition}) are shared as in Equation~(\ref{eqn:prior expansion shared nuisance}); or (ii) hyperparameters ($\boldsymbol{\alpha}$ and $\boldsymbol{\beta}$; Section \ref{sec:IM definition}) are defined which are shared, also as in Equation~(\ref{eqn:prior expansion shared nuisance}). In the IP-paradigm the posterior distribution of \textit{interior} parameters is fundamentally inseparable -- even if all nuisance parameters and all hyperparameters are unshared -- because the EOS parameters $\boldsymbol{\theta}$ are shared by all stars.

Implicit in the ansatz of posterior separability is the ansatz that the prior is separable: this requires both unshared (exterior and nuisance) parameters and unshared hyperparameters. It follows that we must either not define \textit{shared} hyperparameters in the global hierarchical model $\mathcal{M}$, or \textit{fix} shared hyperparameters by conditioning on a hyperprior which exhibits singular support; the latter is more principled because unshared parameters are explicitly drawn \textit{a priori} from an assumedly known population-level distribution described by the fixed shared hyperparameters.

Let us first invoke the ansatz that all nuisance parameters $\boldsymbol{\eta}$ are \textit{unshared} between stars. Equation~(\ref{eqn:prior expansion shared nuisance}) becomes
\begin{equation}
\begin{aligned}
\mathcal{P}(\boldsymbol{z},\boldsymbol{\alpha},\boldsymbol{\eta},\boldsymbol{\beta}\;|\;\mathcal{M},\mathcal{I})
=
\mathcal{P}(\boldsymbol{\alpha}\;|\;\mathcal{M},\mathcal{I})\mathcal{P}(\boldsymbol{\beta}\;|\;\mathcal{M},\mathcal{I})
\mathop{\prod}_{i=1}^{s}
\mathcal{P}(\boldsymbol{z}_{i}\;|\;\boldsymbol{\alpha},\mathcal{M})
\mathcal{P}(\boldsymbol{\eta}_{i}\;|\;\boldsymbol{z}_{i},\boldsymbol{\beta},\mathcal{M},\mathcal{I}_{i});
\end{aligned}
\label{eqn:prior expansion}
\end{equation}
however, (a subset of) hyperparameters remain shared. If we are subjectively ignorant of the nature of population-level distributions and are uninterested in simultaneously learning hyperparameters (including nuisance hyperparameters), we can choose \textit{not} to define hyperparameters. We can then condition on some weakly information or noninformative; such a prior may be viewed as existing in a space of distributions parametrised by hyperparameters. In this case Equation~(\ref{eqn:prior expansion}) becomes
\begin{equation}
\mathcal{P}(\boldsymbol{z},\boldsymbol{\eta}\;|\;\mathcal{M},\mathcal{I})
=
\mathop{\prod}_{i=1}^{s}
\mathcal{P}(\boldsymbol{z}_{i}\;|\;\mathcal{M},\mathcal{I}_{i})
\mathcal{P}(\boldsymbol{\eta}_{i}\;|\;\boldsymbol{z}_{i},\mathcal{M},\mathcal{I}_{i}).
\label{eqn:separable joint prior}
\end{equation}
Provided the data $\mathcal{D}$ are \textit{informative}, the marginal posterior distribution of exterior parameters should by definition be insensitive to the choice of the prior.

Crucially, when shared hyperparameters are fixed, certain parameters associated with different stars are assumed to be drawn from the same underlying prior distribution, and that distribution is fixed; thus we are not permitted the \textit{choice} of different priors for different stars.\footnote{The prior distribution of parameters associated with a given star can itself be a posterior distribution conditioned on independent data sets, provided that the \textit{fundamental} prior conditioned on in the Bayesian chain of posterior updates is consistent with the distribution given by the fixed hyperparameters or is sufficiently noninformative for posterior exterior-parameter inferences to be insensitive to any inconsistencies.} Combining Equations~(\ref{eqn:violate equality by expanding hyperprior in terms of spacetime parameters}),  (\ref{eqn:marginal EP posterior}), and (\ref{eqn:separable joint prior}) yields an approximation to a marginal posterior distribution of the EOS parameters:
\begin{equation}
\begin{aligned}
\mathcal{P}_{\textrm{EP}}\left(\boldsymbol{\theta}\;|\;\mathcal{D},\mathcal{M},\mathcal{I}\right)
&\propto
\mathop{\int}
\left[\mathop{\prod}_{i=1}^{s}\mathop{\int}\mathcal{P}\left(\mathcal{D}_{i}\;|\;\boldsymbol{z}_{i},\boldsymbol{\eta}_{i},\mathcal{M}\right)
\mathcal{P}(\boldsymbol{z}_{i}\;|\;\mathcal{M},\mathcal{I}_{i})
\mathcal{P}(\boldsymbol{\eta}_{i}\;|\;\boldsymbol{z}_{i},\mathcal{M},\mathcal{I}_{i})d\boldsymbol{\eta}_{i}\right]
\Biggl\lvert\det\left(\frac{\partial\boldsymbol{z}}{\partial\boldsymbol{y}}\right)\Biggl\rvert d\boldsymbol{\rho}d\boldsymbol{\Omega}\\
&\propto
\mathop{\int}
\left[\mathop{\prod}_{i=1}^{s}\mathcal{P}\left(\boldsymbol{z}_{i}\;|\;\mathcal{D}_{i},\mathcal{M},\mathcal{I}_{i}\right)\right]
\Biggl\lvert\det\left(\frac{\partial\boldsymbol{z}}{\partial\boldsymbol{y}}\right)\Biggl\rvert d\boldsymbol{\rho}d\boldsymbol{\Omega},
\end{aligned}
\end{equation}
where the marginal posteriors $\mathcal{P}\left(\boldsymbol{z}_{i}\;|\;\mathcal{D}_{i},\mathcal{M},\mathcal{I}_{i}\right)$ may be computed in parallel, with independent compute resources, by one or more groups. The scope for computational parallelisation of posterior evaluation thus arises if the shared hyperparameters $\boldsymbol{\alpha}$ controlling the prior distribution of exterior parameters are fixed.

\begin{figure}
\centering
\includegraphics[width=0.9\textwidth]{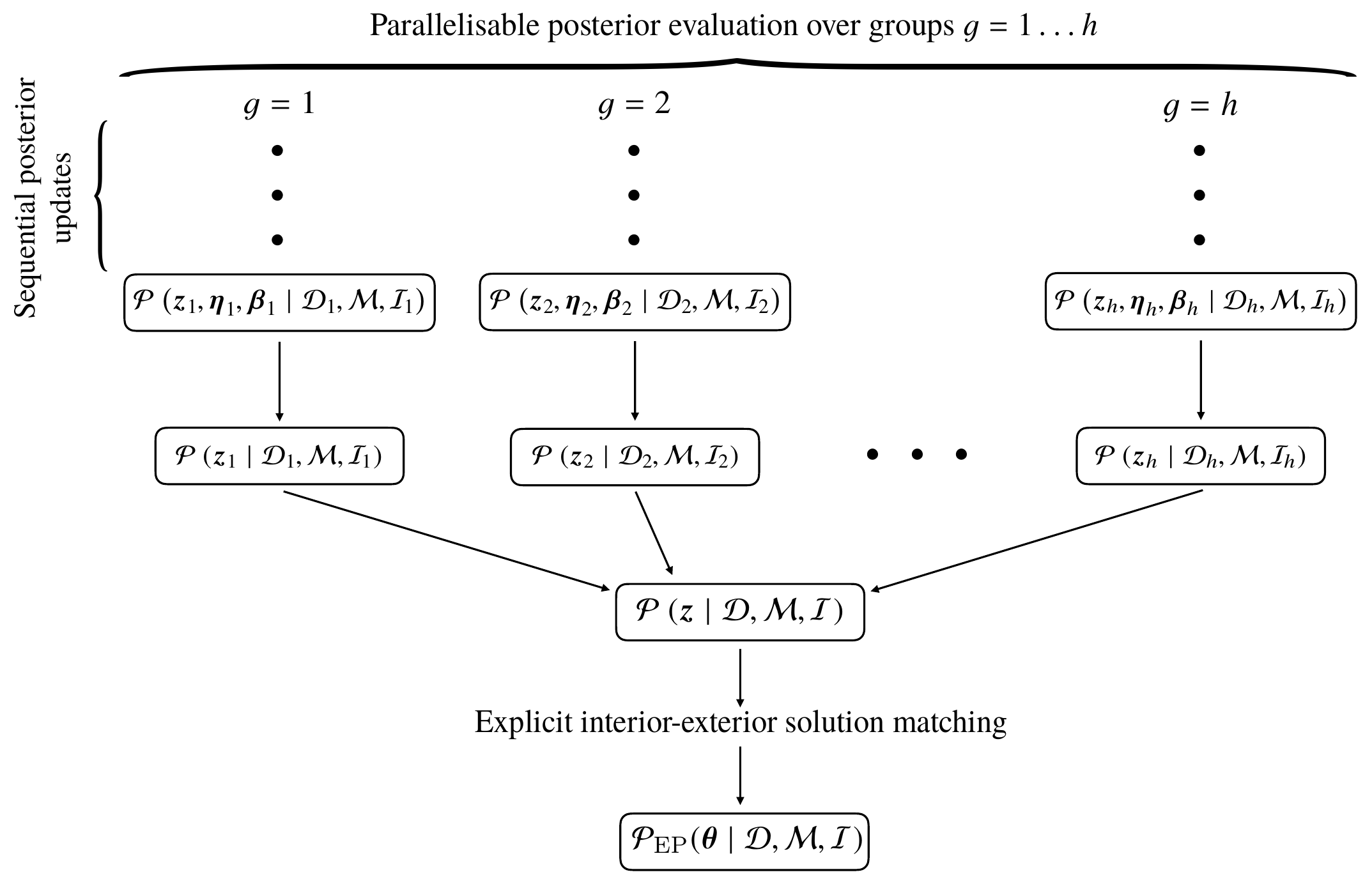}
\caption{The EOS parameter estimation procedure under the EP-paradigm, with fixed hyperparameters $\boldsymbol{\alpha}$. Evaluation of the marginal posterior distribution of the exterior parameters $\boldsymbol{z}$ is distributed amongst research groups in a parallel manner. A single group then post-processes this posterior -- via matching of sets of stable interior solutions to the field equations to sets of exterior solutions -- to define a marginal posterior distribution of the EOS parameters.}
\label{fig:EP diagram}
\end{figure}

Let us also consider a more general case in which subsets $\boldsymbol{z}_{\varg}$ of exterior parameters, subsets $\boldsymbol{\eta}_{\varg}$ of nuisance parameters,  and subsets $\boldsymbol{\beta}_{\varg}$ of hyperparameters are \textit{unshared} by a set of groups enumerated by $\varg\in\mathbb{N}_{>0}$,  but may be shared or unshared between stars. Further, let each group analyse a data subset $\mathcal{D}_{g}$ assumed to originate from one or more stars. In this case Equation~(\ref{eqn:prior expansion shared nuisance}) may instead be written
\begin{equation}
\begin{aligned}
\mathcal{P}(\boldsymbol{z},\boldsymbol{\eta},\boldsymbol{\beta}\;|\;\mathcal{M},\mathcal{I})
=
\mathop{\prod}_{\varg}
\mathcal{P}(\boldsymbol{z}_{\varg}\;|\;\mathcal{M},\mathcal{I}_{\varg})
\mathcal{P}(\boldsymbol{\beta}_{\varg}\;|\;\mathcal{M},\mathcal{I}_{\varg})
\mathcal{P}(\boldsymbol{\eta}_{\varg}\;|\;\boldsymbol{z}_{\varg},\boldsymbol{\beta}_{\varg},\mathcal{M},\mathcal{I}_{\varg}).
\end{aligned}
\label{eqn:prior expansion over groups}
\end{equation}
The justification for this form of the (hyper)prior is that the observational data from subsets of stars may be described with the same underlying type of submodel, and thus nuisance parameters or hyperparameters may be shared; a group may therefore perform exterior-parameter estimation for an appropriate subset of stars. The marginal posterior distribution of EOS parameters is then given by
\begin{equation}
\begin{aligned}
\mathcal{P}_{\textrm{EP}}\left(\boldsymbol{\theta}\;|\;\mathcal{D},\mathcal{M},\mathcal{I}\right)
&\propto
\mathop{\int}
\left[\mathop{\prod}_{\varg}
\mathop{\int}\mathcal{P}\left(\mathcal{D}_{\varg}\;|\;\boldsymbol{z}_{\varg},\boldsymbol{\eta}_{\varg},\mathcal{M}\right)
\mathcal{P}(\boldsymbol{z}_{\varg}\;|\;\mathcal{M},\mathcal{I}_{\varg})
\mathcal{P}(\boldsymbol{\beta}_{\varg}\;|\;\mathcal{M},\mathcal{I}_{\varg})
\mathcal{P}(\boldsymbol{\eta}_{\varg}\;|\;\boldsymbol{z}_{\varg},\boldsymbol{\beta}_{\varg},\mathcal{M},\mathcal{I}_{\varg})
d\boldsymbol{\eta}_{\varg}d\boldsymbol{\beta}_{\varg}\right]
\Biggl\lvert\det\left(\frac{\partial\boldsymbol{z}}{\partial\boldsymbol{y}}\right)\Biggl\rvert d\boldsymbol{\rho}d\boldsymbol{\Omega}\\
&\propto
\mathop{\int}
\left[\mathop{\prod}_{\varg}
\mathcal{P}\left(\boldsymbol{z}_{\varg}\;|\;\mathcal{D}_{\varg},\mathcal{M},\mathcal{I}_{\varg}\right)\right]
\Biggl\lvert\det\left(\frac{\partial\boldsymbol{z}}{\partial\boldsymbol{y}}\right)\Biggl\rvert d\boldsymbol{\rho}d\boldsymbol{\Omega},
\end{aligned}
\label{eqn:EP posterior groups}
\end{equation}
where the marginal posterior distributions under the product may be computed in parallel, with independent compute resources, by the $\varg^{th}$ group. There is also scope for cognitive parallelisation: whilst every group must construct a submodel using \textit{consistent} analytical exterior spacetime solutions (where there is freedom to truncate rotational deformations where appropriate), groups generally have the freedom to define unshared nuisance parameters and unshared hyperparameters in order to model observational phenomena. We illustrate this collaborative analysis in Fig.~\ref{fig:EP diagram}. We note that in general the marginal posterior distribution of exterior parameters $\boldsymbol{z}_{\varg}$ contributed by the $\varg^{th}$ group can naturally be the result of a series of Bayesian distributional updates, as suggested in Fig.~\ref{fig:EP diagram} and in analogy with the IP-paradigm discussed in Section \ref{sec:computational tractability}. These updates may in principle be applied by multiple groups (who, e.g., are experts in modelling different observational phenomena from the same star) who communicate distributions to one another, and these groups may accelerate their respective likelihood evaluation procedures as suggested in Section \ref{sec:computational tractability}. Nevertheless, there are many arrangements of the computation of the marginal posterior distribution of $\boldsymbol{z}$, and thus we do not attempt to enumerate the process beyond use of the label $\varg$.

Finally, given that separability of the marginal posterior distribution (of exterior parameters) over groups has been achieved, we remark on EOS parameter estimation. Each component posterior distribution of exterior parameters (the distribution of $\boldsymbol{z}_{\varg}$) needs to be communicated to the \textit{single} group who are to compute the marginal posterior distribution of EOS parameters by explicitly matching sets of stable interior solutions to sets of exterior solutions of the field equations. In order for the EP-paradigm EOS parameter estimation process to be simplified and accelerated, each communicated posterior distribution of exterior parameters may be some approximation to the full numerical distribution.

As we discuss in Section \ref{sec:EP alternatives}, however, provided a prior on a space of exterior parameters is noninformative relative to the likelihood function, a derived posterior distribution on that space is likelihood-dominated and can thus be used (in an approximative manner) in the IP-paradigm, wherein a prior is instead defined on a space of interior parameters. Therefore, the penultimate marginal distribution depicted in Fig. \ref{fig:EP diagram}, $\mathcal{P}(\boldsymbol{z}\;|\;\mathcal{D},\mathcal{M},\mathcal{I})$, could in principle be fed -- with instructions as discussed in Section~\ref{sec:computational tractability} -- to a \textit{single} group for use in the IP-paradigm as an effective \textit{nuisance-marginalised likelihood function}.

\subsection{Updating equation of state parameter knowledge}\label{sec:updating}
We have argued that the IP-paradigm is principled from the perspective of a Bayesian: inbuilt is the definition of a posterior as both an updated prior, and a prior to be updated. In Section \ref{sec:computational tractability} in particular, it is manifest how we should go about updating \textit{a posteriori} our knowledge of the EOS parameters $\boldsymbol{\theta}$. Conversely, it is not clear how we should update our (approximate) knowledge of $\boldsymbol{\theta}$ if the EP-paradigm is invoked because in general the relevant prior information would be defined on the space of $\boldsymbol{\theta}$.

Let us briefly consider several notable scenarios in which knowledge is to be updated \textit{a posteriori} in a future analysis, given a marginal posterior distribution of $\boldsymbol{\theta}$ \textit{defined} via Equation~(\ref{eqn:EP posterior groups}). First, suppose that some arbitrary subset of the exterior parameters $\boldsymbol{z}$ defined in Equation~(\ref{eqn:EP posterior groups}) are to be updated given some independent data set (see also Fig. \ref{fig:EP diagram}). We advocate that one should \textit{never} attempt to use a posterior distribution of interior parameters $\boldsymbol{y}$ -- the integrand of Equation~(\ref{eqn:EP posterior groups}) -- as a prior on a space of exterior parameters. Instead the existing marginal posterior distribution of the exterior parameters $\boldsymbol{z}$ must first be updated on the relevant subspace of $\mathbb{R}^{d}$, and only then can an ill-defined transformation onto the space $\mathbb{R}^{n}$ of EOS parameters be performed (the last stage illustrated in Fig. \ref{fig:EP diagram}). In other words, we should: (i) update the relevant posterior distributions of the exterior parameters $\boldsymbol{z}_{\varg}$ (refer to Fig. \ref{fig:EP diagram}); (ii) combine those distributions multiplicatively to form the marginal posterior distribution of $\boldsymbol{z}$; and (iii) only then define the marginal posterior distribution of $\boldsymbol{\theta}$ via Equation~(\ref{eqn:EP posterior groups}). It follows that (approximations to) the intermediary posterior distributions always need to be archived so that future updates can be applied where relevant, followed by subsequent (re)definition of the posterior distribution of $\boldsymbol{\theta}$.

Second, suppose that we aim to use constraints on additional exterior parameters which do not comprise the exterior parameters $\boldsymbol{z}$ initially considered; these additional parameters may be associated with an additional star added to the model ensemble to describe new data acquired through observations of a distinct star, for instance. Clearly this necessitates redefinition of the map $\boldsymbol{y}\mapsto\boldsymbol{z}$: the interior and exterior parameter dimensionalities need to match in order to admit a probability density transformation between spaces. Whilst the EOS model can be modified to depend on a greater number of EOS parameters, this notion is not useful if we aim to \textit{update} our knowledge of the previously applied model. On the other hand, if the EOS model is unmodified and the exterior solution of the additional star is to be described by \textit{greater} than two exterior parameters, one of the exterior parameters previously defined needs to be neglected; for instance, a star which previously contributed three exterior parameters must now be omitted from the analysis.

In summary, there is no sensible framework for updating our knowledge of the EOS parameters via the EP-paradigm because a prior exists on a space of interior parameters (either explicitly or implicitly, the latter via an ill-defined transformation) and there does not exist a diffeomorphic transformation between spaces of interior and exterior parameters (see Appendices~\ref{sec:injectivity} through \ref{sec:surjectivity}).

\section{Spectro-temporal X-ray modelling}\label{sec:appendix example}
The discussion in the body of this work has been deliberately general, with little reference to any of the astrophysical observables which encode information about the EOS. In order to assist the reader, in this Appendix we attempt to render the discussion less abstract by providing an example application -- i.e., a broad description of a model denoted by $\mathcal{M}$ (see Section \ref{sec:implicit}) in the context of techniques currently being used for EOS inference. We very briefly suggest what the relevant terminology (e.g., parametrised sampling distributions, nuisance parameters, and hyperparameters) could mean in this context. It is most natural to begin with a description of the data (\ref{sec:B1}). We then proceed to describe a model for the statistical properties of the data which arise due to the action of stochastic processes (\ref{sec:B2}). Finally, we describe typical model facets which relate parameters defined in underlying physical theories to the parameters of the probability distributions from which the data are assumed to be drawn (\ref{sec:B3}).

\subsection{The data}\label{sec:B1}
Let us consider time- and energy-resolved X-ray observations of stars which, under $\mathcal{M}$, are assumed to be compact and either isolated X-ray pulsars, or accreting matter from a low-mass binary companion star. An example of an operational instrument that could perform such observations is NICER \citep{Arzoumanian14}. Such sources are also a major target for future proposed large-area X-ray telescopes such as eXTP \citep[][]{Zhang16} and STROBE-X \citep{STROBEX}. We note that \textit{primary} observations of NICER are of isolated X-ray pulsars (rotation-powered millisecond pulsars), and most NICER papers to date have focussed on estimation of exterior spacetime parameters, namely the Schwarzschild gravitational masses and the equatorial radii of stars.

The raw astronomical data $\mathcal{D}_{i}$ acquired via such observations of the $i^{th}$ real star on the celestial sphere are on-board photon (count) arrival times distributed over a discrete set of output energy channels in the rest-frame of the detector. The set of photon arrival times and energies are collectively considered as a random variate drawn from some joint probability distribution (a joint sampling distribution), as we describe in Appendix~\ref{sec:B2}. Alternatively, the arrival times may be transformed into (rotation) phases conditional on some \textit{fixed} timing model for a periodic signal (known as phase-folding); the phases and output energy channels may be subsequently grouped into some fixed set of two-dimensional bins, and the count numbers across this set may be considered as a random variate drawn from some joint probability distribution (a joint sampling distribution).

\subsection{The sampling distribution of the data}\label{sec:B2}
A generative model for the data described in Appendix~\ref{sec:B1} is a parametrised sampling distribution of the data; in other words, a statistical model for data generation under the action of stochastic processes. In general there will be a sampling distribution for the data acquired from each star. The sampling distribution of the data from each star exhibits distributional moments (e.g., an expectation, variance, and so on); the values of these distributional moments dependent on model parameters (see Appendix~\ref{sec:B3}).

It is first necessary to define the statistical properties of the acquired data $\mathcal{D}_{i}$. Let us assume that a radiation field incident on a detector is quantum optically described as an inhomogeneous Poisson point process, such that individual photons in \textit{phase-space} are statistically independent, and the \textit{discrete} sampling distribution of the number of photons over some finite subdomain of phase-space is a Poisson distribution with expectation given by an integrated photon phase-space number density. The sampling distribution of the number of photon incidence events on a detector is a Poisson distribution whose expectation is an integral of the specific photon intensity over time, energy, solid angle (subtended on the celestial sphere of a point on the detector surface), and spatial coordinates on the detector surface; the specific photon intensity is the photon phase-space number density multiplied by the speed of light.

The response of an X-ray instrument to an incident radiation field does not typically require definition of nuisance parameters which describe the conversion of incident photons into \textit{counts} via generation of a current (any such parameters are typically constrained during detector testing and calibration phases and are subsequently fixed for analysis of science observations). The response is typically described in terms of the expected fractional redistribution of photons in finite-width input energy channels amongst finite-width output energy channels. Compact stars are unresolvable, so the specific photon flux (the integral of the specific intensity over solid angle) is uniform over an instrument-sized plane perpendicular to the direction to the source. The integral of the specific photon flux over input energy channels yields a discrete set of expected photon fluxes which are to be modulated by the instrument response matrix (which also accounts for the energy-dependent effective collecting area of the instrument). The output is the \textit{count} rate in each of a discrete set of energy channels; each count rate, integrated over some finite time interval, is the expectation of the Poissonian sampling distribution for that channel.

The photon arrival times in $\mathcal{D}_{i}$ are countably discrete due to dependence on the imperfect detector time-resolution; NICER, for instance, has exquisite timing capabilities, meaning that the expected count rates need not in general be \textit{integrated} over finite time intervals -- the read-out intervals are sufficiently small to be considered differential. In general, other detector imperfections such as dead-time and pile-up (dependent on time-resolution and expectation of the photon incidence rate over small regions of the detector) distort the Poissonian sampling distribution of the counts in each channel. In principle we may model such effects (with or without nuisance parameters), or we may simply assume that the distortions are negligible and consider a Poissonian sampling distribution.

\subsection{The dependence of the sampling distribution on model parameters}\label{sec:B3}
The sampling distribution of the data is conditional on a vector of model parameters which govern the statistics of the incident radiation field. The sampling distributions appearing in Equation~(\ref{eqn:exact posterior groups no nuisance separation}) are $\mathcal{P}(\mathcal{D}_{\varg}\;|\;\boldsymbol{\theta},\boldsymbol{\rho}_{\varg},\boldsymbol{\Omega}_{\varg},\boldsymbol{\eta}_{\varg},\mathcal{M})$, where the $\varg^{th}$ group models the observational data $\mathcal{D}_{i}$ associated with the $i^{th}$ star. Let us recall that: the $\varg^{th}$ group models the data $\mathcal{D}_{i}$ from the $i^{th}$ observed star, which requires construction of a \textit{model} star; $\boldsymbol{\rho}_{\varg}$ denotes the central \textit{energy} densities of \textit{model} stars considered by the $\varg^{th}$ group; $\boldsymbol{\Omega}_{\varg}$ denotes the coordinate angular rotation frequencies of \textit{model} stars considered by the $\varg^{th}$ group; $\boldsymbol{\eta}_{\varg}$ denotes the nuisance parameters considered by the $\varg^{th}$ group; and $\boldsymbol{\beta}_{\varg}$ denotes the nuisance hyperparameters considered by the $\varg^{th}$ group. Finally, $\boldsymbol{\theta}$ denotes the EOS parameters which are assumed to be shared by all \textit{model} stars.

As stated in Section \ref{sec:solutions for generative models}, an analytic parametrised exterior spacetime solution (which matches to stable interior spacetime solutions) is required for describing the propagation of radiation from the near vicinity of a model star to an instrument. For high-energy radiation we can invoke a geometrical optics approximation: the invariance in vacuum of the photon phase-space density along differential bundles of null geodesics which connect radiating source material to an instrument allows us to straightforwardly map the specific photon intensity at the source to the incident intensity \citep[see, e.g.,][]{Misner1973,Schneider1992}. In Sections~\ref{sec:solutions for generative models} and \ref{sec:mapping definition} we focussed on practical (perturbative) spacetime solutions, such as embedding a rotationally deformed, radiating 2-surface in a spherically symmetric ambient spacetime \citep[][]{Morsink2007}. \citet[][]{Bhattacharyya2005}, \citet[][]{Cadeau2007}, \citet[][]{Bauboeck2012}, \citet[][]{Psaltis2014}, \citet[][]{Nattila2017}, and \citet[][]{Vincent2018} improve on such an approximation, with the latter presenting a fully numerical treatment with the exact field equations. All of these frameworks could be consistently used in the IP-paradigm in principle, but, naturally, at increased computational expense.

Given a spacetime solution, we require a parametrised description of local radiative processes in order to compute incident specific photon intensities. Both exterior parameters (metric and surface parameters) and nuisance parameters will in general be necessary to model radiation transport in plasma both \textit{on} the model star and in its near vicinity. The radiation field at the surface (the photosphere) can be handled in numerous ways. A phenomenological description, for instance, may define nuisance parameters which everywhere control both the motion of photospheric material (relative to some rotation axis and thus a distant instrument) and the local comoving radiative properties at the photosphere. These nuisance parameters, together with exterior parameters, control the transformation of tensorial quantities between comoving orthonormal frames and Eulerian orthonormal frames -- in particular, relativistic beaming of radiation. Exterior parameters also control the null mapping from (spacetime) events at the photosphere to (spacetime) events at the instrument. Examples of relevant nuisance parameters can be found in, e.g., \citet[][]{Miller98}, \citet[][]{Poutanen2003}, \citet[][]{Bogdanov2013}, \citet[][]{Lo2013}, \citet[][]{Miller15}, and \citet[][]{Ozel2016}, who all focus on modelling rotationally pulsed radiative anisotropies at the surfaces of compact stars. Examples include: the inclination of the line-of-sight of the instrument to the rotational axis; the coordinate angular velocity of photospheric material; the distance between star and instrument; and parameters which control the comoving specific intensity emergent from the photosphere, such as those which describe the boundary of radiatively intense regions such as `hot-spots', and those which control the composition and thermodynamic state of radiating material.

A more physically self-consistent description, on the other hand, may treat the accretion process in terms of nuisance parameters \textit{and} exterior parameters, so the local comoving radiation field is also dependent on the spacetime. The additional modes of dependence of the sampling distribution of the data (on the exterior parameters) will in general affect the structure of the marginal likelihood function of the exterior parameters. The computational expense of likelihood evaluations will also in general be affected if the local comoving radiation field is not expressed analytically, but requires numerical evaluation itself.

Nuisance parameters are in general required to model the modulation of time-domain incident intensities due to the orbital dynamics of the source and detector. Additional nuisance parameters may be defined to model nuisance sources of radiation (backgrounds) that either: (i) do not encode information about the exterior parameters; or (ii) encode information about exterior parameters but the nature of the dependence is not sufficiently well understood for generative modelling without risk of statistical biasing (which can nevertheless be caused by poor phenomenological descriptions and thus sensitivity analyses are important). Such nuisance sources may be in the local vicinity of the star; in the local vicinity of the instrument; in the interstellar space between the star and the instrument; or maybe of a galactic or cosmological origin, but coincident with the star in the field of view of the instrument. Furthermore, the effect of interstellar X-ray absorption may be described with nuisance parameters.

Nuisance parameters and hyperparameters may be defined which are shared between stars or between nuisance sources of radiation, and may thus be shared between groups. Further, nuisance parameters may be defined in order to handle multiple data subsets assumed to be generated by different realisations of the same underlying phenomenon \citep[e.g., X-ray oscillations generated by distinct thermonuclear bursts separated in time by accretion phases;][]{Watts2012}; in this case a subset of parameters (including both interior parameters and nuisance parameters) may be shared by the distinct realisations, whilst subsets of nuisance parameters may be unique to a particular realisation, but associated with hyperparameters which are shared between realisations. Finally, nuisance parameters could in principle be tied to a particular instrument during some period of operation in which multiple stars were observed, and thus enter in parametrised sampling distributions applied to data acquired from those stars.


\bsp	
\label{lastpage}
\end{document}